\documentclass[fleqn,10pt]{wlscirep}
\usepackage{amssymb,amsmath,enumerate,amsmath,amssymb}
\usepackage{multirow}
\usepackage{bm}
\usepackage{epsfig}
\usepackage{multicol}        
\usepackage{url}
\usepackage{bbold}
\usepackage{url}
\usepackage{physics}
\usepackage{graphicx}
\usepackage{dcolumn}
\usepackage{bm}%
\usepackage{authblk}
\usepackage{bbold}
\usepackage{tabularx}
\usepackage{lscape}
\newcolumntype{b}{>{\hsize=1.45\hsize}X}
\newcolumntype{s}{>{\hsize=.15\hsize}X}
\newcolumntype{m}{>{\hsize=.7\hsize}X}

\title{Phase separation and scaling in correlation structures of financial markets}

\author[1,2,*]{Anirban Chakraborti}
\author[3]{Hrishidev}
\author[4]{Kiran Sharma}
\author[5]{Hirdesh K. Pharasi}

\affil[1]{School of Computational and Integrative Sciences, Jawaharlal Nehru University, New Delhi-110067, India}
\affil[2]{Centro Internacional de Ciencias, Cuernavaca-62210, M\'{e}xico}
\affil[3]{Indian Institute of Science Education and Research, Pune-411008, India}
\affil[4]{Chemical \& Biological Engineering, Northwestern University, Evanston, Illinois-60208, USA}
\affil[5]{Instituto de Ciencias F\'{i}sicas, Universidad Nacional Aut\'{o}noma de M\'{e}xico, Cuernavaca-62210, M\'{e}xico}
\affil[*]{anirban@jnu.ac.in}

\begin{abstract}

Financial markets, being spectacular examples of complex systems, display rich correlation structures among price returns of different assets. The correlation structures change drastically, akin to phase transitions in physical phenomena, as do the influential stocks (leaders) and sectors (communities), during market events like crashes. It is crucial to detect their signatures for timely intervention or prevention. Here we use \textit{eigenvalue decomposition} and \textit{eigen-entropy}, computed from \textit{eigen-centralities} of different stocks in the cross-correlation matrix, to extract information about the \textit{disorder} in the market. We construct a ‘phase space’, where different market events (bubbles, crashes, etc.) undergo \textit{phase separation} and display \textit{order-disorder transitions}. An entropy functional exhibits \textit{scaling behavior}. We propose a generic indicator that facilitates the continuous monitoring of the internal structure of the market – important for managing risk and stress-testing the financial system. Our methodology would help in understanding and foreseeing tipping points or fluctuation patterns in complex systems.
\end{abstract}

\begin{document}

\flushbottom
\maketitle

\thispagestyle{empty}

\paragraph{Introduction.---}Even before we could completely recover from the long-lasting effects of the global economic downturn in 2007-08  \cite{altman2009great}, we are threatened by another impending economic crisis that has been triggered by the coronavirus (COVID-19) pandemic. The last crisis  had brought us both predicament and hope! Predicament, since the traditional theories in economics could not predict, not even warn, the near complete breakdown of the global financial system. Hope, since one began to witness signs of change in economic and financial thinking, including the very fact that there is deeper (and less understood) link between macroeconomics and finance  \cite{Sharma_2017,Sharma_2019,Chakrabarti2020}, which certainly merits more attention.
Undoubtedly, the financial market serves as an ideal candidate for modeling  a complex system  \cite{boccara2010modeling,foote2007mathematics}, which is generally composed of many constituents of diverse forms and nature but largely interconnected, such that their strong inter-dependencies and emergent behavior  change with time. Thus, it becomes almost impossible to describe the dynamics of the complex system through some simple mathematical equations, and new tools and interdisciplinary approaches are much needed. Historically, financial markets have often exhibited sharp and largely unpredictable drops at a systemic scale-- `market crashes' \cite{Sornette_2004}. Such rapid changes or `phase transitions' (not in the strict thermodynamic sense \cite{stanley_1971,sethna2006statistical}) may be in some cases triggered by  unforeseen stochastic events or \textit{exogenous shocks} (e.g., coronavirus pandemic), or more often, they may be driven by certain  underlying \textit{endogenous processes} (e.g., housing bubble burst). These events are akin to phase transitions \cite{stanley_1971,sethna2006statistical,goldenfeld2018lectures} in physical phenomena, having interesting dynamics.
New insights and concepts, such as systemic risk, tipping points, contagion and network resilience have surfaced in the financial literature, prompting people to better monitor the highly interconnected macroeconomic and financial systems and, thus, anticipate future economic slowdowns or financial crises.

As a spectacular example of a complex system \cite{goldenfeld1999simple, arthur1999complexity}, the financial market \cite{Mantegna_2007,Bouchaud_2003,Sinha_2010,Chakraborti_2011a}  displays rich correlation structures \cite{Pharasi_2018,Pharasi_2019}, among price returns of different assets,  which have often been visualized as correlation-based networks \cite{Mantegna_2003,Onnela_2003,Mantegna_2010} with the identification of dominant stocks as influential leaders and sectors as communities \cite{fortunato2010community,Garlaschelli_prx_2015,Barabasi_2016}.  The correlation structures often change drastically, as do the leaders and communities in the market, especially during market events like crashes and bubbles \cite{Sornette_2004}. 
Therefore, the continuous monitoring of the complex structures of the market correlations becomes very crucial and practical \cite{Pharasi_2018,almog2019,Anindya_2019}. Recently, Pharasi et al.  \cite{Pharasi_2018,Pharasi_2019} used the tools of random matrix theory to determine market states and long-term precursors to crashes, and confirmed that during a market crash all the stocks behaved similarly such that the whole market acted like a single huge cluster or community. In contrast, during a bubble period,  a particular sector got overpriced or over-performed, causing accentuation of disparities among the various sectors or communities. However, there are no existing formal definitions of market crashes or bubbles; in fact, a certain arbitrariness exists in declaring a market event as a crash or bubble. Hence, it is extremely difficult to detect the signatures of these events so that we can timely intervene or prevent them. 

In this paper, we extract information about the \textit{disorder} in the market using the \textit{eigen-entropy} measure \cite{fan2017lifespan}, computed from the \textit{eigen-centralities} (ranks) \cite{Barabasi_2016} of different stocks in the market, and show for the first time that different market events (correlation structures) undergo \textit{phase separation} \cite{mazurin_1984,stanley_nature_1992} in a constructed  `phase space'. For the construction of the phase space we use transformed variables $|H-H_M|$ and $|H-H_{GR}|$, computed from the eigen-entropies [$H,H_M,H_{GR}$] following the eigenvalue decomposition of the correlation matrices ($C$) into the market modes ($C_M$) and the composite group plus random modes ($C_{GR}$). We further show that all market events, characterized by the [$H,H_M,H_{GR}$], are either `business-as-usual' periods (located towards the interior of the phase space) or `near-critical' events (located at the periphery). Thus,  one can see \textit{order-disorder} transitions as market events evolve in the phase space, as observed in critical phenomena of physical systems. For robustness, we chose two different financial markets-- the US S\&P-500 and Japanese Nikkei-225 over a 32-year period, and studied the evolution of the cross-correlation structures and their corresponding eigen-entropies. 
One of the relative entropy measures $H-H_M$ displays \textit{scaling} \cite{stanley1999scaling} behavior with respect to the mean market correlation $\mu$. Further, a functional of the relative entropy measure, $-\ln(H-H_M)$, acts as a good gauge of the market fear (volatility index $VIX$) \cite{vix}. Analogous to the black-hole entropy that reveals about the internal structure of a black-hole, our methodology with eigen-entropy (measure of market disorder) would also reveal the nature of internal market structure.  Further, the phase separation would help us to label the events as anomalies, bubbles, crashes, or other interesting type of events. It  would also provide a few generic indicators that would facilitate the continuous monitoring of the internal structure of the correlation epochs \cite{sandhu_2016,Pharasi_2018,Pharasi_2019,almog2019,Anindya_2019}. We anticipate that this new methodology would help us to better understand the internal market dynamics and characterize the events in different phases as anomalies, bubbles, crashes, etc., which could help in better risk management and portfolio optimization \cite{markowitz1952}. This could also be easily adapted and broadly applied to the studies of other complex systems such as in brain science \cite{fan2017lifespan} or environment \cite{Chakraborti_2020}.


\paragraph{Data.---}
\label{Sec:methods}
We have used the adjusted closure price time series for United States of America (USA) S\&P-500 index and Japan (JPN) Nikkei-225 index, for the period 02-01-1985 to 30-12-2016, from the Yahoo finance database (https://finance.yahoo.co.jp/; accessed on 7th July, 2017). USA has data for the $N=194$ stocks and the period 02-01-1985 to 30-12-2016 ($T=8068$ days). JPN has data for the $N=165$ stocks and the period 04-01-1985 to 30-12-2016 ($T=7998$ days). Note that we have included those stocks in our analyses, which are present in the data for the entire duration, and added zero return entries corresponding to the missing days. 
The list of stocks (along with the sectors) for the two markets and the sectoral abbreviations are given in the \textit{SI  } Tables~S1, and S2. 

We have also used the daily closure Volatility index ($VIX$) of the Chicago Board Options Exchange (CBOE) from Yahoo finance (https://finance.yahoo.co.jp/; accessed on 13th October, 2019) for the period 02-01-1990 to 30-12-2016, for $T=6805$ days. It acts as a popular measure of the expectation of volatility in the stock market implied by the S\&P-500 index options. It is computed and displayed on a real-time basis by the CBOE, and acts as the `fear index' or the `fear gauge' \cite{vix}.  
 
\begin{figure}[]
\centering
\includegraphics[width=0.9\linewidth]{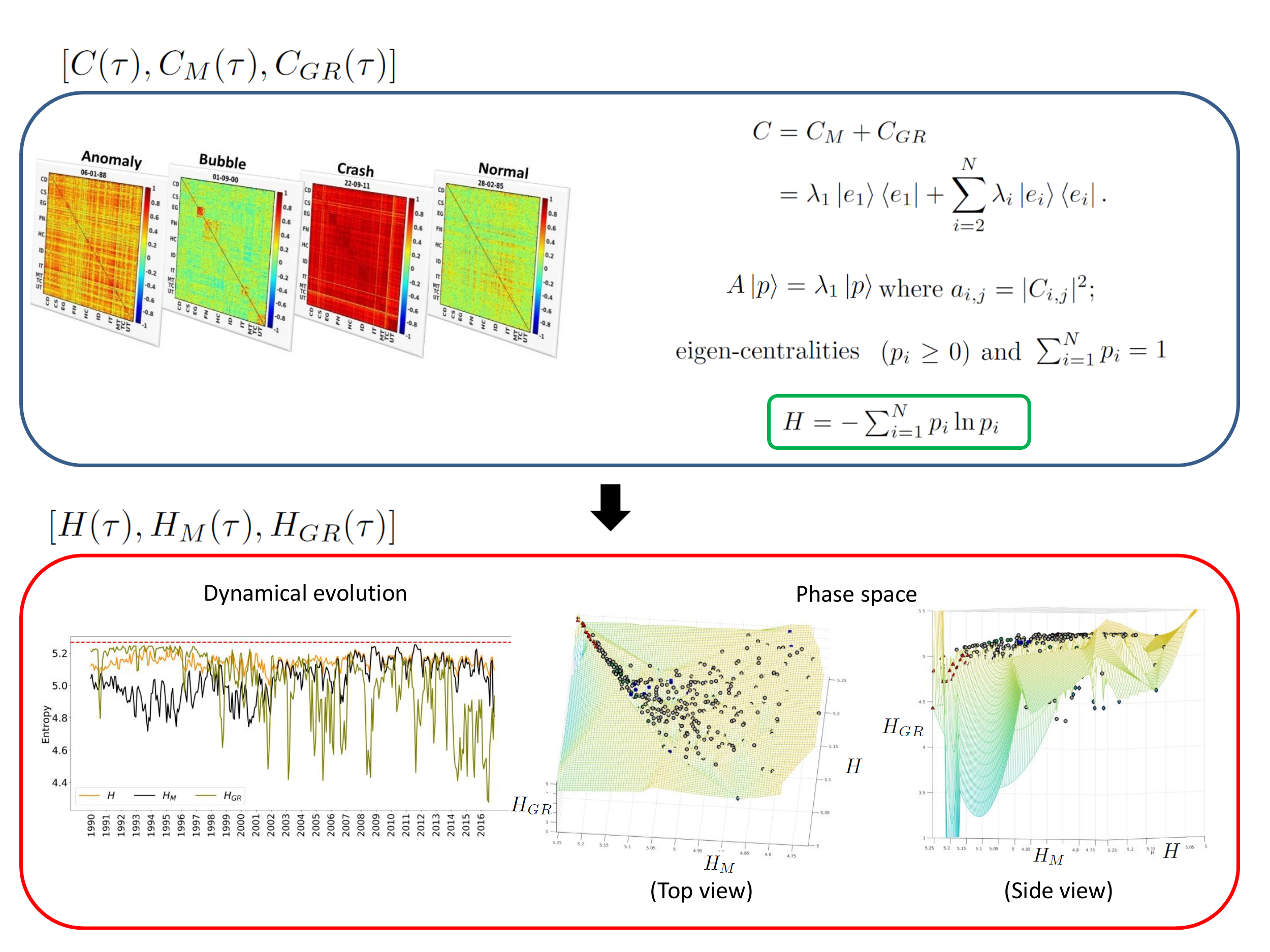}
\caption{\textbf{Schematic diagram of the construction of phase space}. Diagram explains the computation of eigen-entropies [$H, H_M, H_{GR}$], starting from each of the correlation matrices $C$ (four arbitrarily chosen dates). The correlation matrix $C$ is first decomposed to $C_M$ and $C_{GR}$ (see Methodology), and then the eigen-centralities $\lbrace p_i \rbrace $ are computed (from the corresponding maximum eigenvalues for each of these matrices). The eigen-entropies are computed as $-\sum_{i=1}^{N} p_i\ln p_i$. The coordinates [$H, H_M, H_{GR}$] can be used to study the market evolution, or characterization of the market events in the phase space diagrams (top and side views). }
\label{fig:schematic}
\end{figure}

\paragraph{Methodology.---}
The returns series are constructed as
$r_i( \tau ) = \ln P_i( \tau )- \ln P_i(\tau - \Delta)$, where $P_i(\tau)$ is the adjusted closure price of stock $i$ on day $\tau$, and $\Delta$ is the shift in days.
Instead of working with a long time series to determine the correlation matrix for $N$ USA stocks, we work with a \textit{short} time epoch of  $M$ days with a shift of $\Delta$ days. Then, the equal time Pearson correlation coefficients between stocks $i$ and $j$ is defined as
 $ C_{ij} (\tau) = (\langle r_i r_j \rangle - \langle r_i \rangle \langle r_j \rangle)/\sigma_i\sigma_j$, 
where $\langle...\rangle$ represents the expectation value computed over the time-epochs of size $M$ and the day ending on $\tau$, and $\sigma_k$ represents standard deviation of the $k$-th stock evaluated for the same time-epochs. We  use ${C} (\tau)$ to denote the return correlation matrix for the time-epochs ending on day $\tau$ (see e.g., Fig. \ref{fig:schematic}). Here, we show the results for $M = 40$ days with a shift of $\Delta=20$ days (other choices of $M$ and $\Delta$ in \textit{SI  } Fig.~S1).

For any given graph $ G:=(N,E)$ with $ |N|$ nodes and $|E|$ edges, let $ A=(a_{i,j})$ be the adjacency matrix, such that $ a_{i,j}=1$, if node $i$ is linked to node $j$, and $ a_{i,j}=0$ otherwise. The relative centrality $p_i$ score of node $i$ can be defined as:
$$ p_{i}={\frac {1}{\lambda }}\sum _{v\in M(i)}p_{j}={\frac {1}{\lambda }}\sum _{j\in G}a_{i,j}p_{j},$$
where $ M(i)$ is a set of the neighbors of node $i$ and $ \lambda$  is a constant. With a small mathematical rearrangement, this can be written in vector notation as the eigenvector equation ${A}\ket{p} =\lambda \ket{p}$.
In general, there may exist many different eigenvalues $\lambda $  for which a non-zero eigenvector solution $\ket{p}$ exists. 
We use the characteristic equation $ |{A} - \lambda \mathbb{1} | = 0$ to compute the eigenvalues $\lbrace \lambda_1, ..., \lambda_N \rbrace$. 
However, the additional requirement that all the entries in the eigenvector be non-negative ($p_i \geq 0$) implies (by the Perron-Frobenius theorem) that only the maximum eigenvalue ($\lambda_{max}$) results in the desired centrality measure. The $ i^{\text{th}}$ component of the related eigenvector then gives the relative \textit{eigen-centrality} score of the node $ i$ in the network. However, the eigenvector is only defined up to a common factor, so only the ratios of the centralities of the nodes are well defined. To define an absolute score one must \textit{normalise} the eigenvector, such that the sum over all nodes $N$ is unity, i.e., $ \sum_{i=1}^{N} p_i =1$. Furthermore, this can be generalized so that the entries in ${A}$ can be any matrix with real numbers representing the connection strengths. For correlation matrices ${C} (\tau)$, in order to enforce the Perron-Frobenius theorem, we work with $ a_{i,j} = |{C_{i,j}}|^n$, where $i,j=1,\dots,N$ and $n$ is any positive integer (we have used $n=2$ in the paper; other values are discussed in \textit{SI  }; see also \textit{SI  } Fig.~S2).

\begin{figure*}[]
\centering
\includegraphics[width=0.8\linewidth]{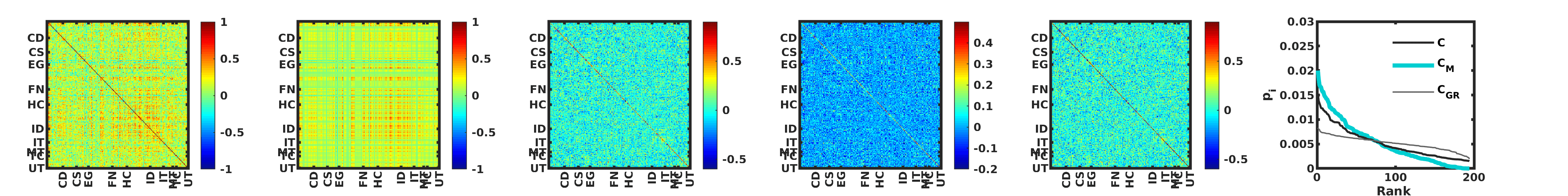}
\llap{\parbox[b]{10.65in}{\textsf{$C$}\\\rule{0ex}{0.7in}}}	\llap{\parbox[b]{8.9in}{\textsf{$C_M$}\\\rule{0ex}{0.7in}}}\llap{\parbox[b]{7.1in}{\textsf{$C_G$}\\\rule{0ex}{0.7in}}}\llap{\parbox[b]{5.25in}{\textsf{$C_R$}\\\rule{0ex}{0.7in}}}\llap{\parbox[b]{3.45in}{\textsf{$C_{GR}$}\\\rule{0ex}{0.7in}}}
\includegraphics[width=0.8\linewidth]{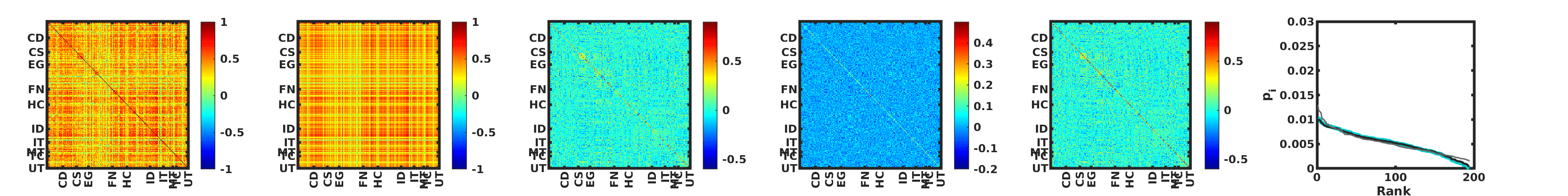}
\includegraphics[width=0.8\linewidth]{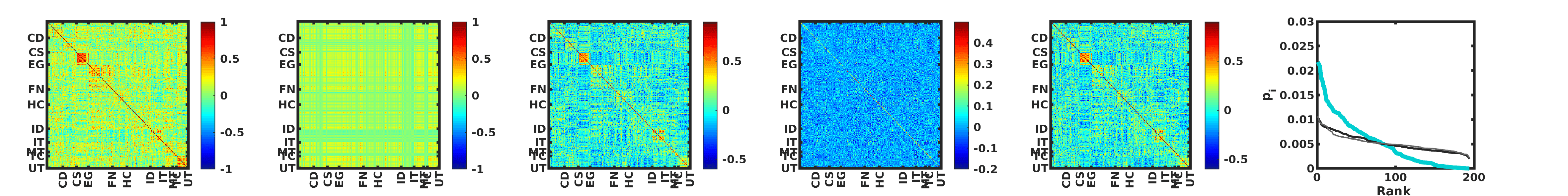}
\includegraphics[width=0.8\linewidth]{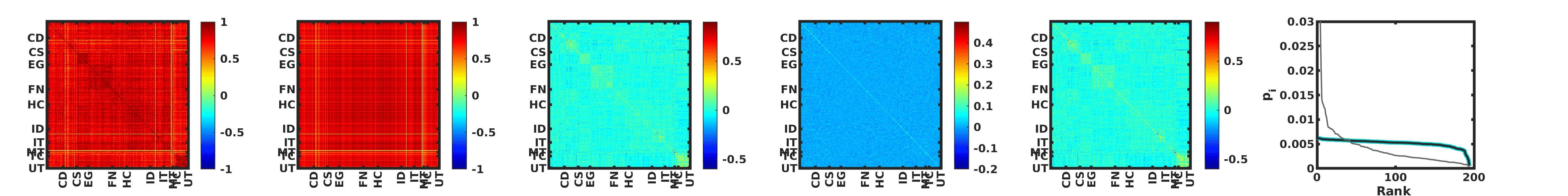}
\includegraphics[width=0.8\linewidth]{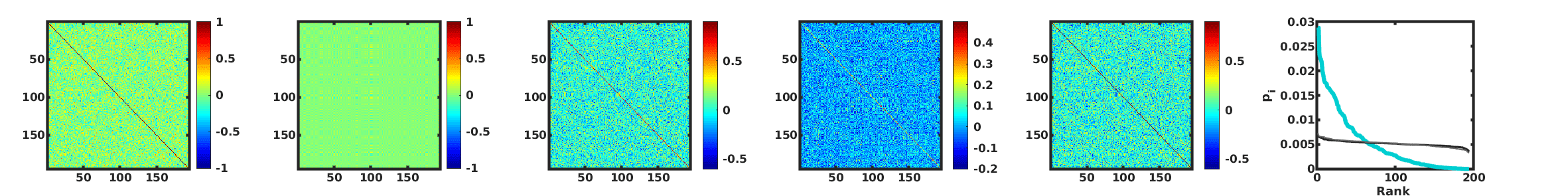}
\caption{\textbf{Eigenvalue decomposition of the correlation matrices, and ranked eigen-centralities.} Plots showing the correlation matrices: (\textit{Left to Right}) full $C$, market mode $C_M$, group mode $C_G$, random mode $C_R$, group-random mode $C_{GR}$ and the ranked eigen-centralities $\lbrace p_i \rbrace $ of the different correlation modes: full ($C$ in black curve), market mode ($C_M$ in turquoise curve) and group-random mode ($C_{GR}$ in grey curve); stocks are arranged sector-wise (names abbreviated). (\textit{Top to Bottom}) Matrices corresponding to normal, anomalous, type-1, crash periods of the financial market, and a random matrix taken from WOE.}
\label{fig:decomposition}
\end{figure*}


The correlation matrix of size $N \times N$ will have $N$ eigenvalues, say $\lbrace \lambda_1, ..., \lambda_N \rbrace$, which may be arranged in descending order of magnitude. Then the maximum eigenvalue $\lambda_1 = \lambda_{max}$ of the correlation matrix $C$, corresponds to a market mode $C_M$ that reflects the aggregate dynamics of the market common across all stocks, and strongly correlated to the mean market correlation $\mu$. The group modes $C_G$ capture the sectoral behavior of the market, which are next few eigenvalues subsequent to the largest eigenvalue of the correlation matrix. Remaining eigenvalues capture the random modes $C_R$ of the market. By using the eigenvalue decomposition, we can thus filter the true correlations (coming from the signal) and the spurious correlations (coming from the random noise). 
Therefore, we may decompose the aggregate correlation matrix as $
C = \sum_{i=1}^{N} \lambda_{i}\ket{e_{i}}\bra{e_{i}} 
$, where $\lambda_{i}$ and $\ket{e_{i}}$ are the eigenvalues and eigenvectors, respectively. 
Traditionally, one decomposes the matrix into three separate components, viz., market mode $C_M$, the group modes $C_G$ and the random modes $C_R$, as shown in Fig. \ref{fig:decomposition}:
\begin{eqnarray}
C &=& C_{M} + C_{G} + C_{R},\\
  &=& \lambda_{1}\ket{e_{1}}\bra{e_{1}} + \sum_{i=2}^{N_{G}} \lambda_{i}\ket{e_{i}}\bra{e_{i}} + \sum_{i=N_{G}+1}^{N} \lambda_{i}\ket{e_{i}}\bra{e_{i}} , \nonumber
\end{eqnarray}
where $N_{G}$ is the number of eigenvalues that satisfy the constraint $\lambda_{+}\leq\lambda_{G}\textless\lambda_1$, with
$\lambda_{+}= \sigma^2 \left(1+ \frac{1}{\sqrt{Q}}\right)^2 .$

For empirical matrices, it is often very difficult to determine the exact value of $\lambda_+$ and hence figure out $N_G$, for which the eigenvectors from 2 to $N_{G}$ would describe the sectoral dynamics. Here, we choose $N_G=20$ arbitrarily for the correlation decomposition (Fig.~\ref{fig:decomposition}), corresponding to the  $20$ largest eigenvalues after the largest one.  
In order to avoid this arbitrary choice of $N_G$, we prefer to  decompose the correlation matrix into the market mode $C_M$ and the composite group plus random mode $C_{GR}$:
\begin{eqnarray}
C &=& C_{M} + C_{GR}\\
  &=& \lambda_{1}\ket{e_{1}}\bra{e_{1}} + \sum_{i=2}^{N} \lambda_{i}\ket{e_{i}}\bra{e_{i}} . \nonumber
\end{eqnarray}

Following the tradition in information theory, we use the eigen-entropy $ H= - \sum_{i=1}^{N} p_i \ln p_i $, since all the normalised eigen-centralities are non-negative ($p_i \geq 0$) and $ \sum_{i=1}^{N} p_i =1$, by construction. The eigen-entropy may be described as kind of measure of disorder in the matrix ${A}$, where $a_{i,j}= |{C_{i,j}}|^2$; higher the eigen-entropy, higher is the disorder in the matrix; the highest being in the case of WOE (Wishart Orthogonal Ensemble) , where $H \sim \ln N$. 
For empirical correlation matrices, the eigen-entropy will be bounded by these two limits [$0,\ln N$].

Similarly, corresponding to $C_{M}$ and $C_{GR}$, we can compute $H_M$ and $H_{GR}$, respectively. Thus, from each cross-correlation matrix $C(\tau)$, we can use the eigenvalue decomposition to construct the set of matrices [$C(\tau),C_M(\tau),C_{GR}(\tau)$], and then construct the set of phase space coordinates [$H(\tau),H_M(\tau),H_{GR}(\tau)$], as illustrated in Fig.\ref{fig:schematic}.


\paragraph{Results.---} 

Fig.~\ref{fig:decomposition} shows the eigenvalue decompositions of the correlation matrices, for: (\textit{Top to Bottom}) normal, anomalous, type-1 event, crash, and WOE. We have denoted the different matrices as:  full correlation $C$, market mode $C_M$, group mode $C_G$, random mode $C_R$, group-random mode $C_{GR}$ and displayed the results in Fig.~\ref{fig:decomposition} (\textit{Left to Right}). The last column shows the results for the ranked eigen-centralities $p_i$ of the different correlation modes: full ($C$ in black curve), market mode ($C_M$ in turquoise curve) and group-random mode ($C_{GR}$ in grey curve). 
Evidently, the internal structure of the cross-correlation matrix changes a lot with time, and causes the change in the importance/hierarchy of the stocks (leaders) and block structures (communities). This further changes the eigen-entropies [$H(\tau), H_M(\tau), H_{GR}(\tau)$] that are used to create a phase space where each frame is represented by a point. As time evolves, different parts of the phase space are occupied and this allows us to identify certain phases (restricted to some regions) and  characterize the market events as crashes, etc. (see Supplementary Videos 1, 2).
Interestingly, for a normal period, the three curves are distinct and there are
hierarchies in ranks in all curves; for the market anomaly, all the three curves almost coincide; in the interesting type-1 period (classified due to the position of the point in certain region of phase space), the curves
corresponding to the full and the group-random modes coincide while there is a strict hierarchy in the eigen-centralities of the
market mode; for crash period, the curves corresponding to the full and the market modes coincide while there is a strict
hierarchy in the eigen-centralities of the group-random mode; and for the WOE (without internal structure), once again the curves corresponding to the full
and the group-random modes coincide while there is a strict hierarchy in the eigen-centralities of the market mode. 

As seen in Fig. \ref{fig:schematic}a, the events in phase space [$H, H_M, H_{GR}$] appear to be scattered on a complicated manifold (created by MATLAB surface interpolation).
However, there appears to be some segregation of events. We plotted the events (see Fig. S4) in another phase space with transformed variables [$H-H_M, H_M-H_{GR}, H-H_{GR}$]; the points  are then seen to lie on a flat surface, and with better segregation. The time-evolution of the transformed variables [$H-H_M, H_M-H_{GR}, H-H_{GR}$] show interesting dynamics.
The characterized events (\textit{SI  } Fig.~S4b, d) are indicated as vertical lines in the time-evolution plots (\textit{SI  } Fig.~S4a, c; Supplementary Videos 1, 2). We found that many anomalies occurred just around the major crashes and intriguing patterns (termed as interesting events of type-1 and type-2, belonging to two distinct regions in the phase space) appeared. The crashes occupy the region in the phase space, where $H - H_M \simeq 0$. During the crashes, the $H$ and $H_M$ almost touch the maximum disorder, $\ln N$ (corresponding to the random WOE).
The events like `Dot-com bubble' that appear in the $H - H_{GR} \simeq 0$ axis are termed as interesting events of type-1. 
The events which lie far away from the origin and both the axes, are termed as interesting events of type-2, which include frames with exogenous shocks (like Hurricane Katrina, etc.). The events lying close to the origin are like anomalies happening right before or right after major crashes.


\begin{figure*}[]
\centering
\includegraphics[width=0.28\linewidth]{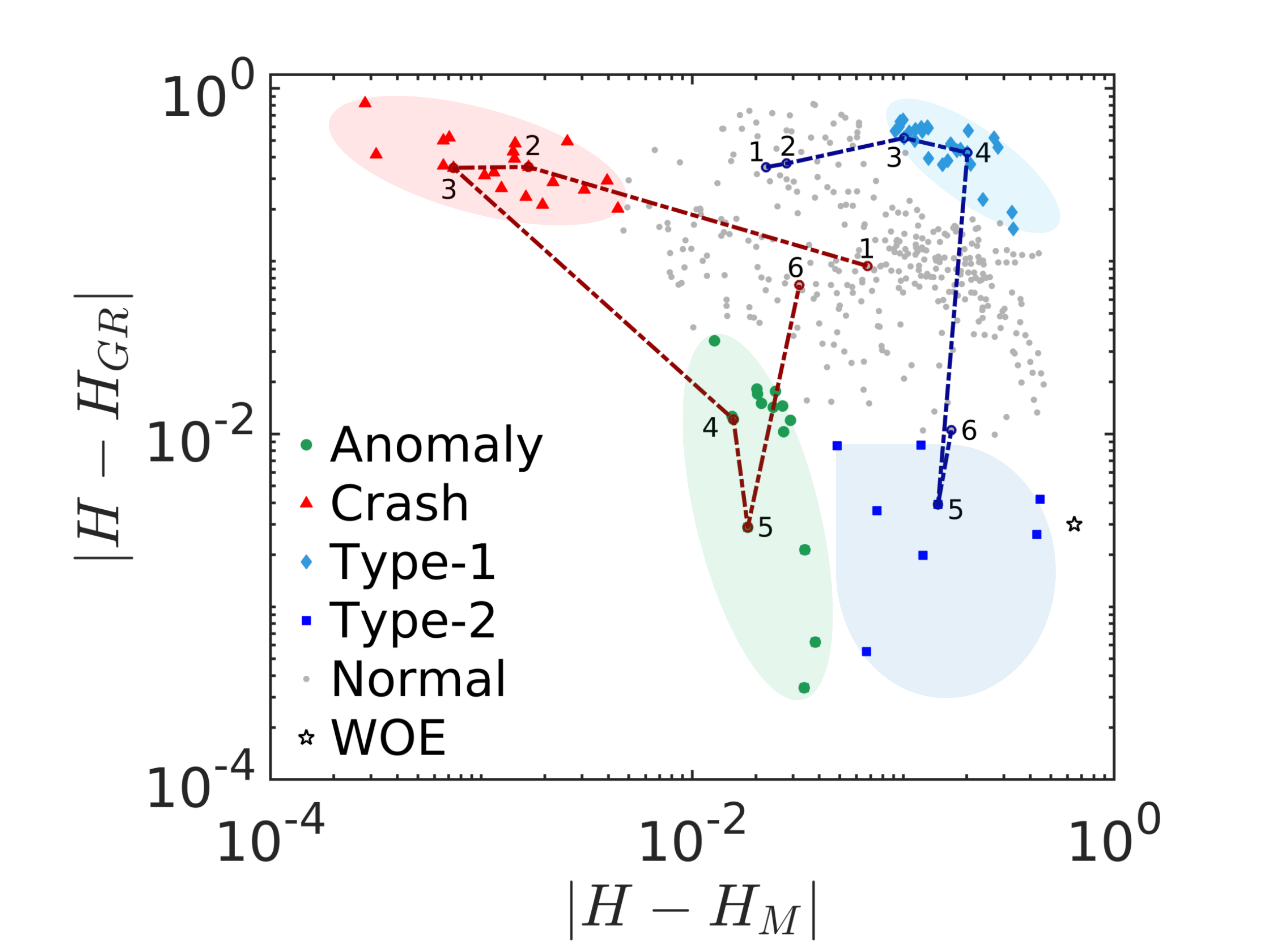}
\llap{\parbox[b]{3.9in}{\textsf{\textbf{a}}\\\rule{0ex}{1.3in}}}
\includegraphics[width=0.28\linewidth]{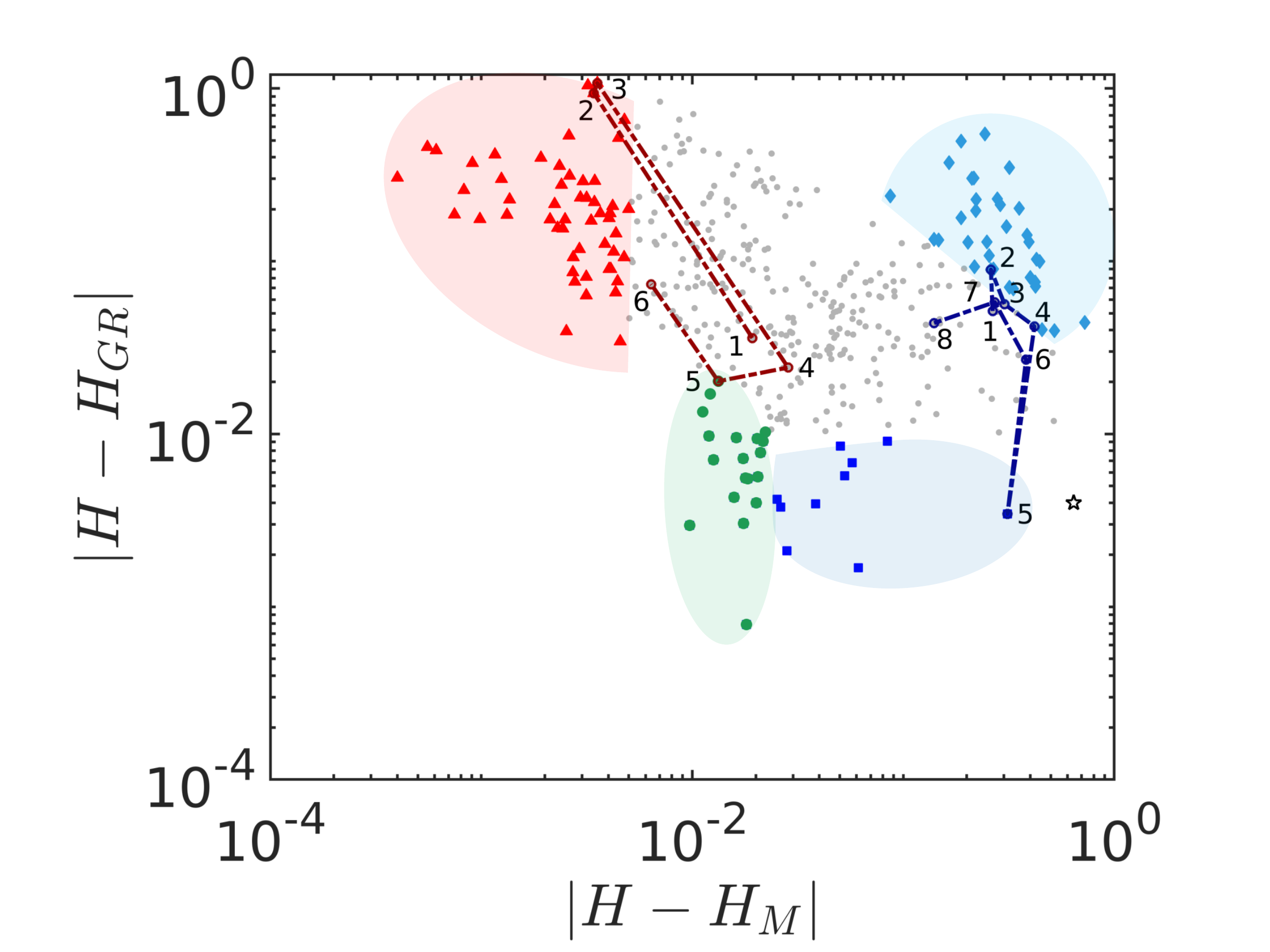}
\llap{\parbox[b]{3.9in}{\textsf{\textbf{b}}\\\rule{0ex}{1.3in}}}
\includegraphics[width=0.28\linewidth]{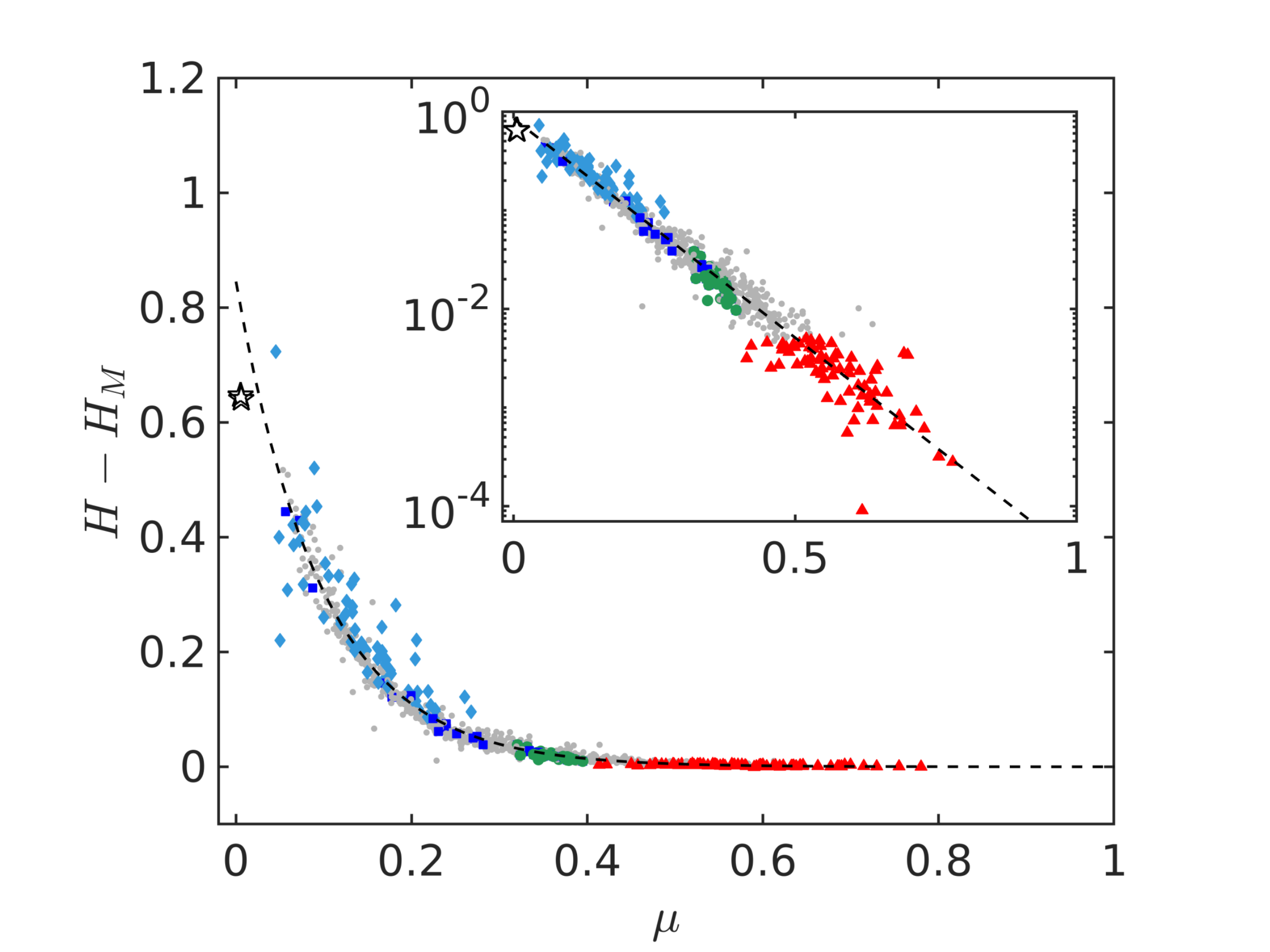}
\llap{\parbox[b]{3.9in}{\textsf{\textbf{c}}\\\rule{0ex}{1.3in}}}
\caption{\textbf{Phase separation, order-disorder transitions and scaling behavior.} Panels \textbf{a-b} show  plots of $|H-H_M|$ and $|H-H_{GR}|$ for S\&P-500 and Nikkei-225 markets, respectively, where the events show clear phase separation with order-disorder transitions (red dash-dot line connecting sequence of events around a crash, and blue dash-dot line connecting sequence of events around a bubble).  Panel \textbf{c} plots $H-H_M$ versus mean market correlation $\mu$  for both markets (Inset: The same in linear-logarithmic scale). The data-collapse indicates a scaling behavior. 
For all panels, the black stars represent WOE.}
\label{fig:scaling}
\end{figure*}

\begin{figure}[]
\centering
\includegraphics[width=0.85\linewidth]{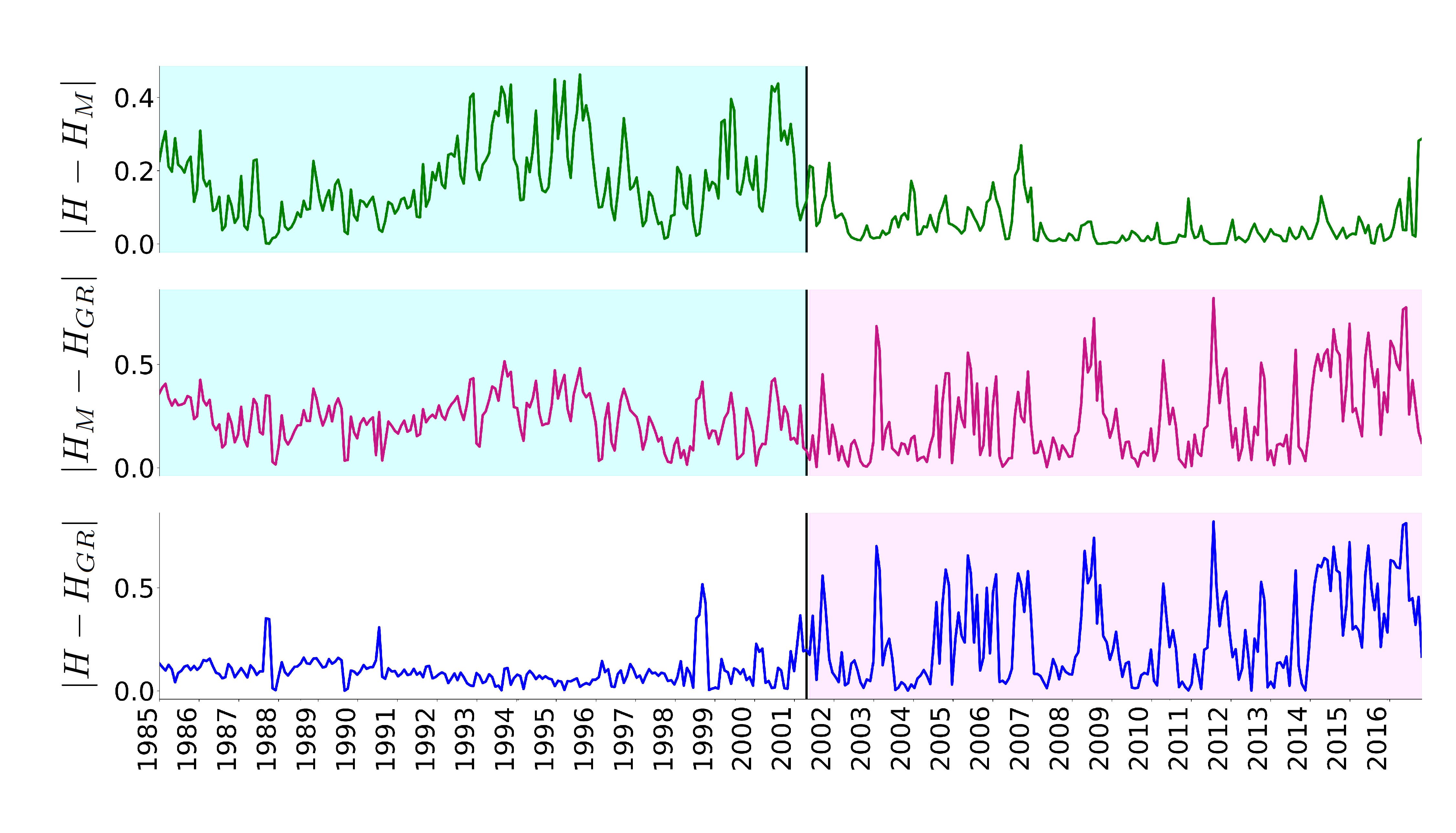}
\caption{\textbf{Evolution of S\&P-500 market}. Plots of:  (\textit{Top to Bottom}) $|H-H_M|$,$|H_M-H_{GR|}$ and $|H-H_{GR}|$. The blue and pink bands show time-periods over which the variables display high correlations ($\sim 0.71$ and $\sim 0.96$, respectively).}
\label{fig:entropy_dynamics}
\end{figure}

The above interesting features led us to try the transformed variables $|H-H_M|$ and $|H-H_{GR}|$ as independent coordinates of phase space. Very interestingly, as evident from Figure~\ref{fig:scaling}a, b, the event frames show clear phase separation-- anomalies (green region), crashes (red region), normal (grey), type-1 (light blue region) and type-2 (deep blue region), for both  S\&P-500 and Nikkei-225 markets.  The order-disorder transitions-- normal (at the central region) to near-critical phases (at the peripheral regions) are intriguing.
We have also studied in detail the sequence of seven frames (three frames before, the event (in black), and three frames after) to follow the order-disorder transitions (\textit{SI  } Figs.~S5, S6) in cases of major crashes and bubbles (\textit{SI  } Table~S1). The similar nature of the order-disorder transitions in all the major crashes and Dot-com bubbles, ten events in USA and thirteen events in JPN, certainly indicate robustness of the method.
Moreover, we found that $(H-H_M) \sim \alpha \exp(- \beta \mu)$, where $\alpha$ and $\beta$ are constants (see Fig.~\ref{fig:scaling}c for USA and JPN). We found that the best-fit line yields $\alpha \simeq 0.85 \pm 0.03$ and $\beta \simeq  -10.22 \pm 0.25$; adjusted $R^2=0.95$. Interestingly, the market event frames segregate into different portions, interspersed by the normal events. This data-collapse on a single curve indicates a scaling behavior \cite{stanley1999scaling}, which implies that the co-movements in price returns for different financial assets and varying across countries are governed by the same statistical law-- certainly non-trivial and striking behavior! 
This suggests that markets have an inherent structure that remains pretty invariant-- it has an average structure with fluctuations (dispersion). The dispersion around the average behavior is slightly more in JPN than USA.  
The phase properties are found to be pretty robust, though the phase boundaries are not very sharp (and may depend on the parameters like window choice, shift, etc.; \textit{SI  } Fig.~S1, S2). All frames in a certain phase have very similar properties (hierarchies in ranks of stocks) and can be averaged over to represent a certain phase (\textit{SI  } Fig.~S7). The properties remain similar across different markets (USA and JPN) and across various periods of time.   In all the panels, the black stars signify results for WOE, which has no inherent structure (see \textit{SI  } Figs. S3, S7). One could also simulate (to be reported elsewhere) various correlation structures  from a correlated WOE with the mean correlation as tuning parameter. The non-trivial inherent market structure (sectors or communities) plays a crucial role in the observed scaling behavior. We also observed from the evolution of the variables $|H-H_M|$,$|H_M-H_{GR|}$ and $|H-H_{GR}|$ (Fig. \ref{fig:entropy_dynamics}), and other market indicators (\textit{SI} Figs.~S8 and S9) that the market behavior has changed radically after 2002 (USA) and 1990 (JPN) corroborating the findings of our earlier work \cite{Pharasi_2018}.

\begin{figure}[]
\centering
\includegraphics[width=0.95\linewidth]{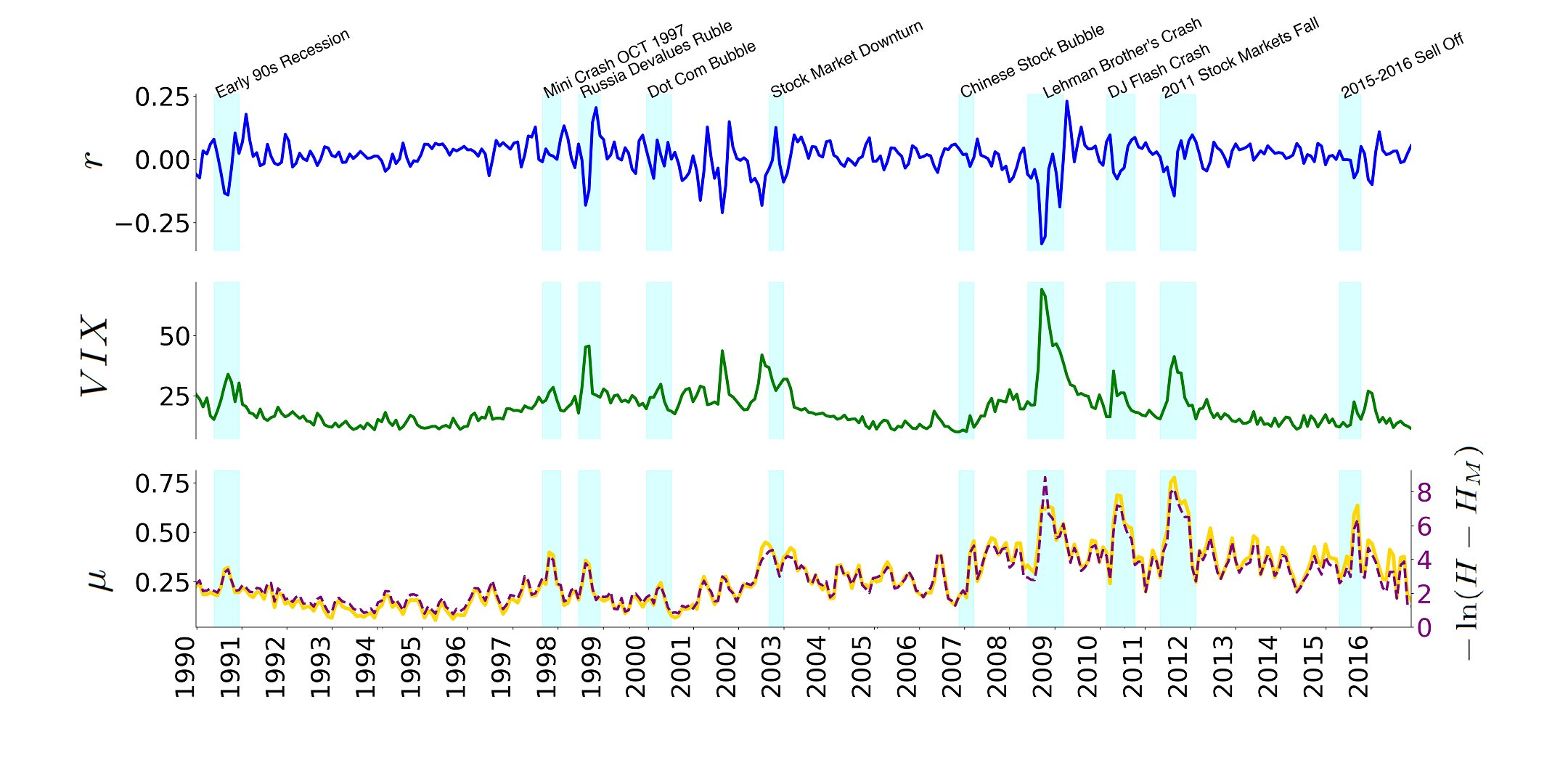}
\caption{\textbf{Evolution of S\&P-500 market indicators}. Plots of: (\textit{Top to Bottom}) index returns $r$, volatility index ($VIX$), mean market correlation $\mu$ and the functional $-\ln(H-H_M)$, for 1990-2016. Light blue bands are representative critical periods (see \textit{SI  } Table~S3).
}
\label{fig:indicator}
\end{figure}

Finally, the functional $-\ln(H-H_M)$ is found to act as a good gauge of the market characteristic ($\mu$) and  market fear ($VIX$) (Fig.~ \ref{fig:indicator}). There exist significant and non-trivial correlations between these variables, and the other market indicators (\textit{SI} Fig.~S10). Hence, this functional $-\ln(H-H_M)$ can serve as a very good generic indicator. 

\paragraph{Discussions.---}
We emphasize that our eigen-entropy measure has a few advantages--uniquely determined, non-arbitrary, computational cheap (low complexity), when compared to existing methods, e.g., structural entropy  \cite{almog2019}. Note that the structural entropy (or any other network-based entropy measures) is very sensitive to the community structure and construction of the network. An algorithm  \cite{Garlaschelli_prx_2015} involves identifying the group mode from the correlation matrix, which may be hard and non-arbitrary (the boundary determined by the eigenvalues of the correlation matrix is not sharp). Finally, we reiterate that this type of phase separation behavior has never been recorded for financial markets; it is very distinct from the two-phase behavior in financial markets reported earlier by Plerou et al.  \cite{Plerou_2003}. The data collapse certainly suggests that the fluctuations in price returns for different financial assets, varying across countries, economic sectors and market parameters, are governed by the same statistical law. This  scaling behavior may motivate us to do further research as to determine which market forces are responsible for driving the market or are important for determining the price co-movements and correlations.
In addition, this may lead to a foundation for understanding scaling in a broader context, and providing us with altogether new concepts not anticipated previously.
Our proposed methodology may further help us to understand the market events and their dynamics, as well as find the time-ordering and appearances of the bubbles (formations or bursts) and crashes, separated by normal periods. This methodology may be generalized and used in other complex systems (to be reported elsewhere) to understand and foresee tipping points and fluctuation patterns.

\section*{Acknowledgements}The authors are grateful to A.S. Chakrabarti, F. Leyvraz and T.H. Seligman for their critical inputs and suggestions. H.K.P. is grateful for financial support provided by UNAM-DGAPA and CONACYT Proyecto Fronteras 952. A.C. and K.S. acknowledge support from the project UNAM-DGAPA-PAPIIT AG 100819 and CONACyT Project FRONTERAS 201.
\bibliography{nature_main}

\begin{thebibliography}{10}
\expandafter\ifx\csname url\endcsname\relax
  \def\url#1{\texttt{#1}}\fi
\expandafter\ifx\csname urlprefix\endcsname\relax\def\urlprefix{URL }\fi
\expandafter\ifx\csname doiprefix\endcsname\relax\def\doiprefix{DOI }\fi
\providecommand{\bibinfo}[2]{#2}
\providecommand{\eprint}[2][]{\url{#2}}

\bibitem{altman2009great}
\bibinfo{author}{Altman, R.~C.}
\newblock \bibinfo{title}{The great crash, 2008}.
\newblock \emph{\bibinfo{journal}{Foreign Aff.}} \textbf{\bibinfo{volume}{88}},
  \bibinfo{pages}{1} (\bibinfo{year}{2009}).

\bibitem{Sharma_2017}
\bibinfo{author}{Sharma, K.}, \bibinfo{author}{Gopalakrishnan, B.},
  \bibinfo{author}{Chakrabarti, A.~S.} \& \bibinfo{author}{Chakraborti, A.}
\newblock \bibinfo{title}{Financial fluctuations anchored to economic
  fundamentals: A mesoscopic network approach}.
\newblock \emph{\bibinfo{journal}{Scientific Reports}}
  \textbf{\bibinfo{volume}{7}}, \bibinfo{pages}{8055} (\bibinfo{year}{2017}).

\bibitem{Sharma_2019}
\bibinfo{author}{Sharma, K.}, \bibinfo{author}{Chakrabarti, A.~S.} \&
  \bibinfo{author}{Chakraborti, A.}
\newblock \bibinfo{title}{Multi-layered network structure: Relationship between
  financial and macroeconomic dynamics}.
\newblock In \emph{\bibinfo{booktitle}{New Perspectives and Challenges in
  Econophysics and Sociophysics}}, \bibinfo{pages}{117--131}
  (\bibinfo{publisher}{Springer}, \bibinfo{year}{2019}).

\bibitem{Chakrabarti2020}
\bibinfo{author}{Chakrabarti, A.~S.}, \bibinfo{author}{Chakraborti, A.} \&
  \bibinfo{author}{Upmanyu, S.}
\newblock \bibinfo{title}{Macroeconomic and financial networks: review of some
  recent developments in parametric and non-parametric approaches}
  (\bibinfo{publisher}{Proceedings of the International Conference of Game
  Theory and Networks-ICGTN 2019, Eds: Borkotokey et al.},
  \bibinfo{year}{2020}).
\newblock \urlprefix\url{https://ssrn.com/abstract=3557742}.

\bibitem{boccara2010modeling}
\bibinfo{author}{Boccara, N.}
\newblock \emph{\bibinfo{title}{Modeling complex systems}}
  (\bibinfo{publisher}{Springer Science \& Business Media},
  \bibinfo{year}{2010}).

\bibitem{foote2007mathematics}
\bibinfo{author}{Foote, R.}
\newblock \bibinfo{title}{Mathematics and complex systems}.
\newblock \emph{\bibinfo{journal}{Science}} \textbf{\bibinfo{volume}{318}},
  \bibinfo{pages}{410--412} (\bibinfo{year}{2007}).

\bibitem{Sornette_2004}
\bibinfo{author}{Sornette, D.}
\newblock \emph{\bibinfo{title}{Why stock markets crash: Critical events in
  complex financial systems}} (\bibinfo{publisher}{Princeton University Press},
  \bibinfo{year}{2004}).

\bibitem{stanley_1971}
\bibinfo{author}{Stanley, H.~E.}
\newblock \emph{\bibinfo{title}{Phase transitions and critical phenomena}}
  (\bibinfo{publisher}{Clarendon Press, Oxford}, \bibinfo{year}{1971}).

\bibitem{sethna2006statistical}
\bibinfo{author}{Sethna, J.}
\newblock \emph{\bibinfo{title}{Statistical mechanics: Entropy, order
  parameters, and complexity}}, vol.~\bibinfo{volume}{14}
  (\bibinfo{publisher}{Oxford University Press}, \bibinfo{year}{2006}).

\bibitem{goldenfeld2018lectures}
\bibinfo{author}{Goldenfeld, N.}
\newblock \emph{\bibinfo{title}{Lectures on phase transitions and the
  renormalization group}} (\bibinfo{publisher}{CRC Press},
  \bibinfo{year}{2018}).

\bibitem{goldenfeld1999simple}
\bibinfo{author}{Goldenfeld, N.} \& \bibinfo{author}{Kadanoff, L.~P.}
\newblock \bibinfo{title}{Simple lessons from complexity}.
\newblock \emph{\bibinfo{journal}{Science}} \textbf{\bibinfo{volume}{284}},
  \bibinfo{pages}{87--89} (\bibinfo{year}{1999}).

\bibitem{arthur1999complexity}
\bibinfo{author}{Arthur, W.~B.}
\newblock \bibinfo{title}{Complexity and the economy}.
\newblock \emph{\bibinfo{journal}{Science}} \textbf{\bibinfo{volume}{284}},
  \bibinfo{pages}{107--109} (\bibinfo{year}{1999}).

\bibitem{Mantegna_2007}
\bibinfo{author}{Mantegna, R.~N.} \& \bibinfo{author}{Stanley, H.~E.}
\newblock \emph{\bibinfo{title}{An introduction to econophysics: Correlations
  and complexity in finance}} (\bibinfo{publisher}{Cambridge University Press,
  Cambridge}, \bibinfo{year}{2007}).

\bibitem{Bouchaud_2003}
\bibinfo{author}{Bouchaud, J.-P.} \& \bibinfo{author}{Potters, M.}
\newblock \emph{\bibinfo{title}{{Theory of financial risk and derivative
  pricing: From statistical physics to risk management}}}
  (\bibinfo{publisher}{Cambridge University Press}, \bibinfo{year}{2003}).

\bibitem{Sinha_2010}
\bibinfo{author}{Sinha, S.}, \bibinfo{author}{Chatterjee, A.},
  \bibinfo{author}{Chakraborti, A.} \& \bibinfo{author}{Chakrabarti, B.~K.}
\newblock \emph{\bibinfo{title}{Econophysics: An introduction}}
  (\bibinfo{publisher}{John Wiley \& Sons}, \bibinfo{year}{2010}).

\bibitem{Chakraborti_2011a}
\bibinfo{author}{Chakraborti, A.}, \bibinfo{author}{Muni~Toke, I.},
  \bibinfo{author}{Patriarca, M.} \& \bibinfo{author}{Abergel, F.}
\newblock \bibinfo{title}{Econophysics review: I. empirical facts}.
\newblock \emph{\bibinfo{journal}{Quantitative Finance}}
  \textbf{\bibinfo{volume}{11}}, \bibinfo{pages}{991--1012}
  (\bibinfo{year}{2011}).

\bibitem{Pharasi_2018}
\bibinfo{author}{Pharasi, H.~K.} \emph{et~al.}
\newblock \bibinfo{title}{Identifying long-term precursors of financial market
  crashes using correlation patterns}.
\newblock \emph{\bibinfo{journal}{New Journal of Physics}}
  \textbf{\bibinfo{volume}{20}}, \bibinfo{pages}{103041}
  (\bibinfo{year}{2018}).

\bibitem{Pharasi_2019}
\bibinfo{author}{Pharasi, H.~K.}, \bibinfo{author}{Sharma, K.},
  \bibinfo{author}{Chakraborti, A.} \& \bibinfo{author}{Seligman, T.~H.}
\newblock \bibinfo{title}{Complex market dynamics in the light of random matrix
  theory}.
\newblock In \emph{\bibinfo{booktitle}{New Perspectives and Challenges in
  Econophysics and Sociophysics}}, \bibinfo{pages}{13--34}
  (\bibinfo{publisher}{Springer International Publishing},
  \bibinfo{address}{Cham}, \bibinfo{year}{2019}).

\bibitem{Mantegna_2003}
\bibinfo{author}{Bonanno, G.}, \bibinfo{author}{Caldarelli, G.},
  \bibinfo{author}{Lillo, F.} \& \bibinfo{author}{Mantegna, R.~N.}
\newblock \bibinfo{title}{Topology of correlation-based minimal spanning trees
  in real and model markets}.
\newblock \emph{\bibinfo{journal}{Physical Review E}}
  \textbf{\bibinfo{volume}{68}}, \bibinfo{pages}{046130}
  (\bibinfo{year}{2003}).

\bibitem{Onnela_2003}
\bibinfo{author}{Onnela, J.-P.}, \bibinfo{author}{Chakraborti, A.},
  \bibinfo{author}{Kaski, K.}, \bibinfo{author}{Kertesz, J.} \&
  \bibinfo{author}{Kanto, A.}
\newblock \bibinfo{title}{Dynamics of market correlations: Taxonomy and
  portfolio analysis}.
\newblock \emph{\bibinfo{journal}{Physical Review E}}
  \textbf{\bibinfo{volume}{68}}, \bibinfo{pages}{056110}
  (\bibinfo{year}{2003}).

\bibitem{Mantegna_2010}
\bibinfo{author}{Tumminello, M.}, \bibinfo{author}{Lillo, F.} \&
  \bibinfo{author}{Mantegna, R.~N.}
\newblock \bibinfo{title}{Correlation, hierarchies, and networks in financial
  markets}.
\newblock \emph{\bibinfo{journal}{Journal of Economic Behavior \&
  Organization}} \textbf{\bibinfo{volume}{75}}, \bibinfo{pages}{40--58}
  (\bibinfo{year}{2010}).

\bibitem{fortunato2010community}
\bibinfo{author}{Fortunato, S.}
\newblock \bibinfo{title}{Community detection in graphs}.
\newblock \emph{\bibinfo{journal}{Physics Reports}}
  \textbf{\bibinfo{volume}{486}}, \bibinfo{pages}{75--174}
  (\bibinfo{year}{2010}).

\bibitem{Garlaschelli_prx_2015}
\bibinfo{author}{MacMahon, M.} \& \bibinfo{author}{Garlaschelli, D.}
\newblock \bibinfo{title}{Community detection for correlation matrices}.
\newblock \emph{\bibinfo{journal}{Physical Review X}}
  \textbf{\bibinfo{volume}{5}}, \bibinfo{pages}{021006} (\bibinfo{year}{2015}).

\bibitem{Barabasi_2016}
\bibinfo{author}{Barab{\'a}si, A.-L.}
\newblock \emph{\bibinfo{title}{Network Science}}
  (\bibinfo{publisher}{Cambridge University Press, Cambridge},
  \bibinfo{year}{2016}).

\bibitem{almog2019}
\bibinfo{author}{Almog, A.} \& \bibinfo{author}{Shmueli, E.}
\newblock \bibinfo{title}{Structural entropy: Monitoring correlation-based
  networks over time with application to financial markets}.
\newblock \emph{\bibinfo{journal}{Scientific Reports}}
  \textbf{\bibinfo{volume}{9}}, \bibinfo{pages}{10832} (\bibinfo{year}{2019}).

\bibitem{Anindya_2019}
\bibinfo{author}{Kuyyamudi, C.}, \bibinfo{author}{Chakrabarti, A.~S.} \&
  \bibinfo{author}{Sinha, S.}
\newblock \bibinfo{title}{Emergence of frustration signals systemic risk}.
\newblock \emph{\bibinfo{journal}{Physical Review E}}
  \textbf{\bibinfo{volume}{99}}, \bibinfo{pages}{052306}
  (\bibinfo{year}{2019}).

\bibitem{fan2017lifespan}
\bibinfo{author}{Fan, Y.} \emph{et~al.}
\newblock \bibinfo{title}{Lifespan development of the human brain revealed by
  large-scale network eigen-entropy}.
\newblock \emph{\bibinfo{journal}{Entropy}} \textbf{\bibinfo{volume}{19}},
  \bibinfo{pages}{471} (\bibinfo{year}{2017}).

\bibitem{mazurin_1984}
\bibinfo{author}{Mazurin, O.~V.} \& \bibinfo{author}{Porai-Koshits, E.}
\newblock \emph{\bibinfo{title}{Phase separation in glass}}
  (\bibinfo{publisher}{Elsevier}, \bibinfo{year}{1984}).

\bibitem{stanley_nature_1992}
\bibinfo{author}{Poole, P.~H.}, \bibinfo{author}{Sciortino, F.},
  \bibinfo{author}{Essmann, U.} \& \bibinfo{author}{Stanley, H.~E.}
\newblock \bibinfo{title}{Phase behaviour of metastable water}.
\newblock \emph{\bibinfo{journal}{Nature}} \textbf{\bibinfo{volume}{360}},
  \bibinfo{pages}{324} (\bibinfo{year}{1992}).

\bibitem{stanley1999scaling}
\bibinfo{author}{Stanley, H.~E.}
\newblock \bibinfo{title}{Scaling, universality, and renormalization: Three
  pillars of modern critical phenomena}.
\newblock \emph{\bibinfo{journal}{Reviews of Modern Physics}}
  \textbf{\bibinfo{volume}{71}}, \bibinfo{pages}{S358} (\bibinfo{year}{1999}).

\bibitem{vix}
\bibinfo{title}{Volatility index (vix)}.
\newblock \bibinfo{howpublished}{\url{https://en.wikipedia.org/wiki/VIX}}
  (\bibinfo{year}{2020}).
\newblock \bibinfo{note}{Accessed on 7th July, 2019.}

\bibitem{sandhu_2016}
\bibinfo{author}{Sandhu, R.~S.}, \bibinfo{author}{Georgiou, T.~T.} \&
  \bibinfo{author}{Tannenbaum, A.~R.}
\newblock \bibinfo{title}{Ricci curvature: An economic indicator for market
  fragility and systemic risk}.
\newblock \emph{\bibinfo{journal}{Science Advances}}
  \textbf{\bibinfo{volume}{2}}, \bibinfo{pages}{e1501495}
  (\bibinfo{year}{2016}).

\bibitem{markowitz1952}
\bibinfo{author}{Markowitz, H.}
\newblock \bibinfo{title}{Portfolio selection}.
\newblock \emph{\bibinfo{journal}{The Journal of Finance}}
  \textbf{\bibinfo{volume}{7}}, \bibinfo{pages}{77--91} (\bibinfo{year}{1952}).

\bibitem{Chakraborti_2020}
\bibinfo{author}{Chakraborti, A.} \emph{et~al.}
\newblock \bibinfo{title}{Emerging spectra characterization of catastrophic
  instabilities in complex systems}.
\newblock \emph{\bibinfo{journal}{New Journal of Physics}}
  \textbf{\bibinfo{volume}{22}}, \bibinfo{pages}{063043}
  (\bibinfo{year}{2020}).

\bibitem{Plerou_2003}
\bibinfo{author}{Plerou, V.}, \bibinfo{author}{Gopikrishnan, P.} \&
  \bibinfo{author}{Stanley, H.~E.}
\newblock \bibinfo{title}{Two-phase behaviour of financial markets}.
\newblock \emph{\bibinfo{journal}{Nature}} \textbf{\bibinfo{volume}{421}},
  \bibinfo{pages}{130} (\bibinfo{year}{2003}).

\bibitem{jpn_zerorate}
\bibinfo{title}{Japan ends zero-rate policy}.
\newblock
  \bibinfo{howpublished}{\url{https://www.ft.com/content/902a0d58-12f0-11db-aecf-0000779e2340/}}
  (\bibinfo{year}{2020}).
\newblock \bibinfo{note}{Accessed on 20th July, 2020.}

\bibitem{fed_move}
\bibinfo{title}{Wall st. soars on fed move}.
\newblock
  \bibinfo{howpublished}{\url{https://money.cnn.com/2001/04/18/markets/markets_newyork/}}
  (\bibinfo{year}{2020}).
\newblock \bibinfo{note}{Accessed on 20th July, 2020.}

\bibitem{reszat2003japan}
\bibinfo{author}{Reszat, B.}
\newblock \bibinfo{title}{Japan's financial markets: The lost decade}
  (\bibinfo{year}{2003}).

\bibitem{hayashi20021990s}
\bibinfo{author}{Hayashi, F.} \& \bibinfo{author}{Prescott, E.~C.}
\newblock \bibinfo{title}{The 1990s in japan: A lost decade}.
\newblock \emph{\bibinfo{journal}{Review of Economic Dynamics}}
  \textbf{\bibinfo{volume}{5}}, \bibinfo{pages}{206--235}
  (\bibinfo{year}{2002}).

\bibitem{list}
\bibinfo{title}{List of stock market crashes and bear markets}.
\newblock
  \bibinfo{howpublished}{\url{https://en.wikipedia.org/wiki/List_of_stock_market_crashes_and_bear_markets}}
  (\bibinfo{year}{2019}).
\newblock \bibinfo{note}{Accessed on 7th July, 2019.}

\bibitem{bullmarkets}
\bibinfo{title}{Bull markets}.
\newblock
  \bibinfo{howpublished}{\url{https://bullmarkets.co/u-s-stock-market-in-1996/}}
  (\bibinfo{year}{2019}).
\newblock \bibinfo{note}{Accessed on 7th July, 2019.}

\bibitem{ushousing}
\bibinfo{title}{United states housing bubble}.
\newblock
  \bibinfo{howpublished}{\url{https://en.wikipedia.org/wiki/United_States_housing_bubble}}
  (\bibinfo{year}{2019}).
\newblock \bibinfo{note}{Accessed on 7th July, 2019.}

\bibitem{history_crashes}
\bibinfo{title}{A short history of stock market crashes}.
\newblock
  \bibinfo{howpublished}{\url{https://www.cnbc.com/2016/08/24/a-short-history-of-stock-market-crashes.html}}
  (\bibinfo{year}{2019}).
\newblock \bibinfo{note}{Accessed on 7th July, 2019.}

\bibitem{selloff}
\bibinfo{title}{Stock market selloff}.
\newblock
  \bibinfo{howpublished}{\url{https://en.wikipedia.org/wiki/2015-16_stock_market_selloff}}
  (\bibinfo{year}{2019}).
\newblock \bibinfo{note}{Accessed on 7th July, 2019.}

\end{thebibliography}

\setcounter{figure}{0}
\setcounter{table}{0}
\renewcommand{\thefigure}{S\arabic{figure}}
\renewcommand{\thetable}{S\arabic{table}}

\section*{Supplementary Information}

Two videos showing the continuous monitoring of the correlation patterns ($C$, $C_M$ and $C_{GR}$) along with their corresponding eigen-centralities (ranked), entropy measures and positions of the correlation frame in the phase separation diagrams for USA (SI Video 1) and JPN (SI Video 2), respectively, are uploaded.

\textbf{Video 1 Legend}: Continuous monitoring of the correlation patterns ($C$, $C_M$ and $C_{GR}$) along with their corresponding eigen-centralities (ranked), entropy measures and positions of the correlation frame in the phase separation diagrams for USA. Link to high resolution video: \url{https://drive.google.com/file/d/1VIdfo9LnB-grHxloq8BmUJ5jYyVYH50c/view?usp=sharing}

\textbf{Video 2 Legend}: Continuous monitoring of the correlation patterns ($C$, $C_M$ and $C_{GR}$) along with their corresponding eigen-centralities (ranked), entropy measures and positions of the correlation frame in the phase separation diagrams for JPN. Link to high resolution video: \url{https://drive.google.com/file/d/1gcOYT1QR51nrqRCTp4XLDhJEM1QDCFPu/view?usp=sharing}

\subsection*{Data}
We have used the adjusted closure price time series for United States of America (USA) S\&P-500 index and Japan (JPN) Nikkei-225 index, for the period 02-01-1985 to 30-12-2016, from the Yahoo finance database (https://finance.yahoo.co.jp/; accessed on 7th July, 2017). USA has data for the $N=194$ stocks and the period 02-01-1985 to 30-12-2016 ($T=8068$ days). JPN has data for the $N=165$ stocks and the period 04-01-1985 to 30-12-2016 ($T=7998$ days). Note that we have included those stocks in our analyses, which are present in the data for the entire duration, and added zero return entries corresponding to the missing days. 

The list of stocks (along with the sectors) for the two markets are given in Tables~ \ref{USA_Table} and   \ref{JPN_Table}.
The sectoral abbreviations are given below: \\
{CD} -- Consumer Discretionary; \\
{CS} -- Consumer Staples;      \\ 
{EG} -- Energy;             \\    
{FN} -- Financial;           \\  
{HC} -- Health Care;         \\   
{ID} -- Industrials;          \\ 
{IT} -- Information Technology; \\
{MT} -- Materials;        \\      
{TC} -- Telecommunication Services; and   \\          
{UT} -- Utilities.

\subsection*{Methodology}

\begin{figure*}[b!]
\includegraphics[width=0.5\linewidth]{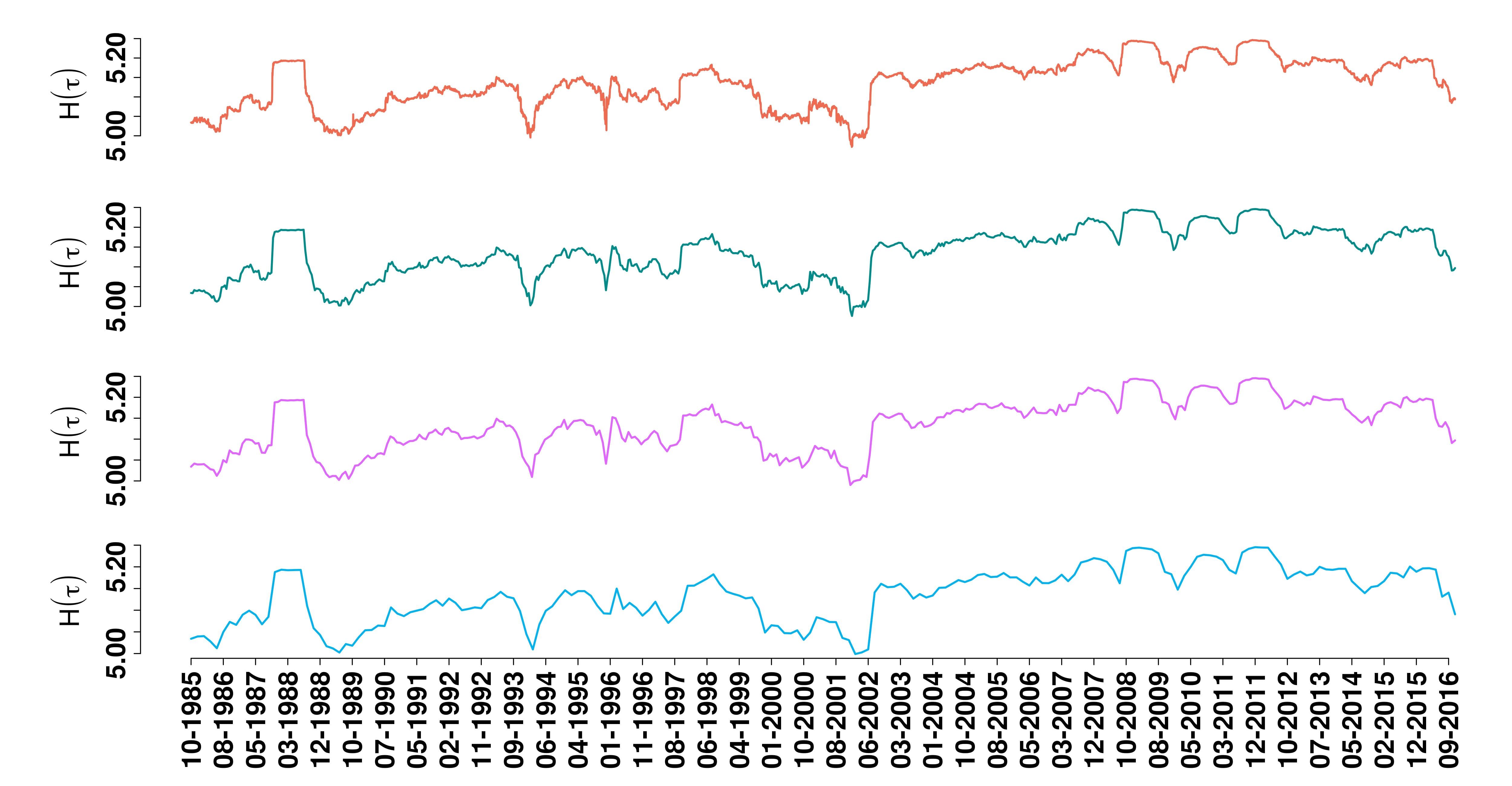}\llap{\parbox[b]{3.4in}{\textsf{\textbf{a}}\\\rule{0ex}{1.9in}}}\llap{\parbox[b]{3.1in}{\textsf{i}\\\rule{0ex}{1.65in}}}\llap{\parbox[b]{3.1in}{\textsf{ii}\\\rule{0ex}{1.28in}}}\llap{\parbox[b]{3.1in}{\textsf{iii}\\\rule{0ex}{0.88in}}}\llap{\parbox[b]{3.1in}{\textsf{iv}\\\rule{0ex}{0.5in}}}
	\includegraphics[width=0.5\linewidth]{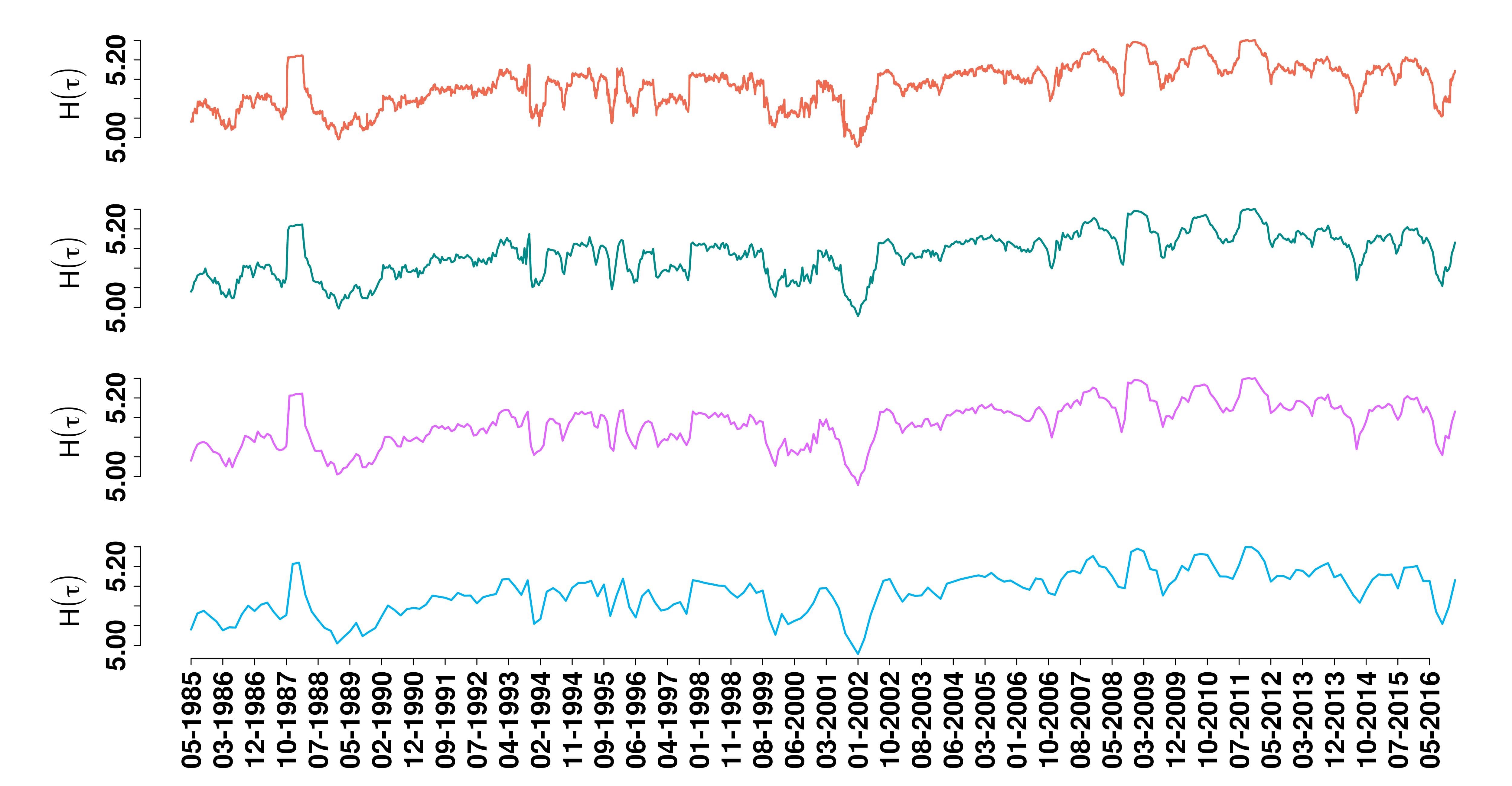}\llap{\parbox[b]{3.4in}{\textsf{\textbf{b}}\\\rule{0ex}{1.9in}}}\llap{\parbox[b]{3.1in}{\textsf{i}\\\rule{0ex}{1.65in}}}\llap{\parbox[b]{3.1in}{\textsf{ii}\\\rule{0ex}{1.28in}}}\llap{\parbox[b]{3.1in}{\textsf{iii}\\\rule{0ex}{0.88in}}}\llap{\parbox[b]{3.1in}{\textsf{iv}\\\rule{0ex}{0.5in}}}
	\includegraphics[width=0.5\linewidth]{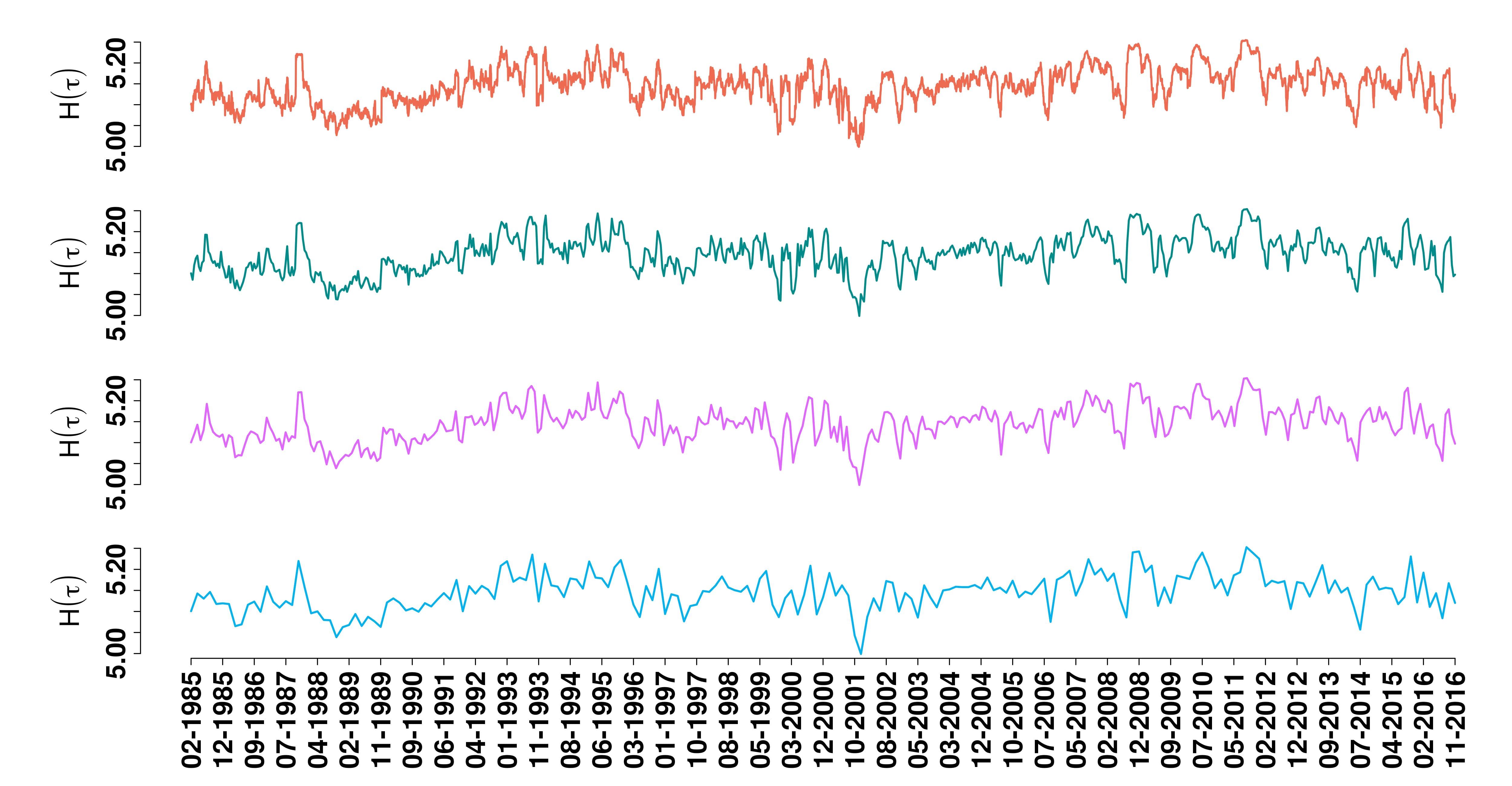}\llap{\parbox[b]{3.4in}{\textsf{\textbf{c}}\\\rule{0ex}{1.9in}}}\llap{\parbox[b]{3.1in}{\textsf{i}\\\rule{0ex}{1.65in}}}\llap{\parbox[b]{3.1in}{\textsf{ii}\\\rule{0ex}{1.28in}}}\llap{\parbox[b]{3.1in}{\textsf{iii}\\\rule{0ex}{0.88in}}}\llap{\parbox[b]{3.1in}{\textsf{iv}\\\rule{0ex}{0.5in}}}
	\includegraphics[width=0.5\linewidth]{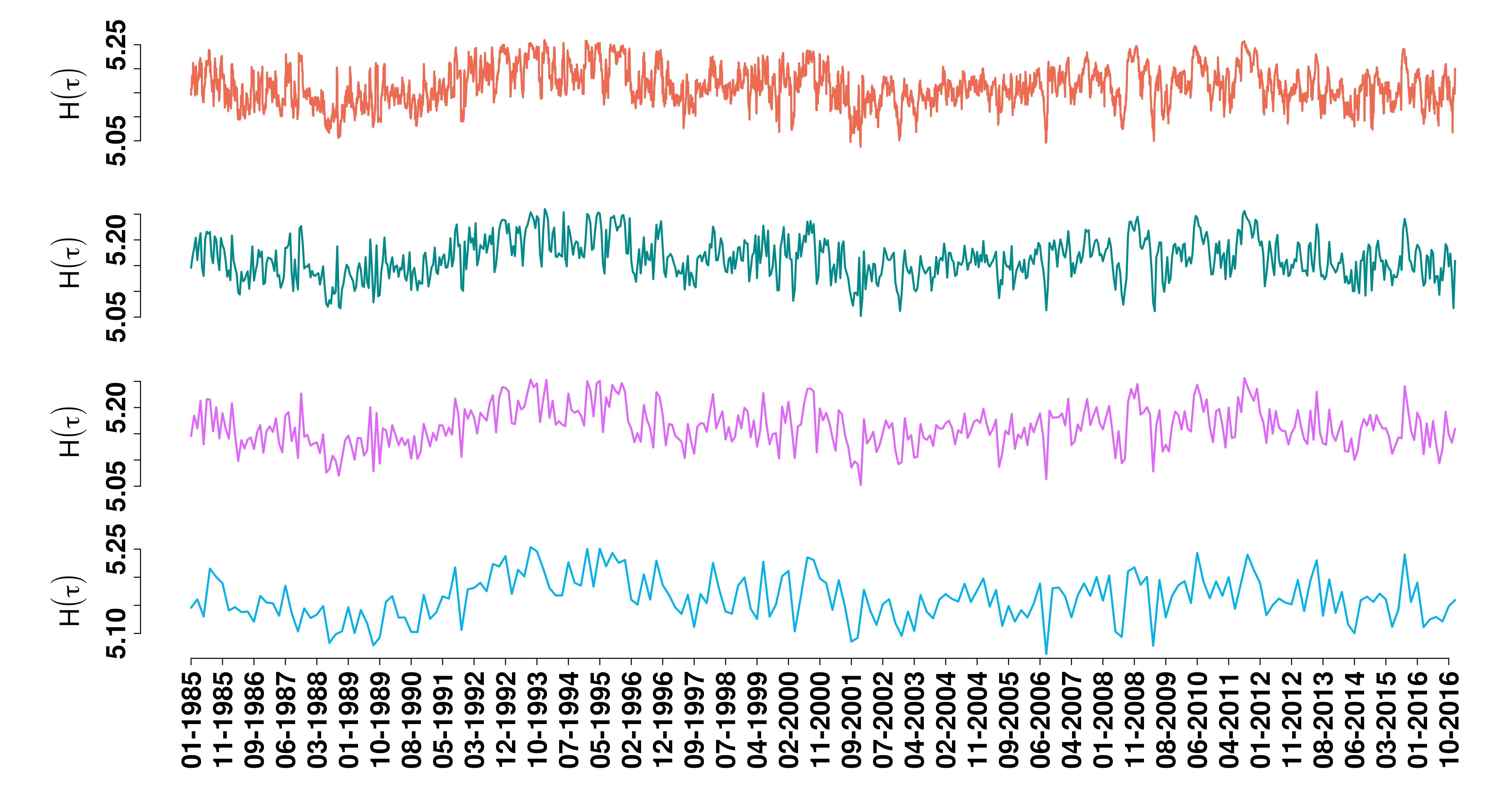}\llap{\parbox[b]{3.4in}{\textsf{\textbf{d}}\\\rule{0ex}{1.9in}}}\llap{\parbox[b]{3.1in}{\textsf{i}\\\rule{0ex}{1.65in}}}\llap{\parbox[b]{3.1in}{\textsf{ii}\\\rule{0ex}{1.28in}}}\llap{\parbox[b]{3.1in}{\textsf{iii}\\\rule{0ex}{0.88in}}}\llap{\parbox[b]{3.1in}{\textsf{iv}\\\rule{0ex}{0.5in}}}
	\caption{\textbf{Effects of epoch size $M$ and shift $\Delta$ on the time series of eigen-entropy $H$.} The evolution of eigen-entropy $H$ is calculated from correlation matrices corresponding to four different time epochs (\textbf{a}) $M=200$, (\textbf{b}) $M=100$, (\textbf{c}) $M=40$, and  (\textbf{d}) $M=20$ days and each with four different shifts (\textsf{i}) $\Delta=1$ day, (\textsf{ii}) $\Delta=10$ days, (\textsf{iii}) $\Delta=20$ days, and (\textsf{iv}) $\Delta=40$ days over a period of 1985-2016. The fluctuations (local) of the eigen-entropy $H$ are smoothened (smaller) for  bigger shifts $\Delta$.}
\label{WindowShift}	
\end{figure*}

\subsubsection*{Effects of the variation of the epoch size $M$ and shift $\Delta$}

The continuous monitoring of the market can be done by dividing the total time series data into smaller epochs of size $M$. The corresponding correlation matrices generated from these smaller epochs are used for calculating the eigen-entropy $H$. In Fig.~\ref{WindowShift}, we investigate the effects of the variation of parameters, epoch size $M$ and shift $\Delta$.
We have observed that either the increase in the epoch $M$ or shift  $\Delta$ makes the time series plot of $H$ more smooth (fewer fluctuations), and vice versa. The choice of these parameters are thus arbitrary to some extent, depending on the research questions and time scale we are interested.

\subsubsection*{Effects of the variation in the element-wise powers of correlation matrices $|C|^n$}
\begin{figure*} 
	\centering
	\includegraphics[width=0.48\linewidth]{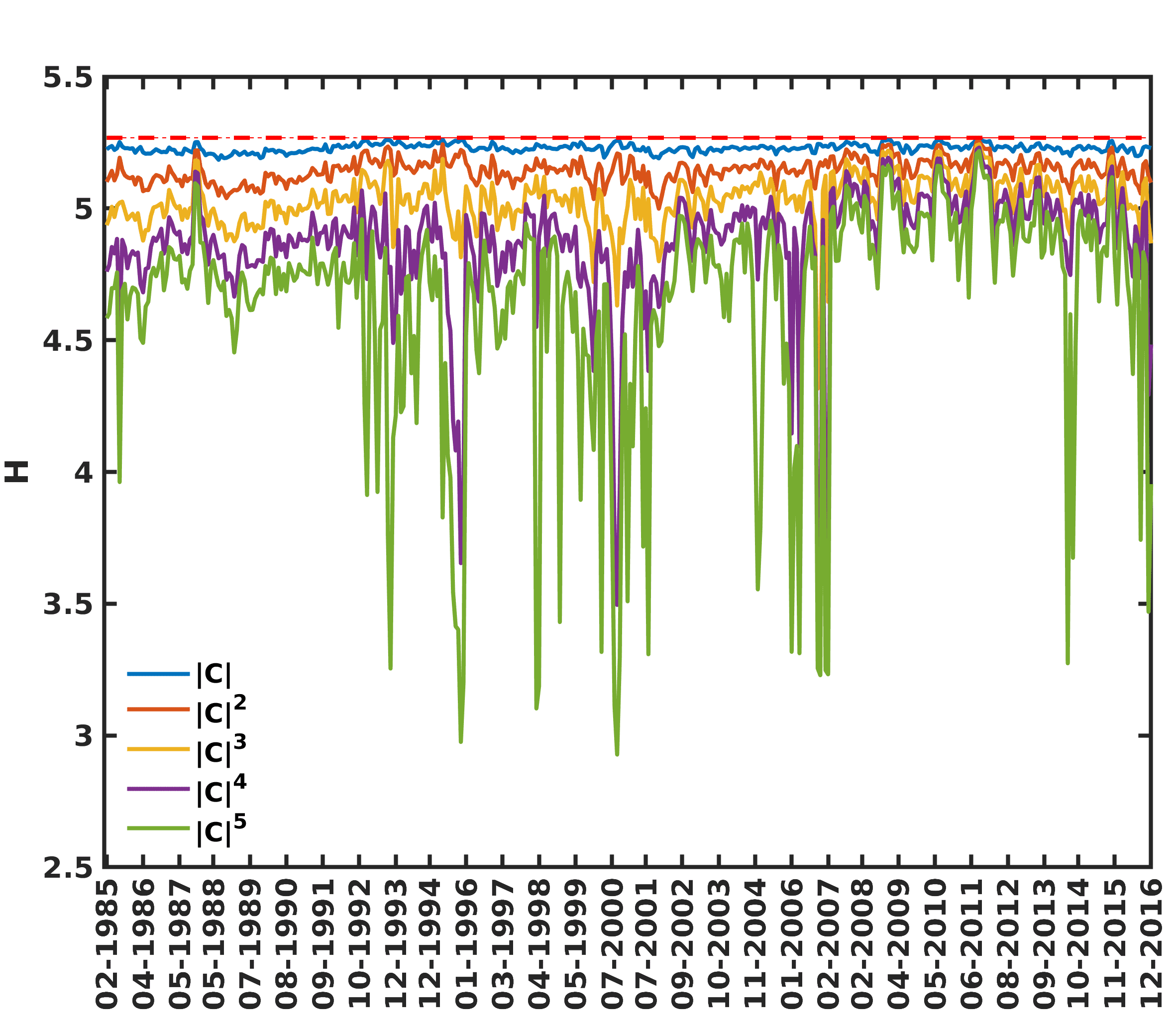}\llap{\parbox[b]{3.4in}{\textsf{\textbf{a}}\\\rule{0ex}{2.8in}}}
	\includegraphics[width=0.48\linewidth]{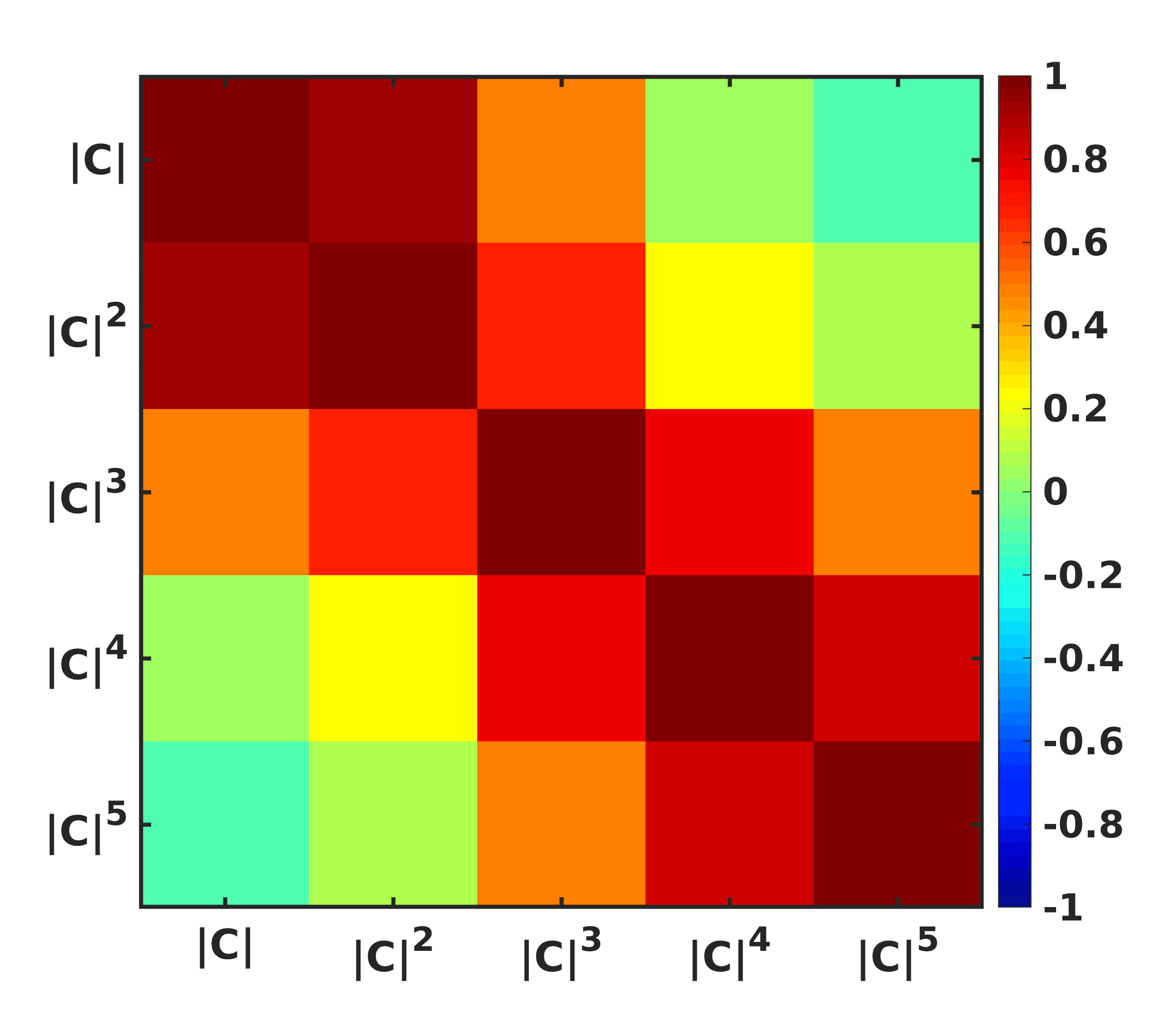}\llap{\parbox[b]{3.2in}{\textsf{\textbf{b}}\\\rule{0ex}{2.8in}}}
	\caption{\textbf{Comparison of the variation of $n$ for $|C|^n$.} The eigen-entropy $H$ is calculated for different powers $n$ of correlation matrix $C$ by raising the elements of $C$ to even powers or the absolute value of $C$ to odd powers. (\textbf{a}) shows the time series of the eigen-entropies $H$ of the correlation matrices of epoch $M=40$ days and $\Delta=20$ days for five different powers upto $n=5$. The correlations among these five time series of eigen-entropy $H$ is shown in (\textbf{b}). }	
	\label{PwrCmpr}
\end{figure*}

Instead of taking the square of individual elements of the correlation matrix $C$, to make all the elements non-negative, we can also use the even powers or the odd powers of absolute values to accomplish the same. The effect of the same is shown in Fig.~\ref{PwrCmpr}. As observed the values of eigen-entropy $H$ differ with the variation of the power $n$ of correlation matrices. This is due to the fact that with the increase in power, the dissimilarities in the elements of the correlation matrix are amplified, which will then in turn change the centrality of the matrix. For very high powers of $n$ the transformed correlation matrices will act like an adjacency matrix with very high values (close to $1$) and very low values (close to $0$). 
It is also interesting to note that, depending on the problem, we can decide the range of correlations to focus on by adjusting the power of the elements of the correlation matrix.

\subsection*{Results}

\subsubsection*{Correlation matrix decomposition and Wishart orthogonal ensemble Results}

\begin{figure*}[h!]
\centering
	\includegraphics[width=0.3\linewidth]{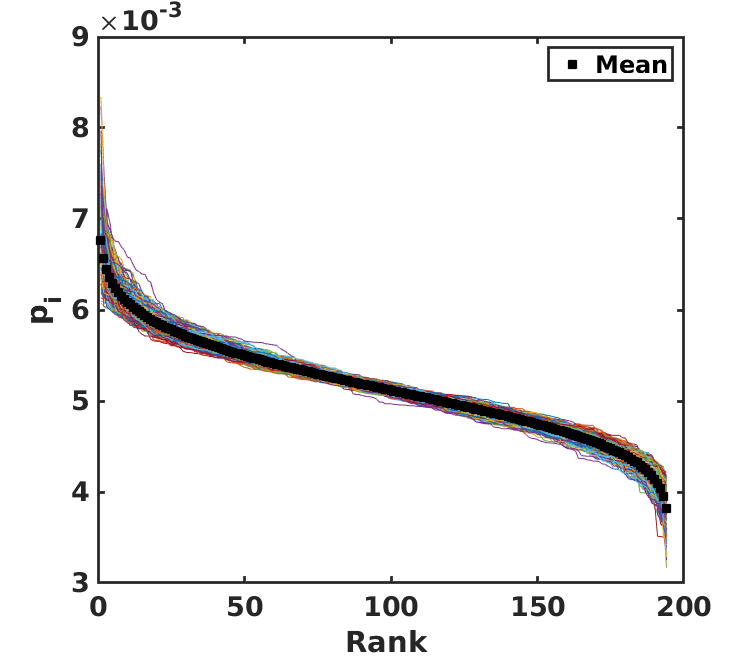}
	\llap{\parbox[b]{2.2in}{{\large \textsf{\textbf{a}}}\\\\\\\\\\\\\\\\\\\\\\}}
	\includegraphics[width=0.3\linewidth]{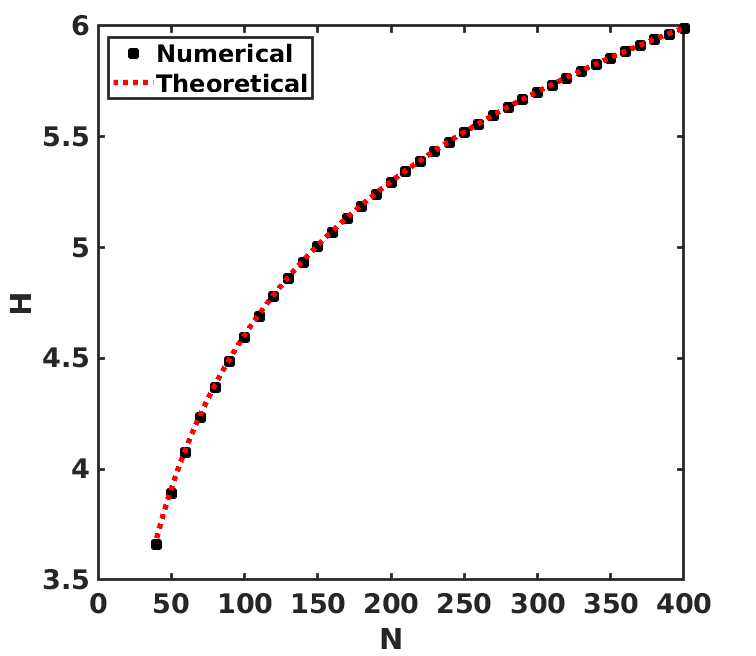}
	\llap{\parbox[b]{2.2in}{{\large \textsf{\textbf{b}}}\\\\\\\\\\\\\\\\\\\\\\}}
\caption{\textbf{Eigen-centralities (ranks) and eigen-entropy.} (\textbf{a}) Plots of sorted eigen-centralities $p_i$ against rank, computed from the normalized eigenvectors corresponding to the maximum eigenvalues for $1000$ independent realizations of the WOE. Filled black squares represent the mean eigen-centralities computed from $1000$ independent realizations of the WOE, that serves as a reference (the maximum disorder or randomness) in the market correlation with $N=194$. (\textbf{b}) Plot shows the variation of eigen-entropy $H$ as a function of system size (correlation matrix size) $N$, where each point represents a mean computed from $1000$ independent realizations of a WOE. The theoretical curve (red dash) shows the variation $\sim \ln N$. 
}
\label{fig:entropy_WOE}
\end{figure*}

Fig.~\ref{fig:entropy_WOE}a shows the plot of sorted eigen-centralities $p_i$ against rank, computed from the normalized eigenvectors corresponding to the maximum eigenvalues for $1000$ independent realizations of a Wishart orthogonal ensemble (WOE). Filled black squares represent the mean eigen-centralities computed from $1000$ independent realizations of the  WOE, that serves as a reference (the maximum disorder or randomness) in the market correlation with $N=194$. Fig.~\ref{fig:entropy_WOE}b shows the plot of the variation of eigen-entropy $H$ as a function of system size (correlation matrix size) $N$, where each point represents a mean computed from $1000$ independent realizations of a WOE. The theoretical curve (red dash) shows the variation $\sim \ln N$.

\subsubsection*{Phase space construction and phase separation}

\begin{figure*}
\centering
\includegraphics[width=0.7\linewidth]{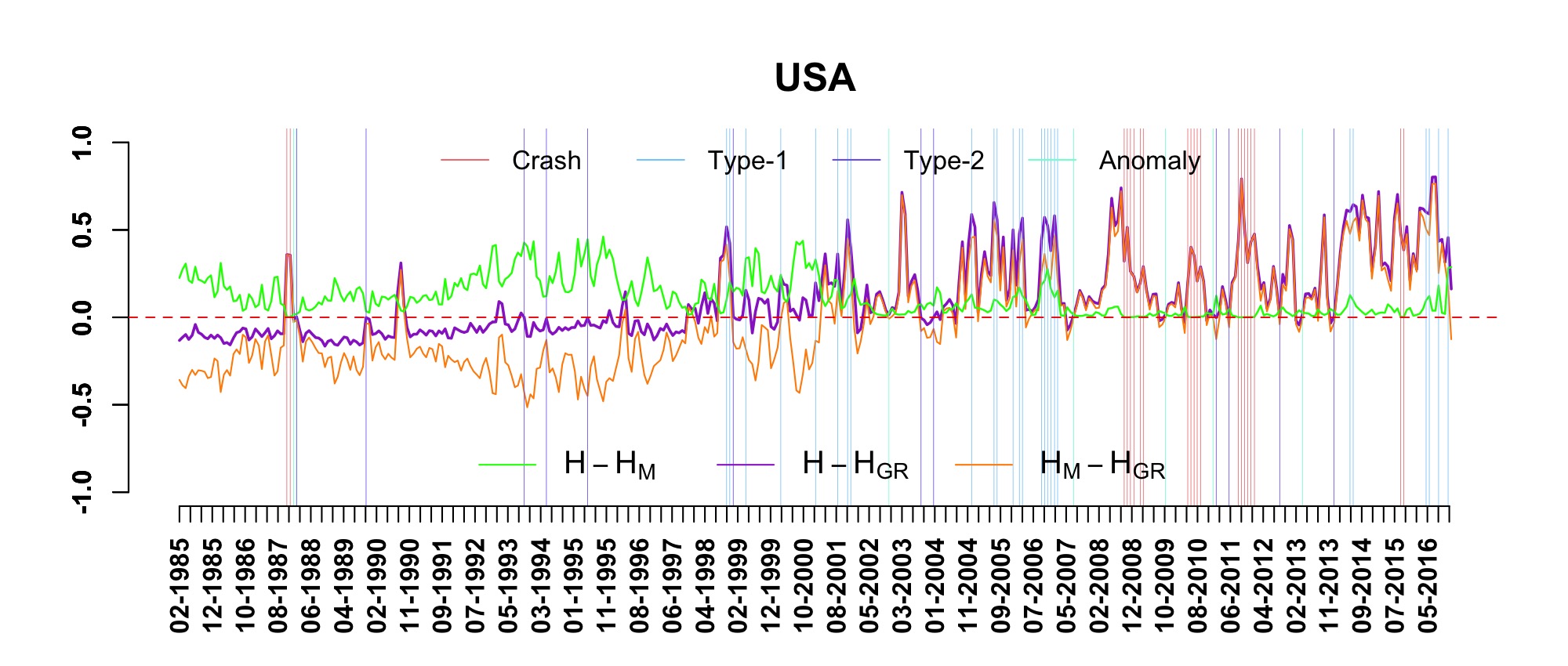}
\llap{\parbox[b]{5in}{{\large\textsf{\textbf{a}}}\\\rule{0ex}{1.6in}}}
\includegraphics[width=0.28\linewidth]{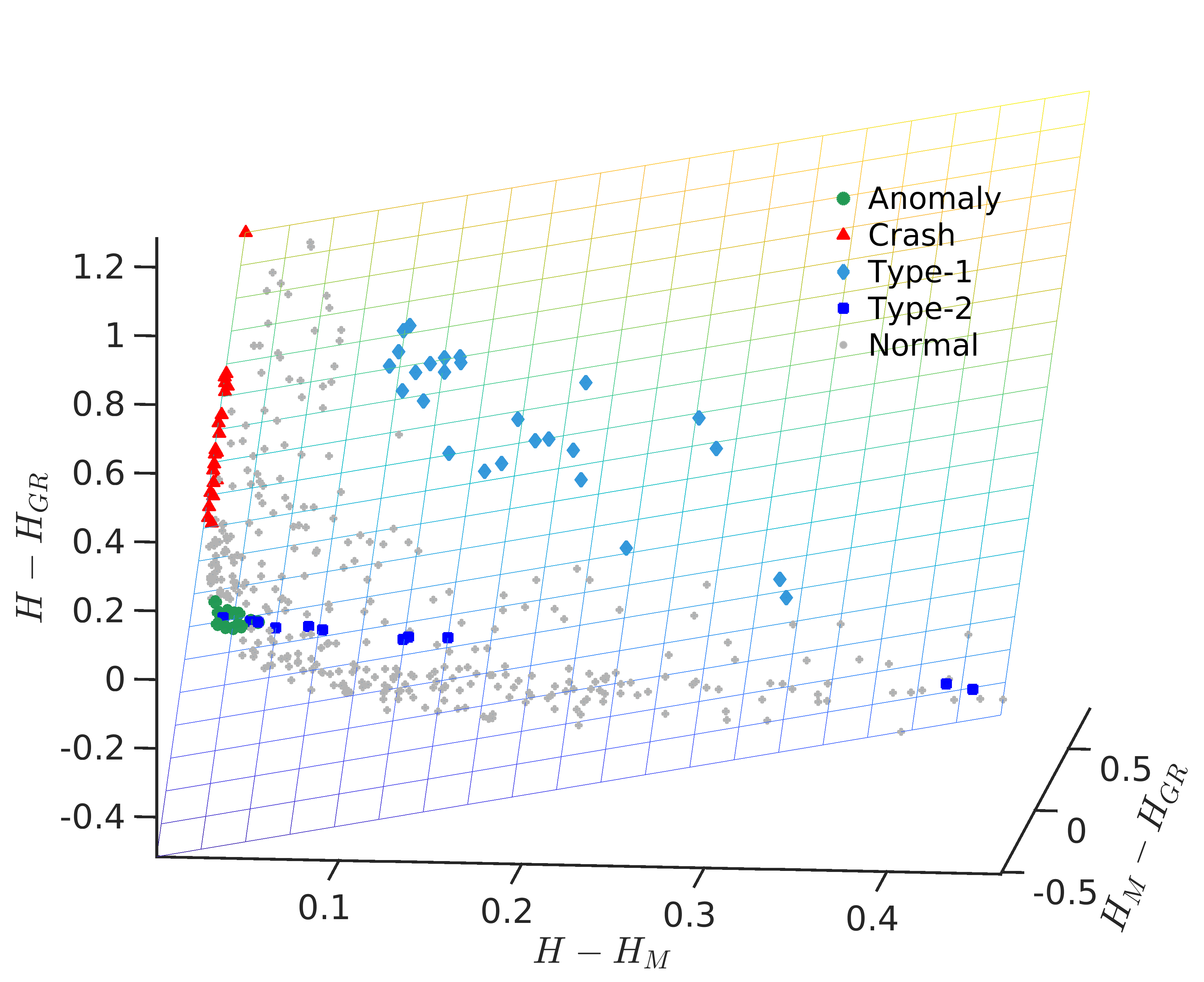}
\llap{\parbox[b]{2in}{{\large\textsf{\textbf{b}}}\\\rule{0ex}{1.6in}}}
\includegraphics[width=0.7\linewidth]{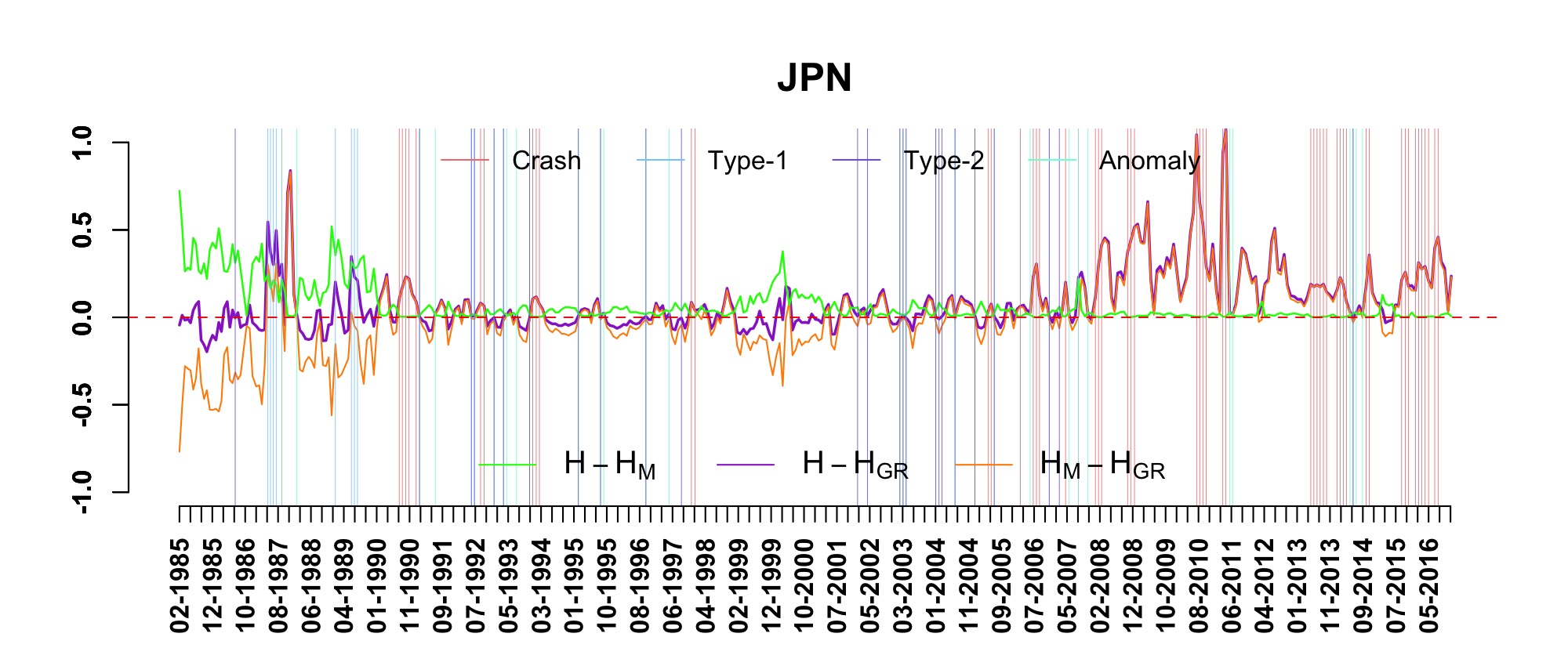}
\llap{\parbox[b]{5in}{{\large\textsf{\textbf{c}}}\\\rule{0ex}{1.6in}}}
\includegraphics[width=0.28\linewidth]{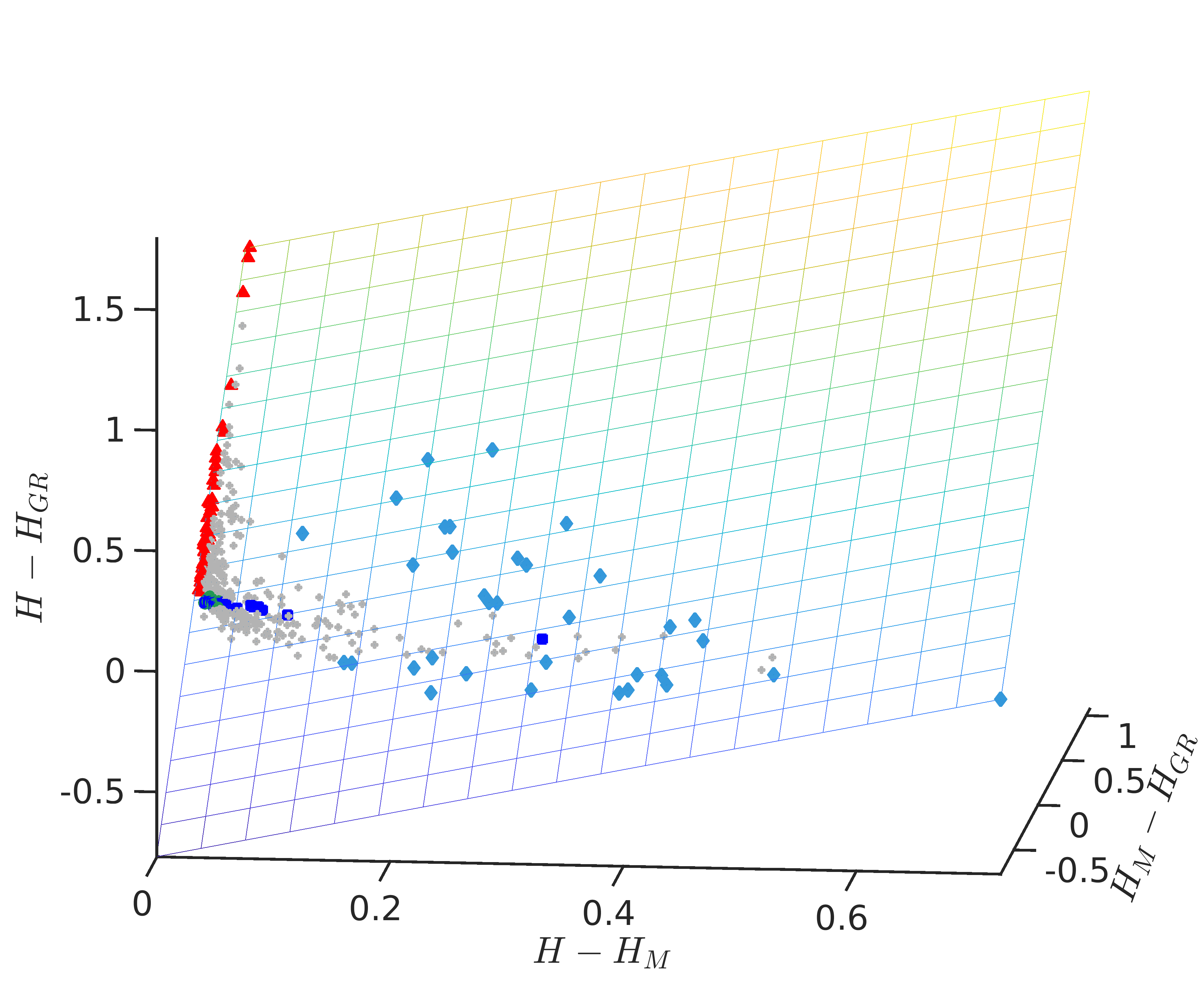}
\llap{\parbox[b]{2in}{{\large\textsf{\textbf{d}}}\\\rule{0ex}{1.6in}}}
\caption{\textbf{Evolution and phase separation of relative-entropies.} For (\textbf{a}) S\&P-500 and (\textbf{c}) Nikkei-225 markets, the relative-entropies  $H-H_M$, $H-H_{GR}$, \&  $H_M-H_{GR}$ are evaluated from full, market and group-random mode to characterize and identify the different market events as anomalies, type-1 events, type-2 events, crashes and normal periods. \textbf{Phase separation.} The 3D-plots of the phase space using relative-entropies $H-H_M$, $H-H_{GR}$, \&  $H_M-H_{GR}$, for (\textbf{b}) S\&P-500, and (\textbf{d}) Nikkei-225 markets. 
The event frames show ``phase separation'' -- segregation of different market events: anomalies, type-1 events, type-2 events, crashes and normal. Note that all the points (market events) actually lie on a plane (see surface-grid).
}
\label{fig:entropy_diff}
\end{figure*}

Figure~\ref{fig:entropy_diff} shows evolution of relative-entropies [$H-H_M$,  $H-H_{GR}$, and $H_M-H_{GR}$] and phase separation. The characterized events Fig.~\ref{fig:entropy_diff}b, d are indicated as vertical lines in the time-evolution plots Fig.~\ref{fig:entropy_diff}a, c (also see Supplementary Videos 1, 2). We found that many anomalies occurred just around the major crashes and intriguing patterns (termed as interesting events of type-1 and type-2, belonging to two distinct regions in the phase space) appeared. 

\subsubsection*{Study of critical events}
\begin{figure*}
	\centering
	\includegraphics[width=0.325\linewidth]{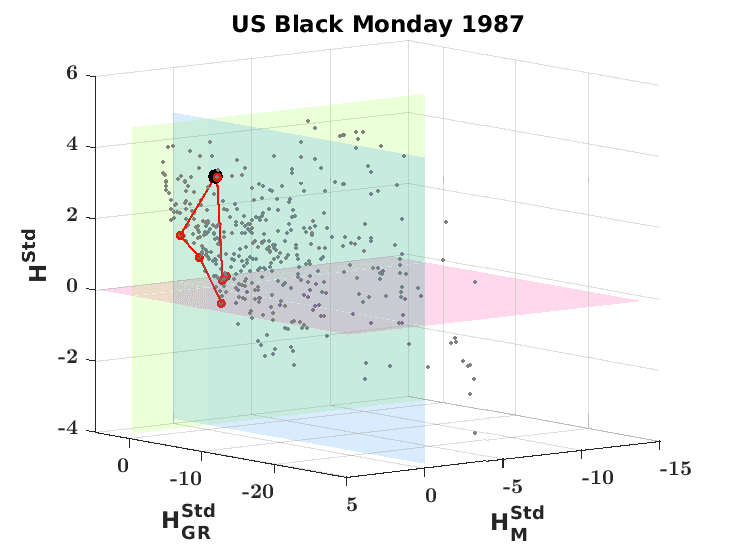}
	\includegraphics[width=0.325\linewidth]{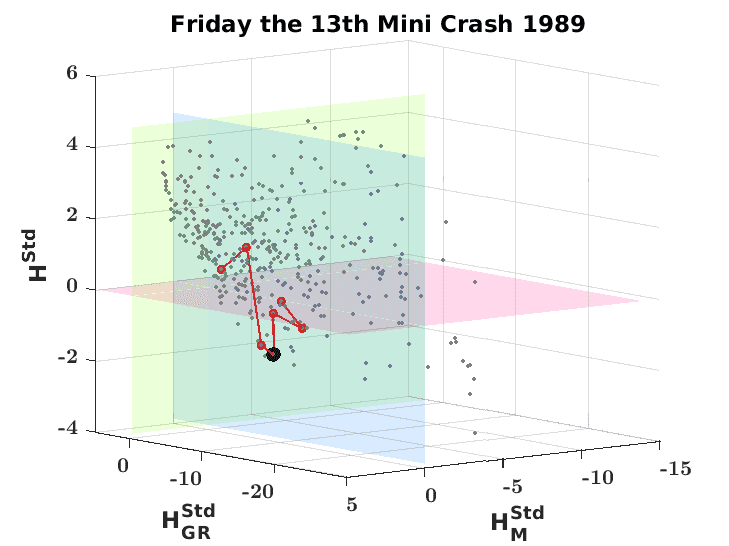}
	\includegraphics[width=0.325\linewidth]{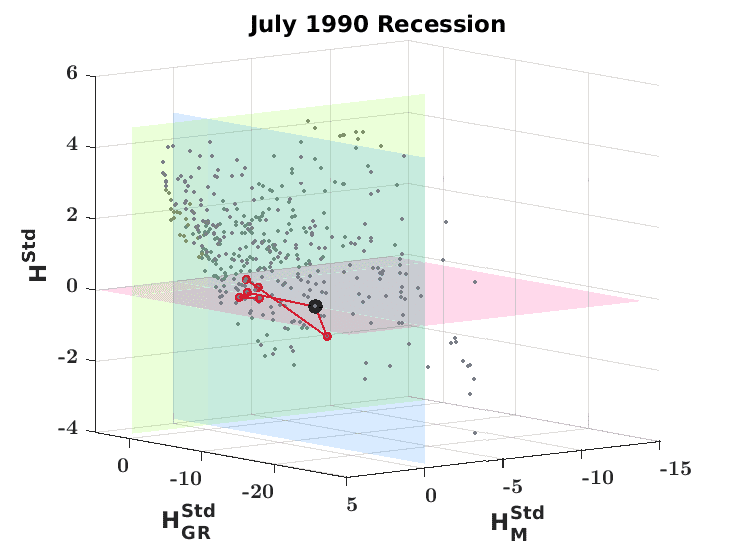}
	\includegraphics[width=0.325\linewidth]{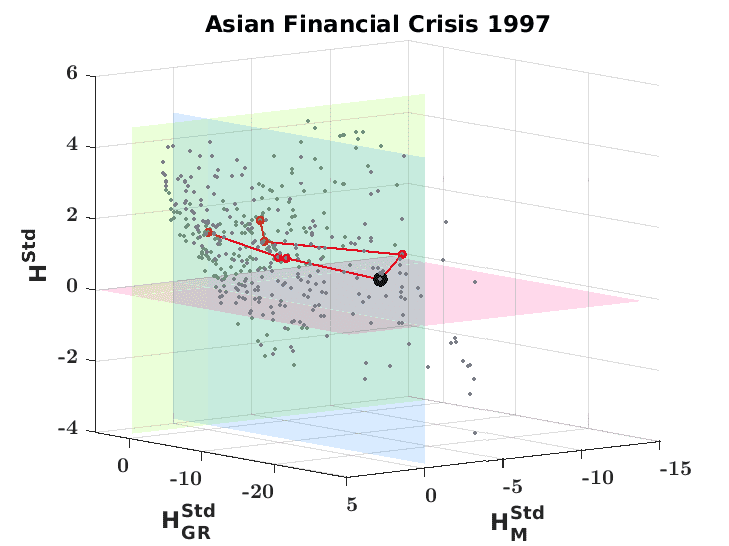}
	\includegraphics[width=0.325\linewidth]{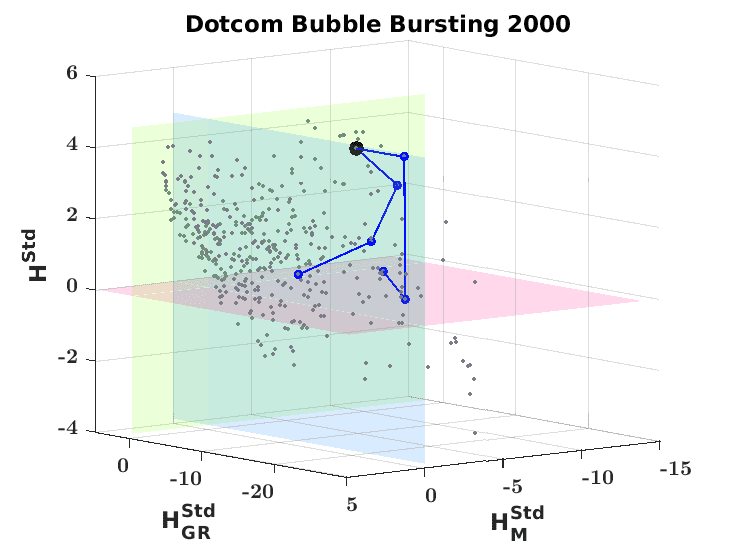}
	\includegraphics[width=0.325\linewidth]{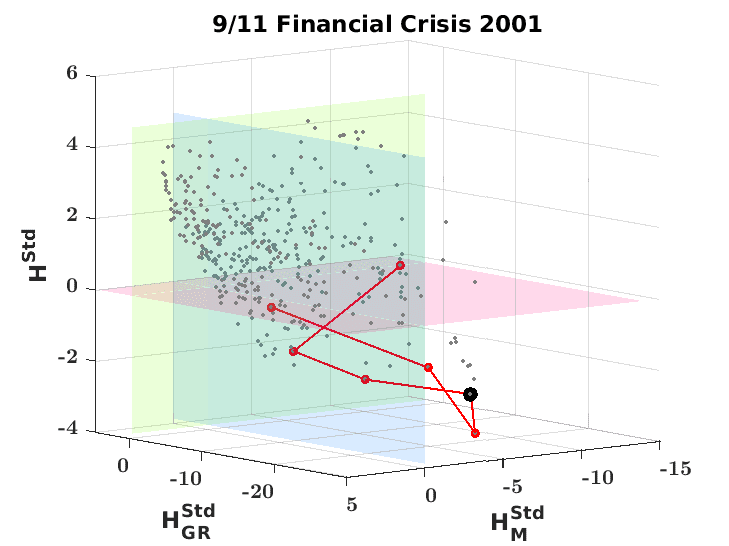}
	\includegraphics[width=0.325\linewidth]{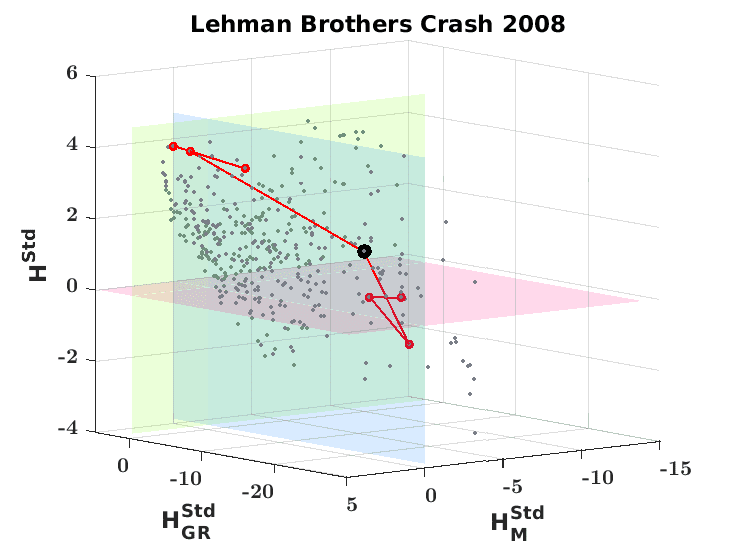}
	\includegraphics[width=0.325\linewidth]{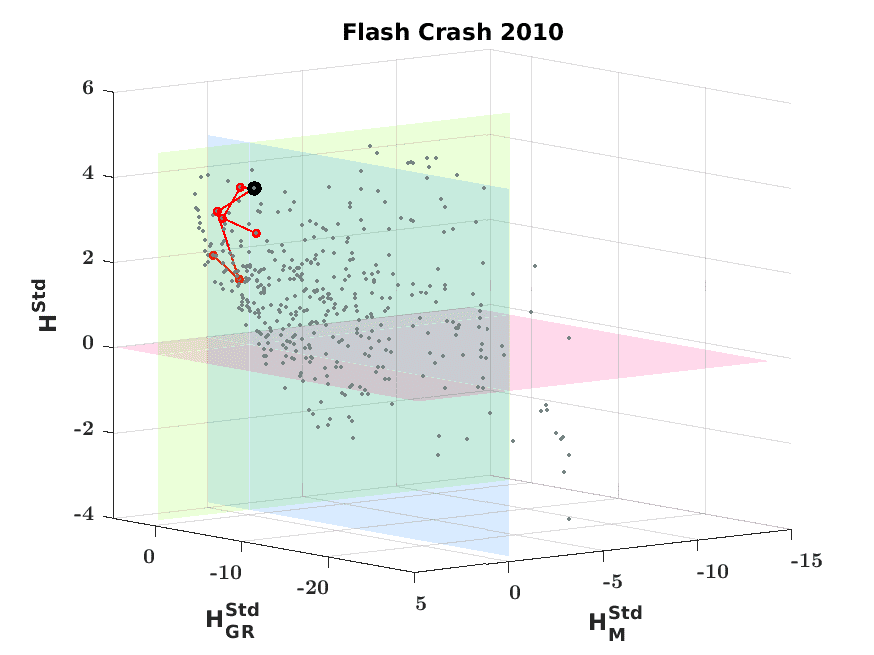}
	\includegraphics[width=0.325\linewidth]{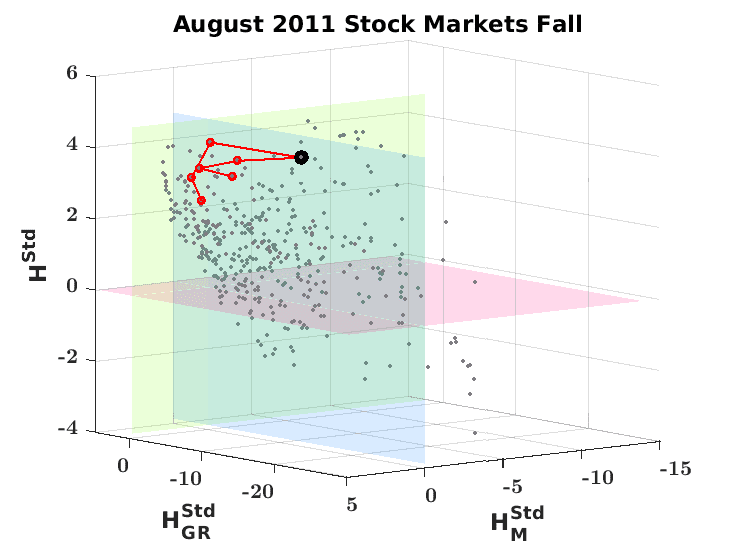}
	\includegraphics[width=0.325\linewidth]{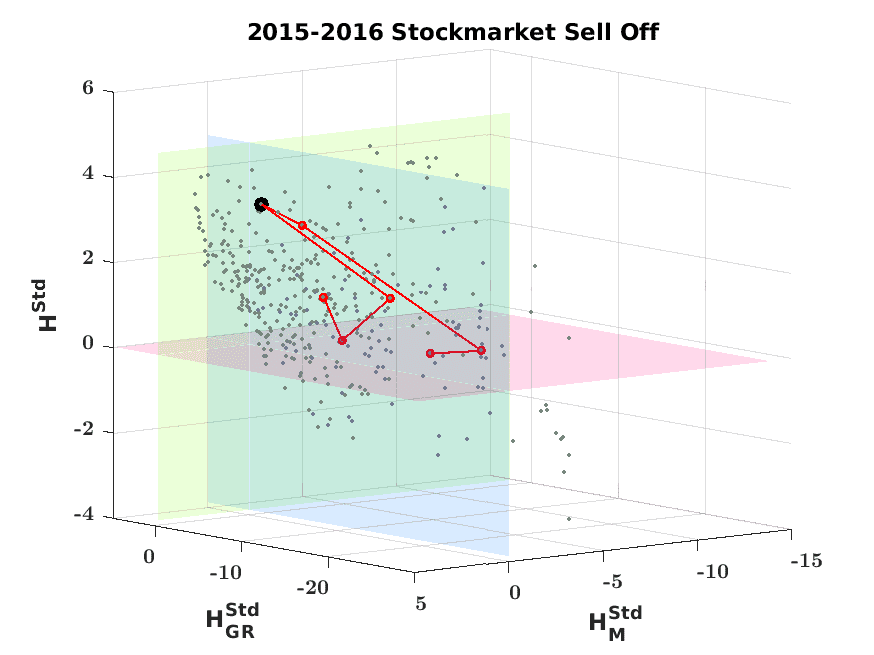}
	\caption{\textbf{Evolution around the important events in USA market.} Eigen-entropy $H$ calculated from the correlation matrices: full $C$, market mode $C_M$ and group-random mode $C_{GR}$ for all the frames (epoch $M=40$ days and shift $\Delta=20$ days) over a period of 1985-2016 of USA (S\&P-500). After standardizing the variables with moving average and moving standard deviation, each frame (grey dot) is embedded in a 3-D space with axes $H^{std}$, $H^{std}_M$ and $H^{std}_{GR}$. Ten important events each with seven frames around these important events (three before and three after the event) were taken from the history and shown in the plots. Critical events are marked with red lines except for the Dot-com bubble burst which is marked with blue line. The frame containing the important event is marked with black circle for better visibility.}\label{fig:usa_events}
\end{figure*}

\begin{figure*}
	\centering
	\includegraphics[width=0.325\linewidth]{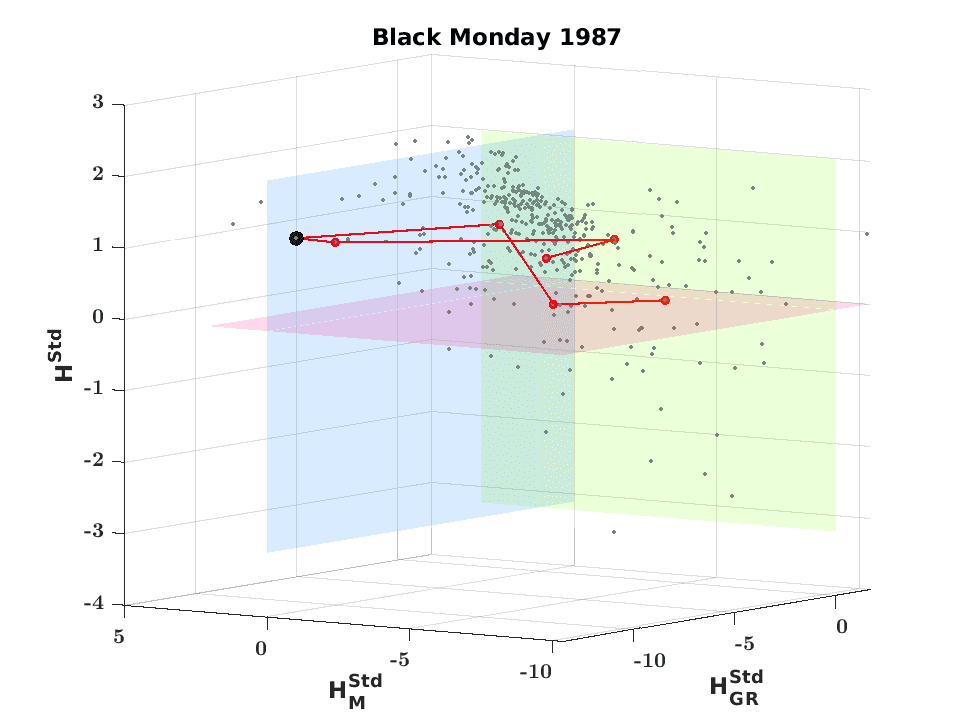}
	\includegraphics[width=0.325\linewidth]{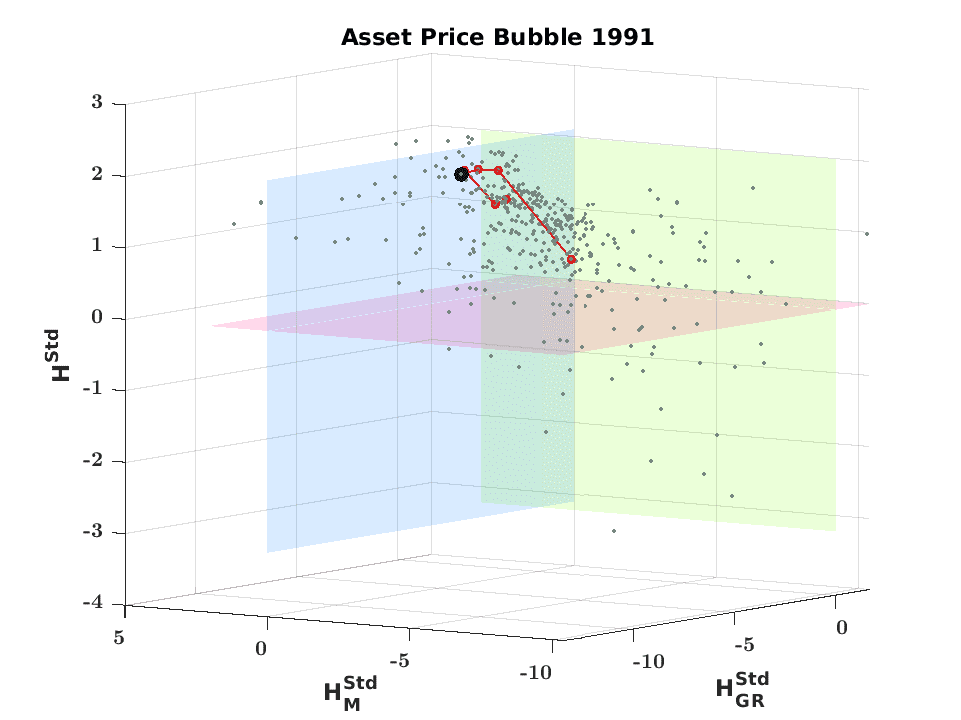}
	\includegraphics[width=0.325\linewidth]{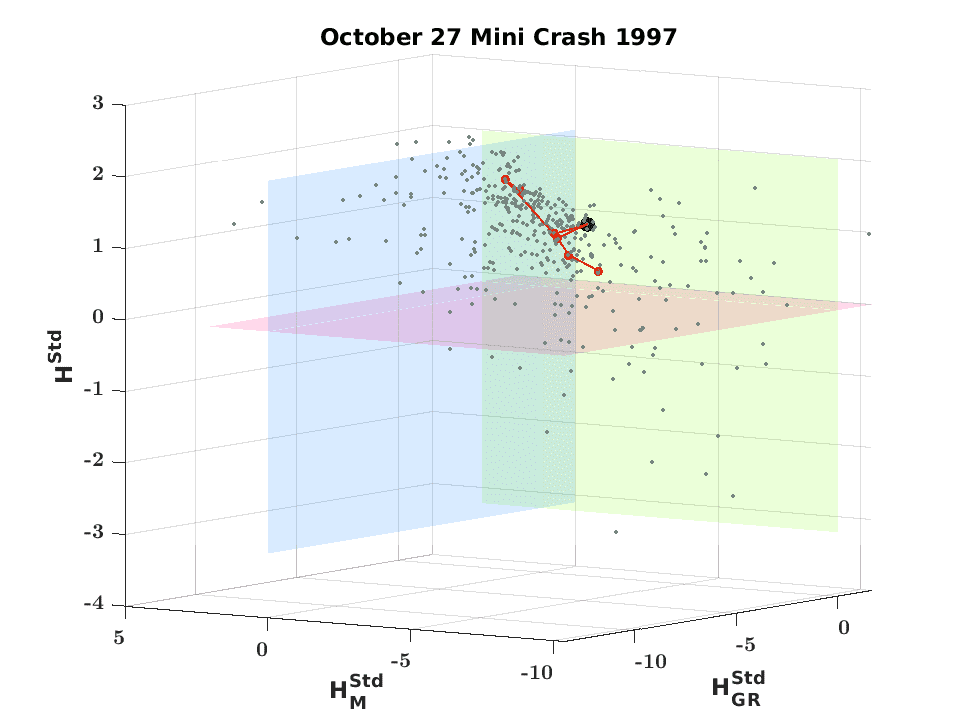}
	\includegraphics[width=0.325\linewidth]{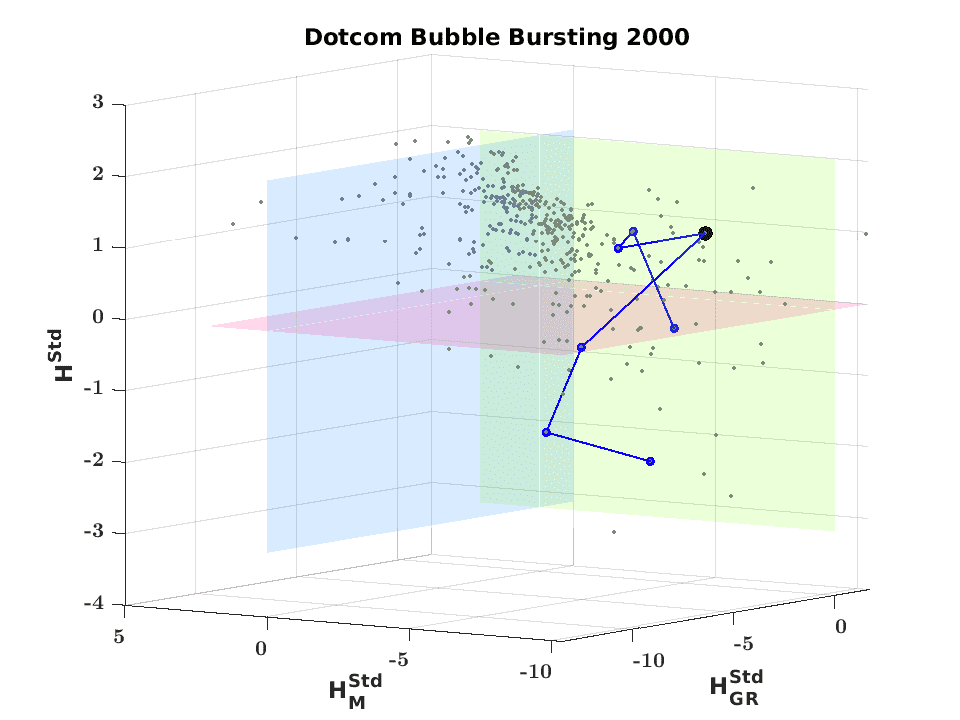}
	\includegraphics[width=0.325\linewidth]{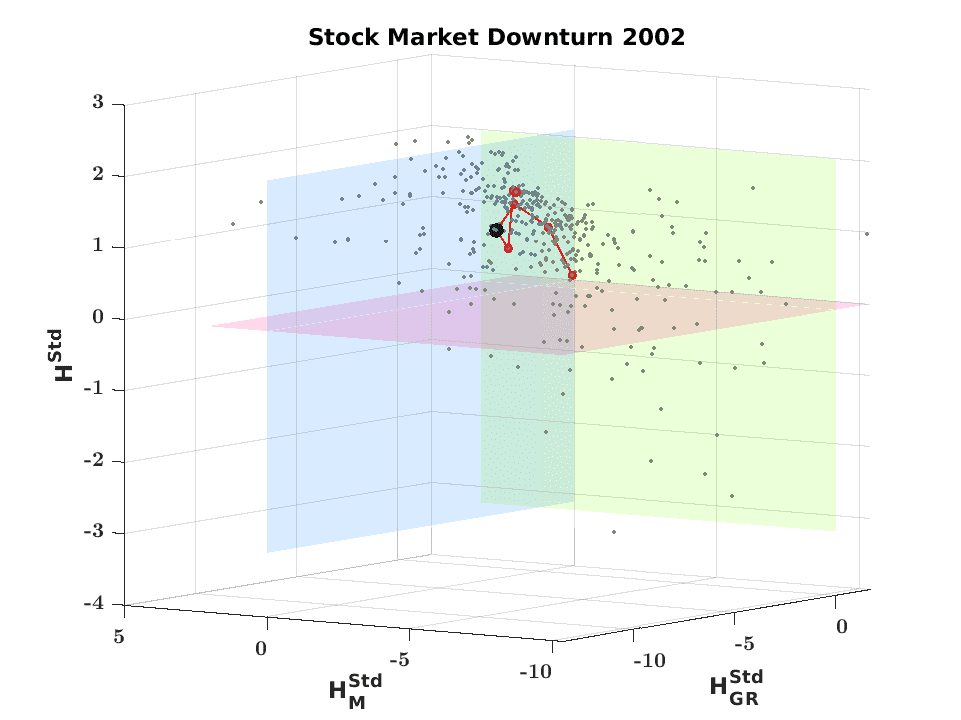}
	\includegraphics[width=0.325\linewidth]{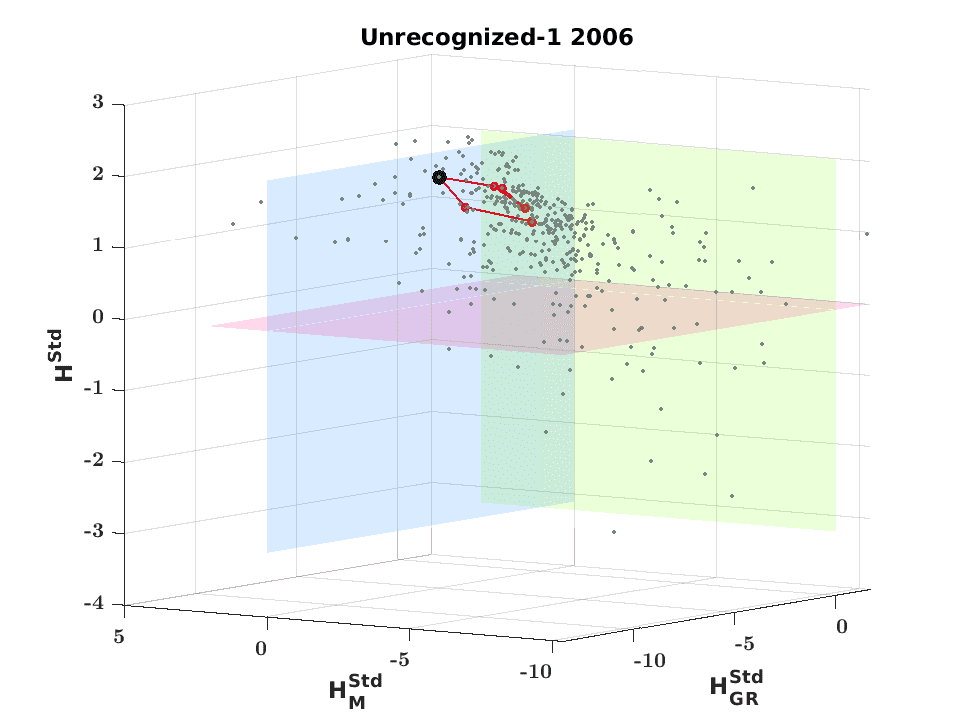}
	\includegraphics[width=0.325\linewidth]{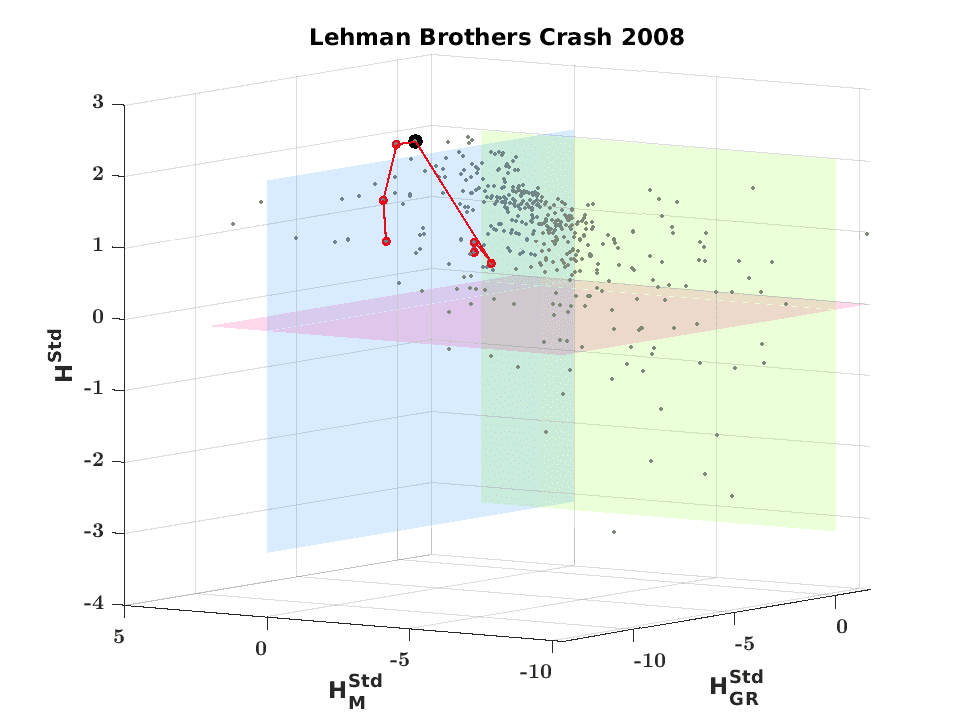}
	\includegraphics[width=0.325\linewidth]{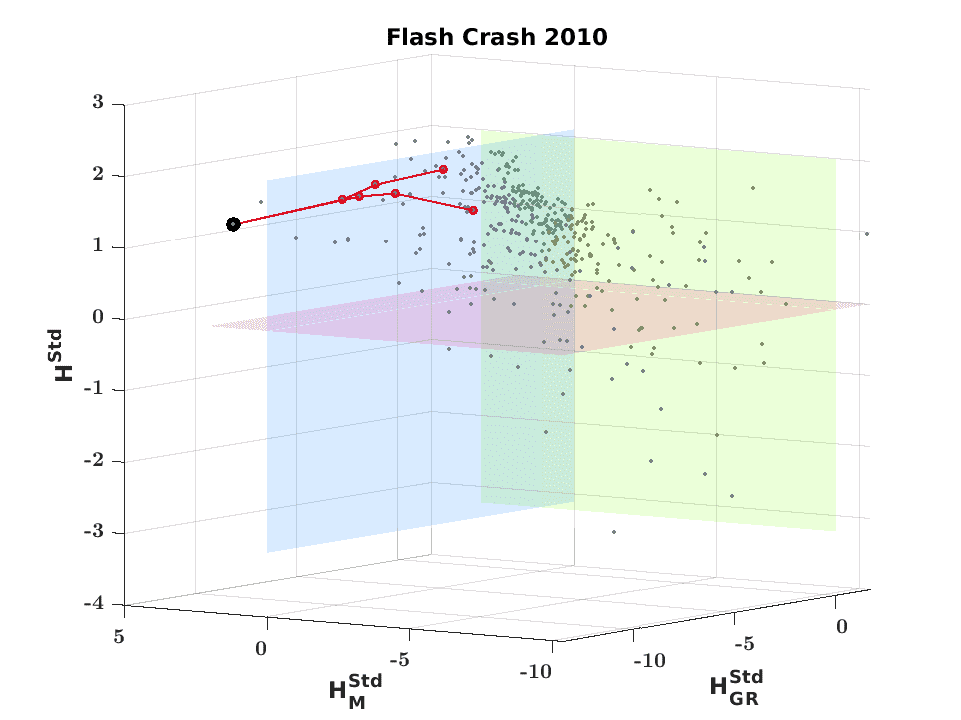}
	\includegraphics[width=0.325\linewidth]{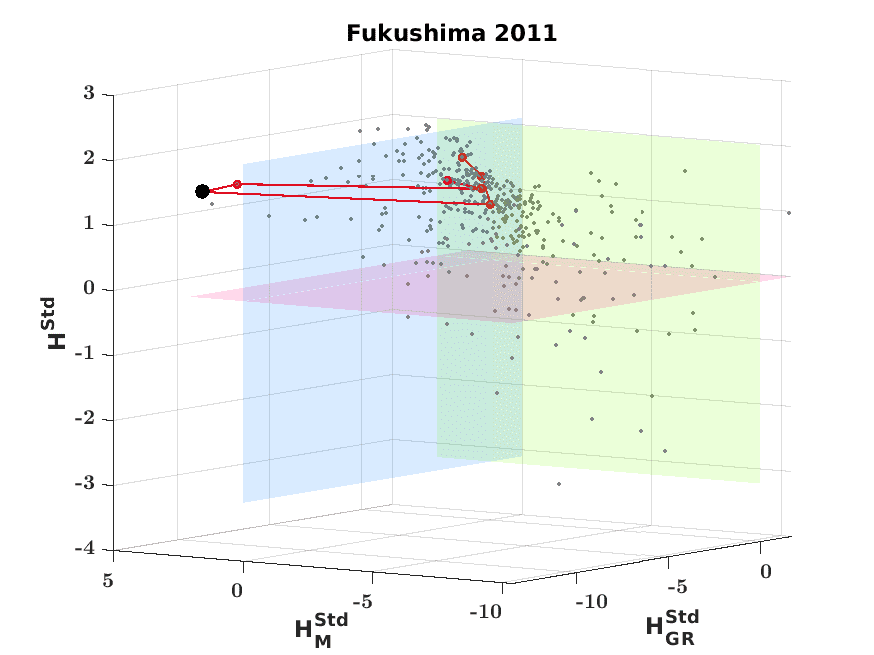}
	\includegraphics[width=0.325\linewidth]{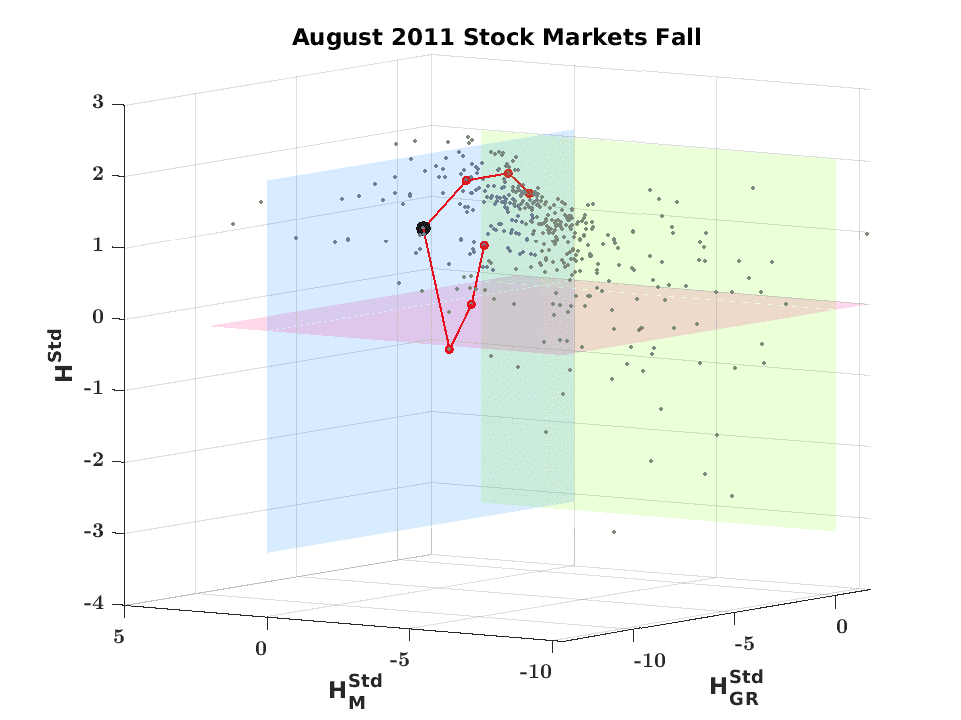}
	\includegraphics[width=0.325\linewidth]{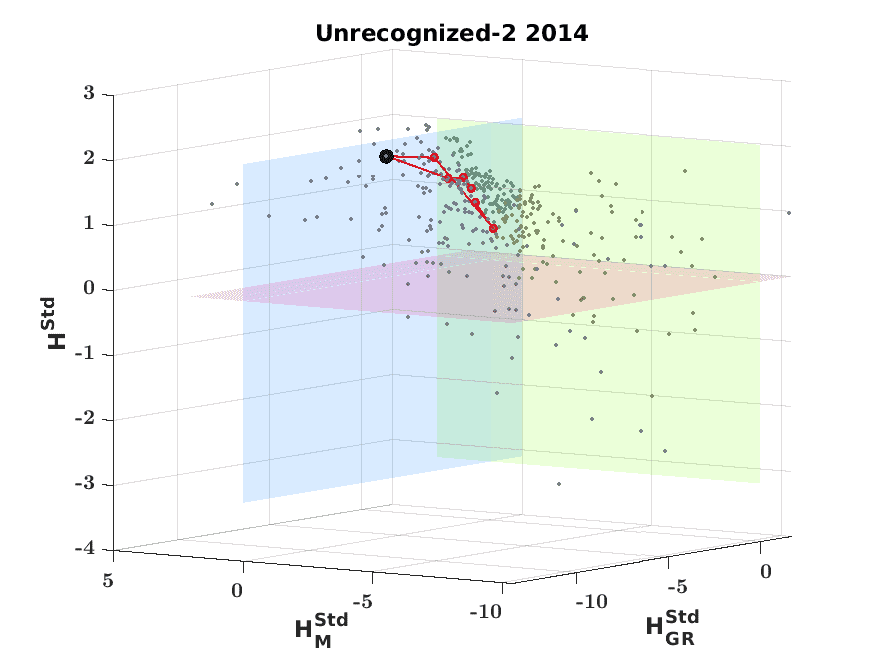}
	\includegraphics[width=0.325\linewidth]{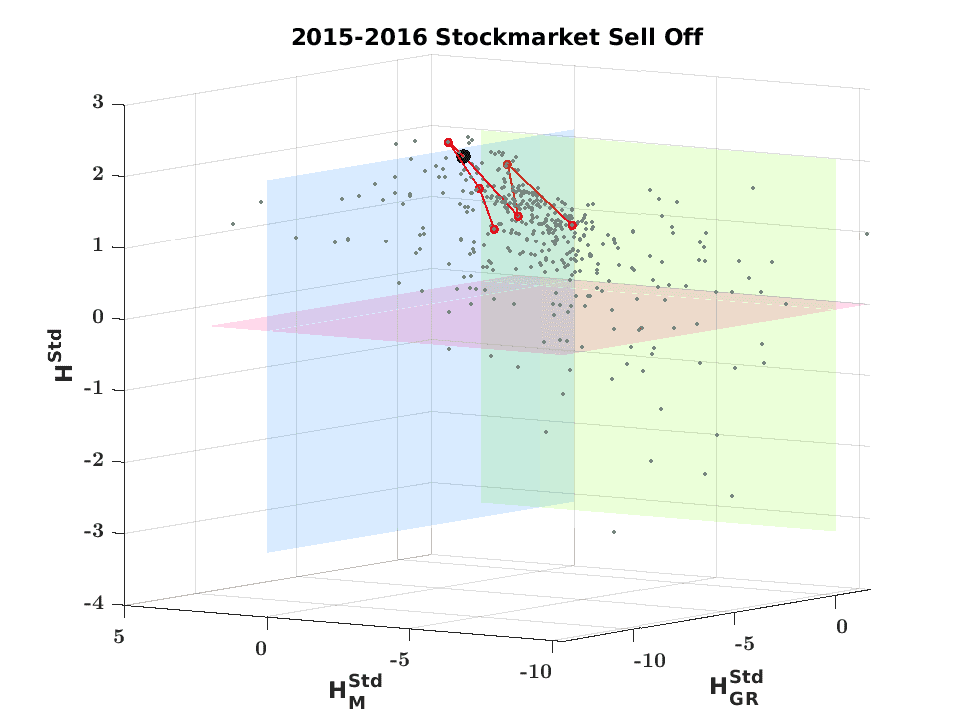}
	\includegraphics[width=0.325\linewidth]{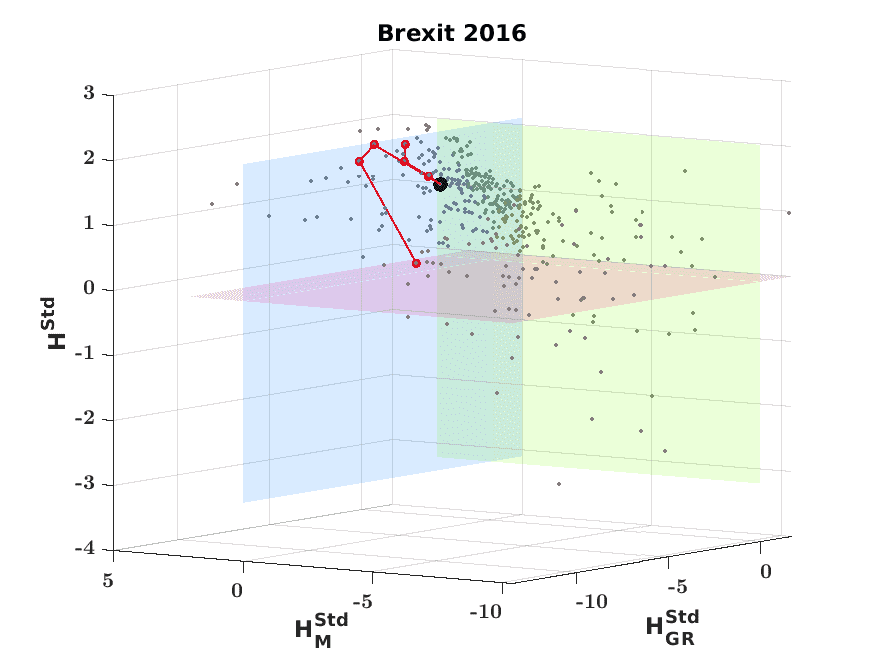}
	\caption{\textbf{Evolution around the important events in JPN market.} Eigen-entropy $H$ calculated from the correlation matrices: full $C$, market mode $C_M$ and group-random mode $C_{GR}$ for all the frames (epoch $M=40$ days and shift $\Delta=20$ days) over a period of 1985-2016 of JPN (Nikkei-225). Three co-ordinates axes $H^{std}$, $H^{std}_M$ and $H^{std}_{GR}$ are the standardized variables, same as Fig.~\ref{fig:usa_events}. Plots show thirteen important events from history. Critical events are marked with red lines except for the Dot-com bubble burst which is marked with blue line. The frame containing the important event is marked with black circle for better visibility.}\label{fig:jpn_events}
\end{figure*}

For the events listed in Table~\ref{table:crashes}, we look at the frames around that particular event and see how it moves around in the phase space in Fig.~\ref{fig:usa_events} and Fig.~\ref{fig:jpn_events}. 
We used a \textit{rolling mean} and \textit{rolling standard deviation} to study the \textit{standardized} eigen-entropies [$H^{Std}$, $H_{M}^{Std}$ and $H_{GR}^{Std}$]. The sequence of seven frames (three frames before, the event (in black), and three frames after) displayed order-disorder transitions in cases of major crashes and bubbles. The similar nature of the order-disorder transitions in all the major crashes and Dot-com bubbles, ten events in USA and thirteen events in JPN, certainly indicate robustness of the method.


\begin{figure*}
\centering
	\includegraphics[width=0.3\linewidth]{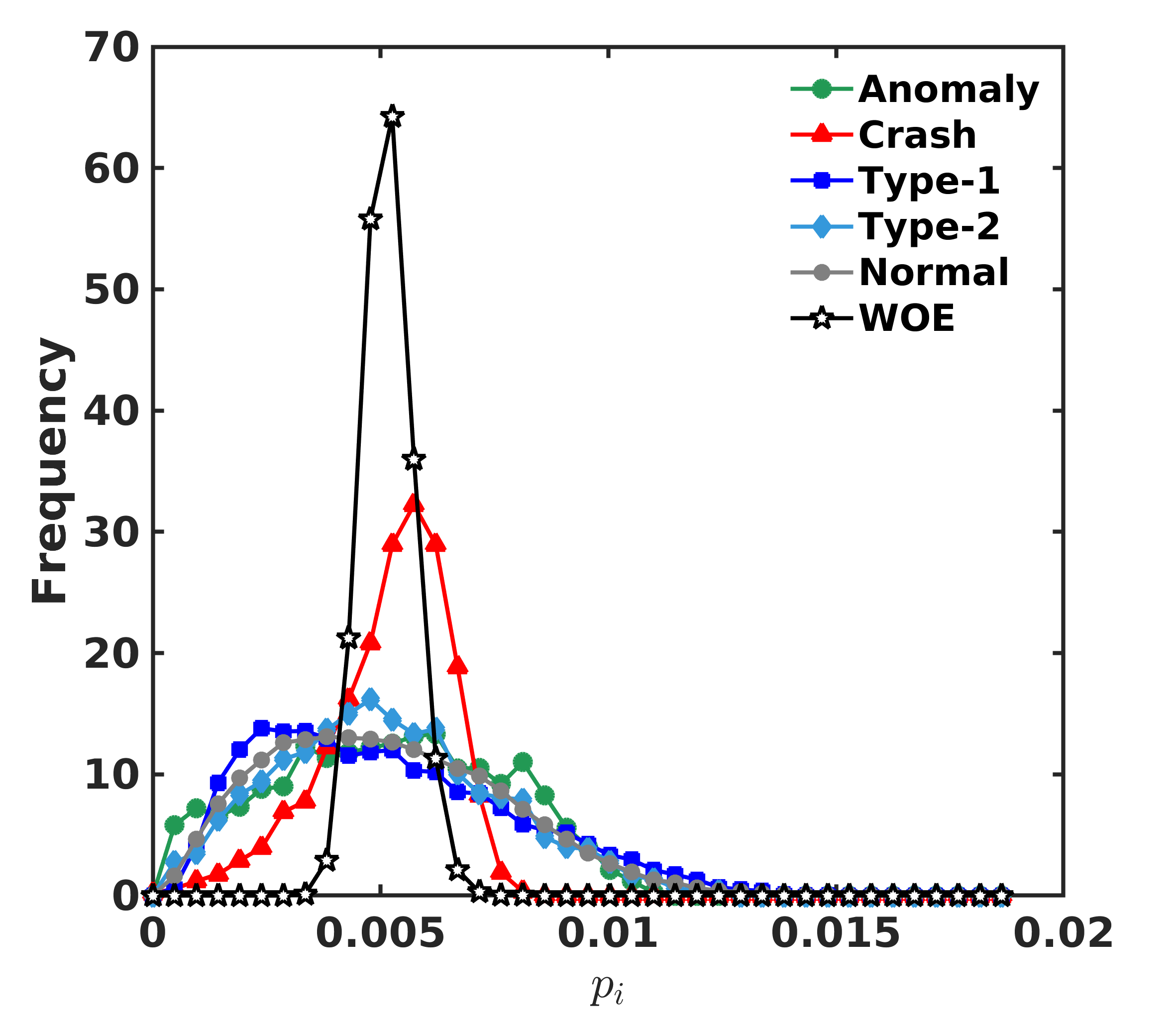}\llap{\parbox[b]{2.2in}{{\large\textsf{\textbf{a}}}\\\rule{0ex}{1.8in}}}
		\includegraphics[width=0.3\linewidth]{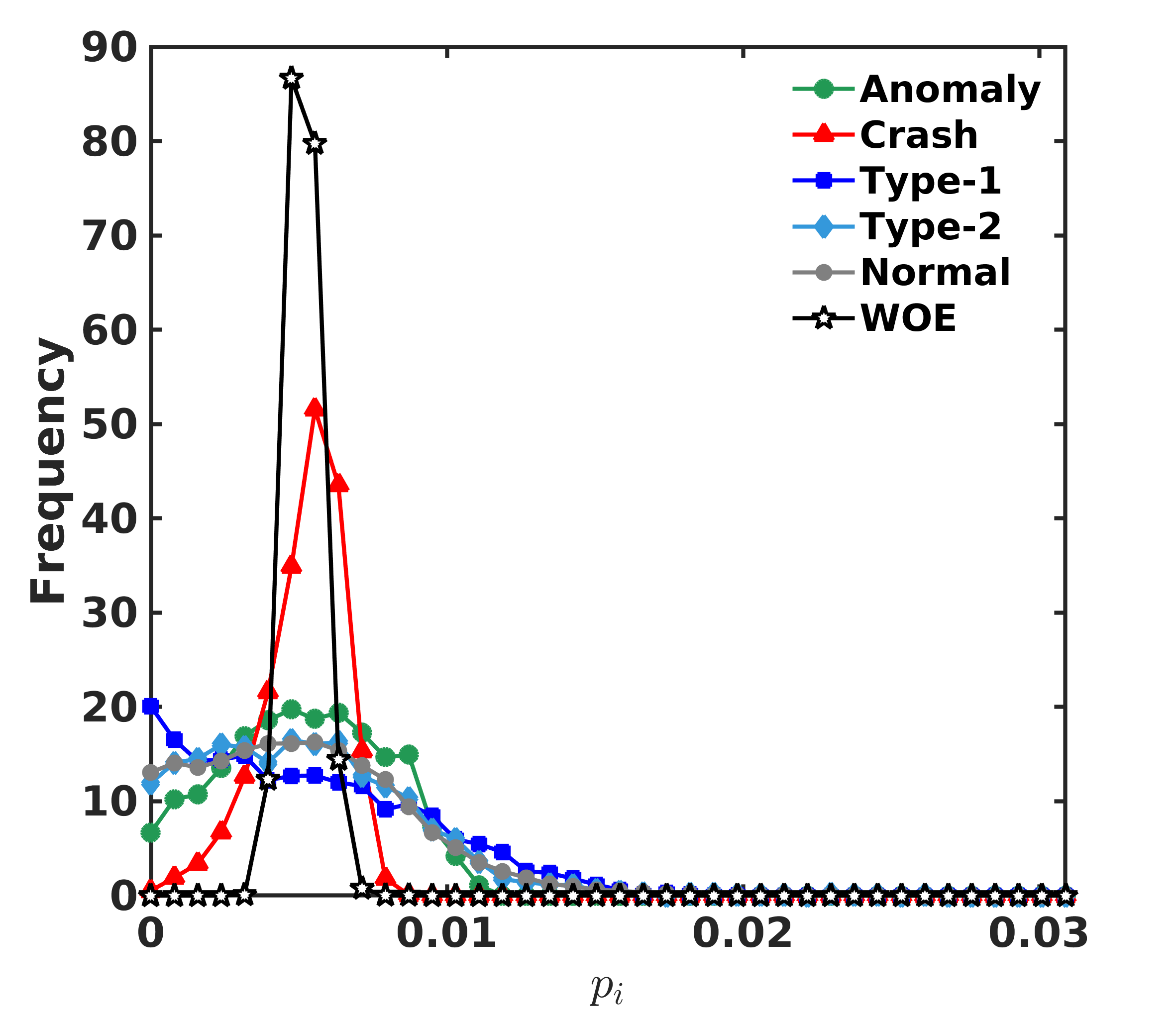}\llap{\parbox[b]{2.2in}{{\large\textsf{\textbf{b}}}\\\rule{0ex}{1.8in}}}
		\includegraphics[width=0.3\linewidth]{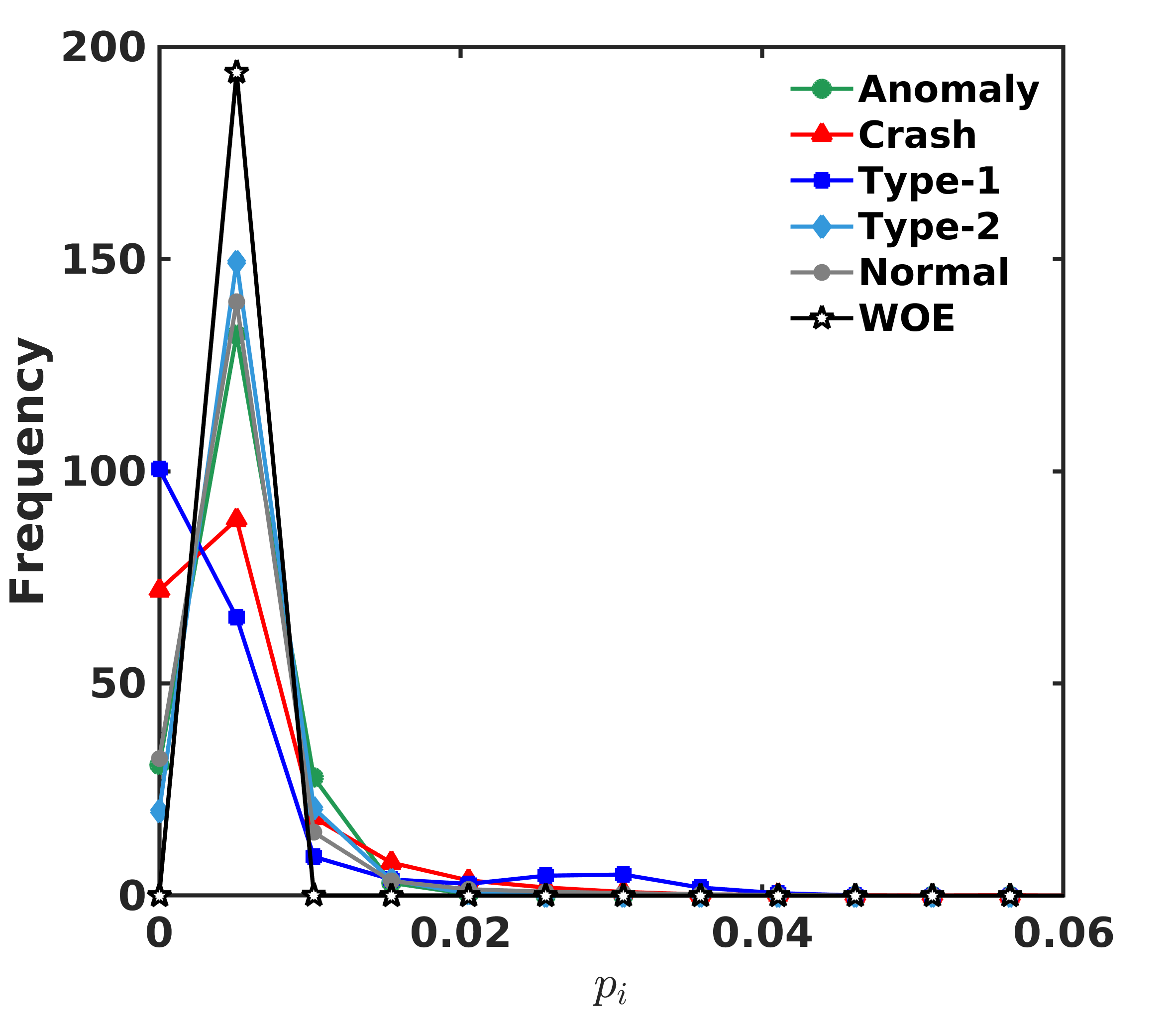}\llap{\parbox[b]{2.2in}{{\large\textsf{\textbf{c}}}\\\rule{0ex}{1.8in}}}
		\includegraphics[width=0.3\linewidth]{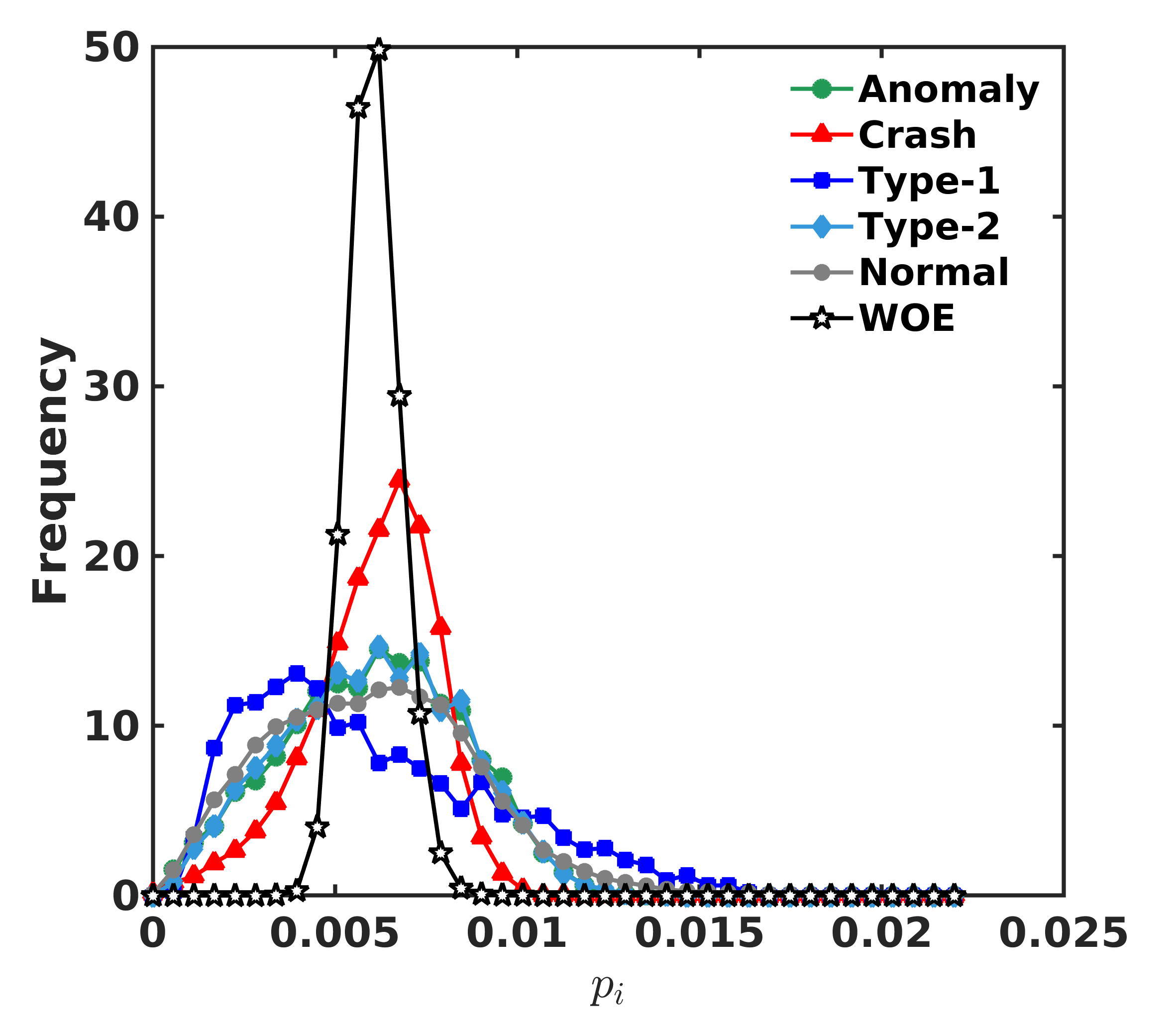}\llap{\parbox[b]{2.2in}{{\large\textsf{\textbf{d}}}\\\rule{0ex}{1.8in}}}
		\includegraphics[width=0.3\linewidth]{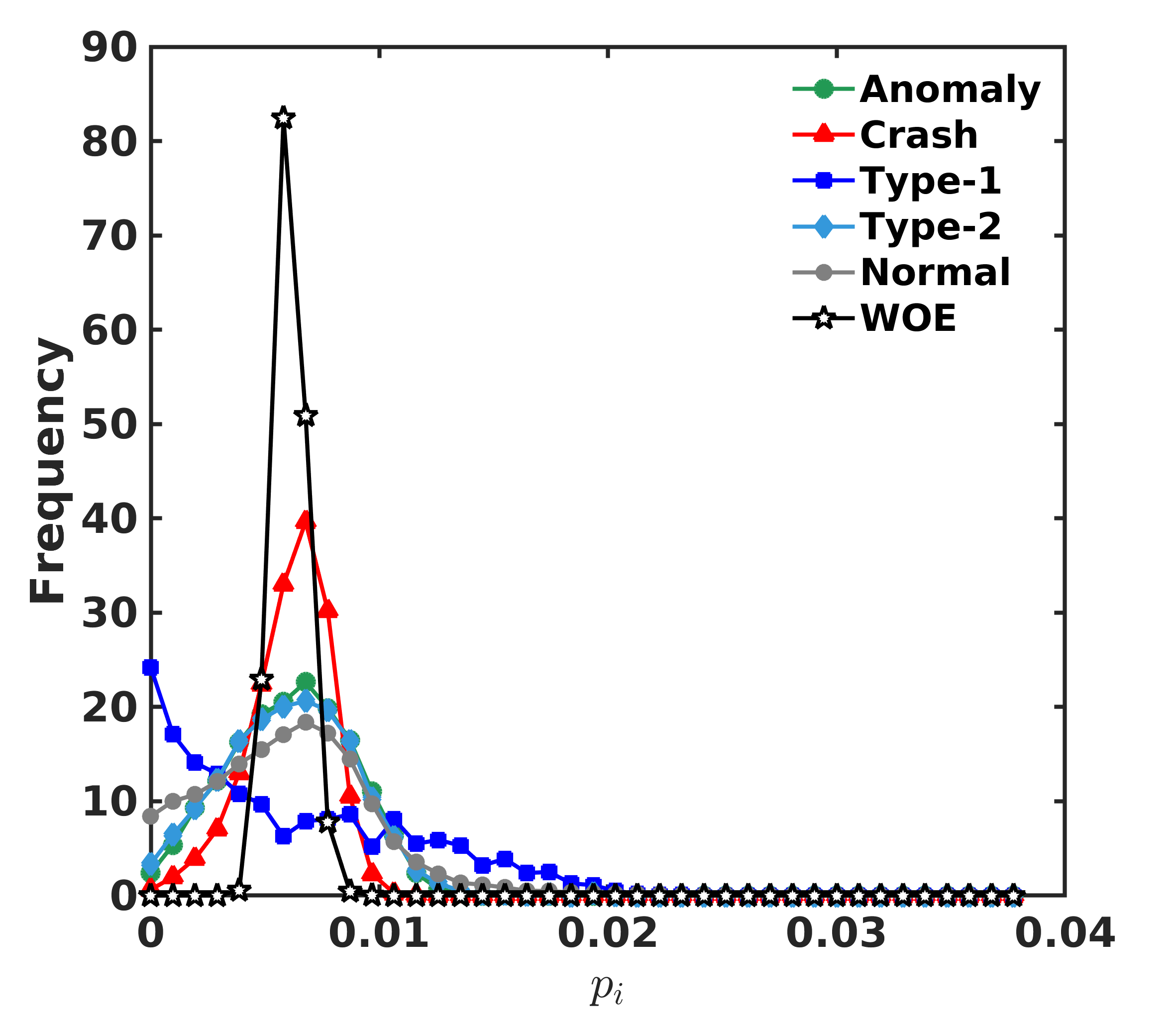}\llap{\parbox[b]{2.2in}{{\large\textsf{\textbf{e}}}\\\rule{0ex}{1.8in}}}
		\includegraphics[width=0.3\linewidth]{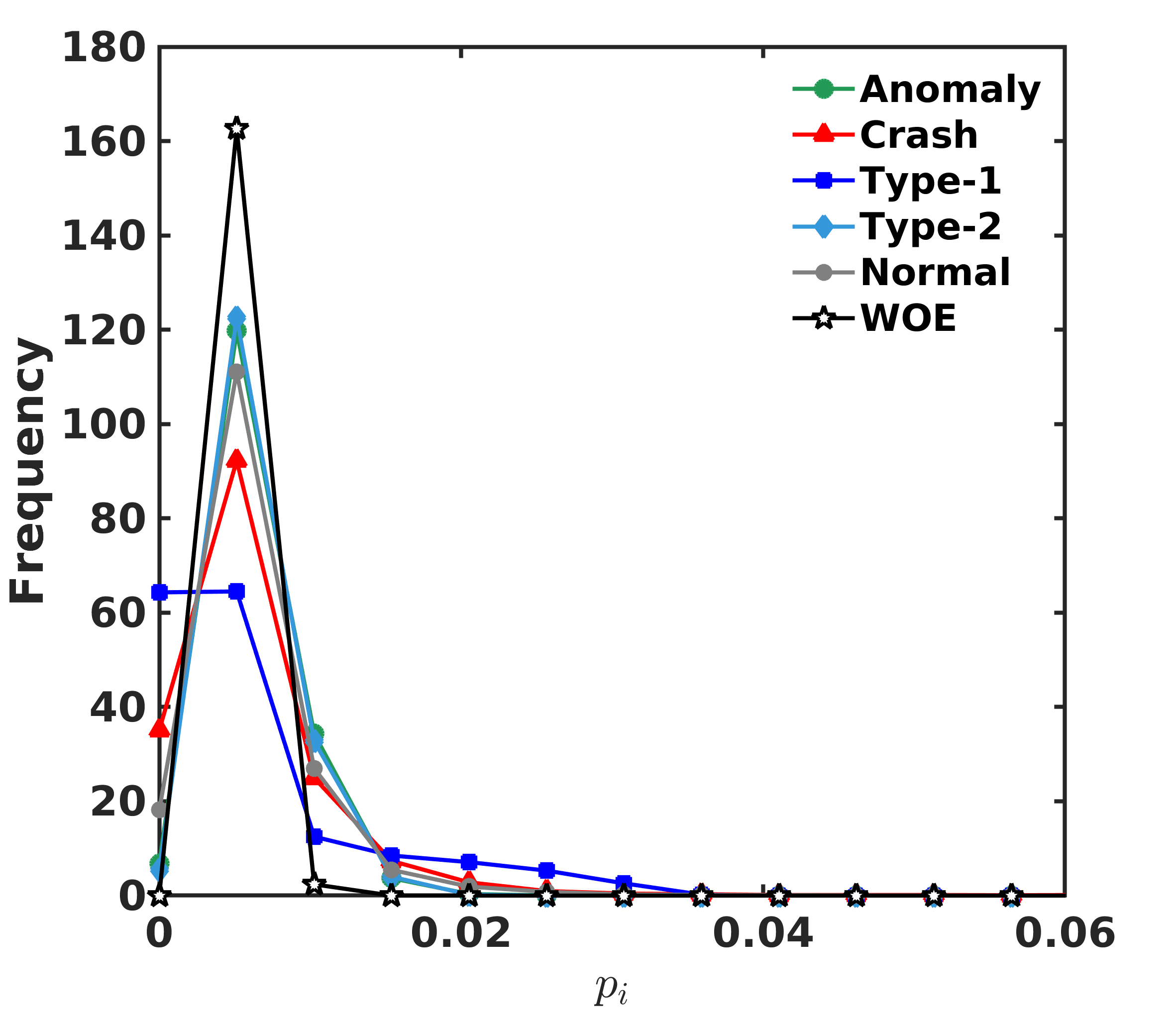}\llap{\parbox[b]{2.2in}{{\large\textsf{\textbf{f}}}\\\rule{0ex}{1.8in}}}
\caption{\textbf{Averaged distributions of the eigen-centralities, showing self-averaging properties.} Histograms of the eigen-centralities $p_i$ for anomalies (green circles), type-1 events (light blue diamonds), type-2 events (blue sqaures), crashes (red triangles) and normal (grey stars) and WOE (black squares), averaged over the respective ensembles  for USA (top row) and for JPN (bottom row). Histograms are evaluated using (\textbf{a} and \textbf{d}) full correlation matrices $C$  and decomposed correlation matrices of (\textbf{b} and \textbf{e}) market mode $C_M$, and (\textbf{c} and \textbf{f}) group and random mode $C_{GR}$.}
\label{fig:Comparison} 
\end{figure*}
Once we are able to characterize the epochs (event frames) into different ``phases'', we can create the different ensembles of anomalies, type-1 events, type-2 events, crashes and normal events. For each type of event, we find that eigen-centralities have distinct ranges of values and the sorted eigen-centrality curves have interesting features (hierarchies) in the eigenmodes. The eigen-entropies actually quantify these features appropriately. For the S\&P-500 and Nikkei-225 markets, we compute the histograms of the eigen-centralities $p_i$.
Fig.~\ref{fig:Comparison} shows  the  histograms (for S\&P-500 (\textit{Top}) and Nikkei-225 (\textit{Bottom}))  for all the characterized anomalies (green circles), type-1 events (light blue diamonds), type-2 events (blue squares), crashes (red triangles),  normal events (grey stars), averaged over the respective ensembles, for the full/decomposed matrices. For comparison, we also plot the results for the WOE (black squares). This helps us understand what actually happens in the market, during these different types of events (phases) and what type of hierarchies exist within the stocks's eigen-centralities. This would shed new light into the understanding of formation of type-1 events, their development and crashes, etc.


\subsubsection*{Cross-correlogram of the market indicators}

\begin{figure*}[h!]
\centering
\includegraphics[width=0.9\linewidth]{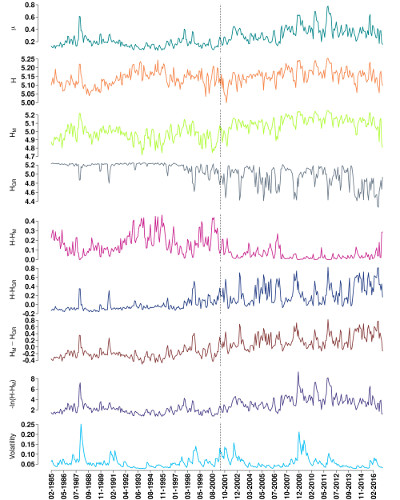}
\caption{\textbf{Temporal evolution of the USA market indicators.} Market behaves differently before and after the vertical dash line at April 2001~\cite{fed_move}.}
\label{fig:MS_TS_USA}
\end{figure*}

\begin{figure*}[h!]
\centering
\includegraphics[width=0.9\linewidth]{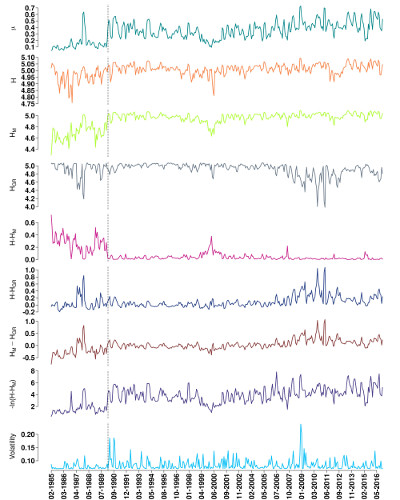}
\caption{\textbf{Temporal evolution of the JPN market indicators.} Market behaves differently before and after the vertical dash line at Feb 1990~\cite{reszat2003japan,hayashi20021990s}.}
\label{fig:MS_TS_JPN}
\end{figure*}

\begin{figure*}[h!]
\centering
\includegraphics[width=0.45\linewidth]{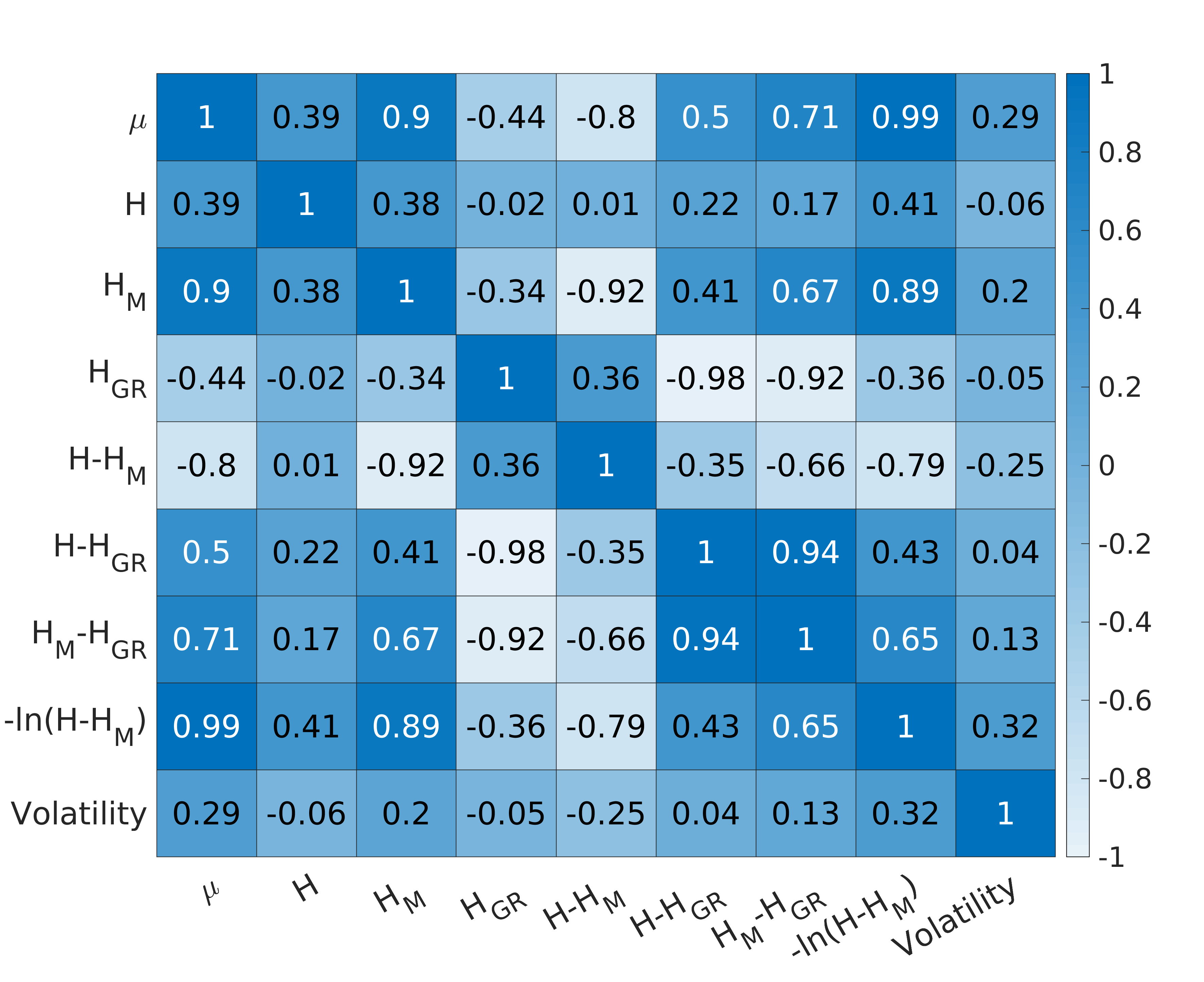}\llap{\parbox[b]{2.9in}{{\large\textsf{\textbf{a}}}\\\rule{0ex}{2.4in}}}
\includegraphics[width=0.45\linewidth]{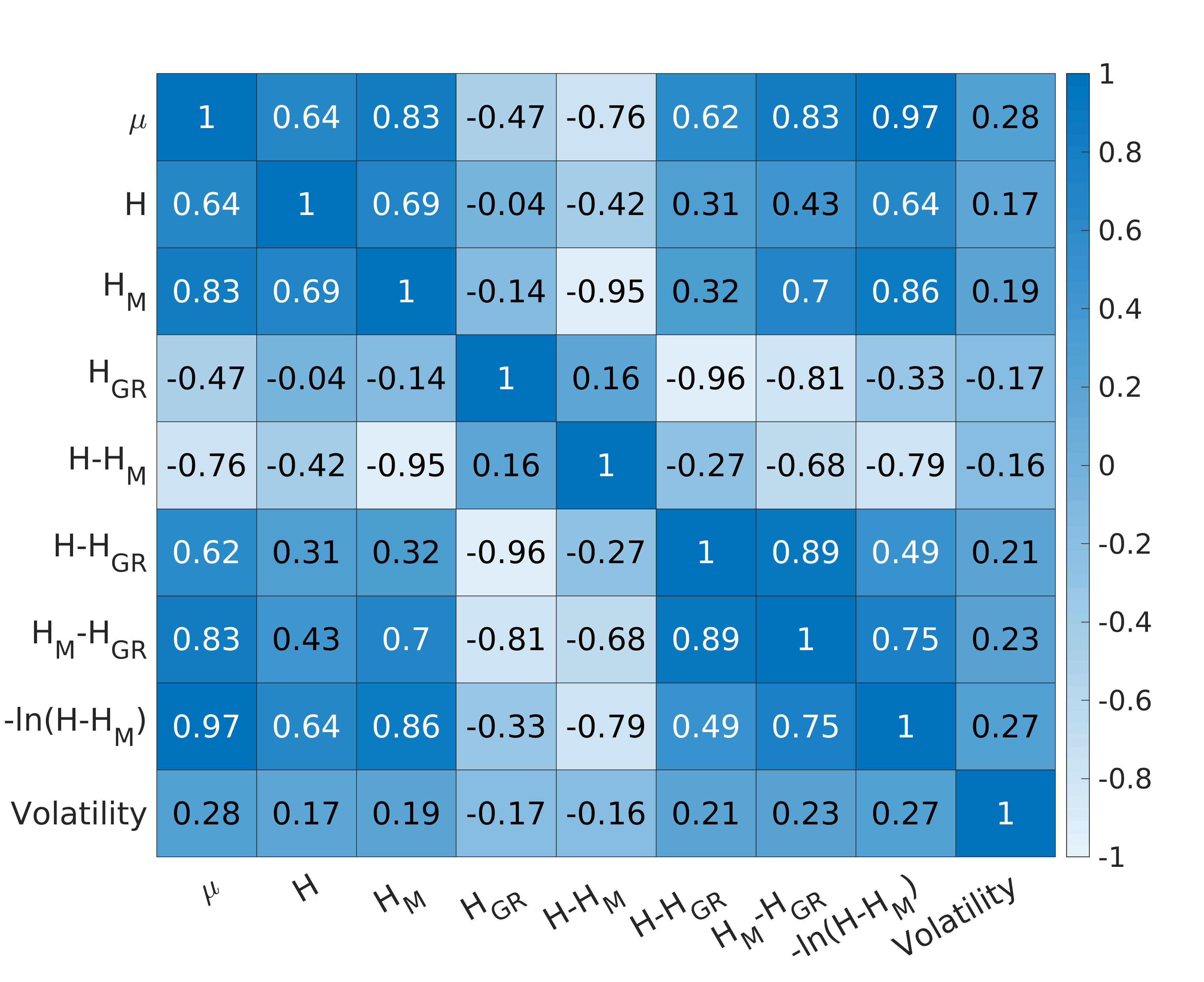}\llap{\parbox[b]{2.9in}{{\large\textsf{\textbf{b}}}\\\rule{0ex}{2.4in}}}
\caption{\textbf{Cross-correlograms of the market indicators}. For (a) USA and (b) JPN.}
\label{fig:correlogram}
\end{figure*}

We have studied the time evolution of several market indicators of USA and JPN markets, shown in Figs.~\ref{fig:MS_TS_USA} and ~\ref{fig:MS_TS_JPN}, respectively,  that can be used for the continuous monitoring of the financial markets. The corresponding cross-correlograms of the market indicators are plotted for USA and JPN in Fig. \ref{fig:correlogram}. Note that in JPN market, only once $H-H_M$ takes a negative value (ignored while plotting); interestingly, this coincides with the period when Japan ended zero-rate policy on $14-07-2006$~\cite{jpn_zerorate}.

\begin{table}[h]
\centering
\caption{List of all stocks of USA market (S\&P-500) considered for the analysis. The first column has the serial number,
the second column has the abbreviation, the third column has the full name of the stock, and the
fourth column specifies the sector as given in the S\&P-500.}
\label{USA_Table}
\begin{tabular}{|l|l|l|l|l|}
\hline
\textbf{S.No.} & \textbf{Code} & \textbf{Company Name} & \textbf{Sector} & \textbf{Abbrv} \\ \hline
1 & CMCSA & Comcast Corp. & Consumer Discretionary & CD \\ \hline
2 & DIS & The Walt Disney Company & Consumer Discretionary & CD \\ \hline
3 & F & Ford Motor & Consumer Discretionary & CD \\ \hline
4 & GPC & Genuine Parts & Consumer Discretionary & CD \\ \hline
5 & GPS & Gap Inc. & Consumer Discretionary & CD \\ \hline
6 & GT & Goodyear Tire \& Rubber & Consumer Discretionary & CD \\ \hline
7 & HAS & Hasbro Inc. & Consumer Discretionary & CD \\ \hline
8 & HD & Home Depot & Consumer Discretionary & CD \\ \hline
9 & HRB & Block H\&R & Consumer Discretionary & CD \\ \hline
10 & IPG & Interpublic Group & Consumer Discretionary & CD \\ \hline
11 & JCP & J. C. Penney Company, Inc. & Consumer Discretionary & CD \\ \hline
12 & JWN & Nordstrom & Consumer Discretionary & CD \\ \hline
13 & LEG & Leggett \& Platt & Consumer Discretionary & CD \\ \hline
14 & LEN & Lennar Corp. & Consumer Discretionary & CD \\ \hline
15 & LOW & Lowe's Cos. & Consumer Discretionary & CD \\ \hline
16 & MAT & Mattel Inc. & Consumer Discretionary & CD \\ \hline
17 & MCD & McDonald's Corp. & Consumer Discretionary & CD \\ \hline
18 & NKE & Nike & Consumer Discretionary & CD \\ \hline
19 & SHW & Sherwin-Williams & Consumer Discretionary & CD \\ \hline
20 & TGT & Target Corp. & Consumer Discretionary & CD \\ \hline
21 & VFC & V.F. Corp. & Consumer Discretionary & CD \\ \hline
22 & WHR & Whirlpool Corp. & Consumer Discretionary & CD \\ \hline
23 & ADM & Archer-Daniels-Midland Co & Consumer Staples & CS \\ \hline
24 & AVP & Avon Products, Inc. & Consumer Staples & CS \\ \hline
25 & CAG & Conagra Brands & Consumer Staples & CS \\ \hline
26 & CL & Colgate-Palmolive & Consumer Staples & CS \\ \hline
27 & CPB & Campbell Soup & Consumer Staples & CS \\ \hline
28 & CVS & CVS Health & Consumer Staples & CS \\ \hline
29 & GIS & General Mills & Consumer Staples & CS \\ \hline
30 & HRL & Hormel Foods Corp. & Consumer Staples & CS \\ \hline
31 & HSY & The Hershey Company & Consumer Staples & CS \\ \hline
32 & K & Kellogg Co. & Consumer Staples & CS \\ \hline
33 & KMB & Kimberly-Clark & Consumer Staples & CS \\ \hline
34 & KO & Coca-Cola Company (The) & Consumer Staples & CS \\ \hline
35 & KR & Kroger Co. & Consumer Staples & CS \\ \hline
36 & MKC & McCormick \& Co. & Consumer Staples & CS \\ \hline
37 & MO & Altria Group Inc & Consumer Staples & CS \\ \hline
38 & SYY & Sysco Corp. & Consumer Staples & CS \\ \hline
39 & TAP & Molson Coors Brewing Company & Consumer Staples & CS \\ \hline
40 & TSN & Tyson Foods & Consumer Staples & CS \\ \hline
41 & WMT & Wal-Mart Stores & Consumer Staples & CS \\ \hline
42 & APA & Apache Corporation & Energy & EG \\ \hline
43 & COP & ConocoPhillips & Energy & EG \\ \hline
44 & CVX & Chevron Corp. & Energy & EG \\ \hline
45 & ESV & Ensco plc & Energy & EG \\ \hline
46 & HAL & Halliburton Co. & Energy & EG \\ \hline
47 & HES & Hess Corporation & Energy & EG \\ \hline
48 & HP & Helmerich \& Payne & Energy & EG \\ \hline
49 & MRO & Marathon Oil Corp. & Energy & EG \\ \hline
50 & MUR & Murphy Oil Corporation & Energy & EG \\ \hline

\end{tabular}
\end{table}
\begin{table}[]
\centering
\begin{tabular}{|l|l|l|l|l|}
\hline
51 & NBL & Noble Energy Inc & Energy & EG \\ \hline
52 & NBR & Nabors Industries Ltd. & Energy & EG \\ \hline
53 & SLB & Schlumberger Ltd. & Energy & EG \\ \hline
54 & TSO & Tesoro Corp & Energy & EG \\ \hline
55 & VLO & Valero Energy & Energy & EG \\ \hline
56 & WMB & Williams Cos. & Energy & EG \\ \hline
57 & XOM & Exxon Mobil Corp. & Energy & EG \\ \hline
58 & AFL & AFLAC Inc & Financials & FN \\ \hline
59 & AIG & American International Group, Inc. & Financials & FN \\ \hline
60 & AON & Aon plc & Financials & FN \\ \hline
61 & AXP & American Express Co & Financials & FN \\ \hline
62 & BAC & Bank of America Corp & Financials & FN \\ \hline
63 & BBT & BB\&T Corporation & Financials & FN \\ \hline
64 & BEN & Franklin Resources & Financials & FN \\ \hline
65 & BK & The Bank of New York Mellon Corp. & Financials & FN \\ \hline
66 & C & Citigroup Inc. & Financials & FN \\ \hline
67 & CB & Chubb Limited & Financials & FN \\ \hline
68 & CINF & Cincinnati Financial & Financials & FN \\ \hline
69 & CMA & Comerica Inc. & Financials & FN \\ \hline
70 & EFX & Equifax Inc. & Financials & FN \\ \hline
71 & FHN & First Horizon National Corporation & Financials & FN \\ \hline
72 & HBAN & Huntington Bancshares & Financials & FN \\ \hline
73 & HCN & Welltower Inc. & Financials & FN \\ \hline
74 & HST & Host Hotels \& Resorts, Inc. & Financials & FN \\ \hline
75 & JPM & JPMorgan Chase \& Co. & Financials & FN \\ \hline
76 & L & Loews Corp. & Financials & FN \\ \hline
77 & LM & Legg Mason, Inc. & Financials & FN \\ \hline
78 & LNC & Lincoln National & Financials & FN \\ \hline
79 & LUK & Leucadia National Corp. & Financials & FN \\ \hline
80 & MMC & Marsh \& McLennan & Financials & FN \\ \hline
81 & MTB & M\&T Bank Corp. & Financials & FN \\ \hline
82 & PSA & Public Storage & Financials & FN \\ \hline
83 & SLM & SLM Corporation & Financials & FN \\ \hline
84 & TMK & Torchmark Corp. & Financials & FN \\ \hline
85 & TRV & The Travelers Companies Inc. & Financials & FN \\ \hline
86 & USB & U.S. Bancorp & Financials & FN \\ \hline
87 & VNO & Vornado Realty Trust & Financials & FN \\ \hline
88 & WFC & Wells Fargo & Financials & FN \\ \hline
89 & WY & Weyerhaeuser Corp. & Financials & FN \\ \hline
90 & ZION & Zions Bancorp & Financials & FN \\ \hline
91 & ABT & Abbott Laboratories & Health Care & HC \\ \hline
92 & AET & Aetna Inc & Health Care & HC \\ \hline
93 & AMGN & Amgen Inc & Health Care & HC \\ \hline
94 & BAX & Baxter International Inc. & Health Care & HC \\ \hline
95 & BCR & Bard (C.R.) Inc. & Health Care & HC \\ \hline
96 & BDX & Becton Dickinson & Health Care & HC \\ \hline
97 & BMY & Bristol-Myers Squibb & Health Care & HC \\ \hline
98 & CAH & Cardinal Health Inc. & Health Care & HC \\ \hline
99 & CI & CIGNA Corp. & Health Care & HC \\ \hline
100 & HUM & Humana Inc. & Health Care & HC \\ \hline

\end{tabular}
\end{table}

\begin{table}[]
\centering
\begin{tabular}{|l|l|l|l|l|}
\hline
101 & JNJ & Johnson \& Johnson & Health Care & HC \\ \hline
102 & LLY & Lilly (Eli) \& Co. & Health Care & HC \\ \hline
103 & MDT & Medtronic plc & Health Care & HC \\ \hline
104 & MRK & Merck \& Co. & Health Care & HC \\ \hline
105 & MYL & Mylan N.V. & Health Care & HC \\ \hline
106 & SYK & Stryker Corp. & Health Care & HC \\ \hline
107 & THC & Tenet Healthcare Corp & Health Care & HC \\ \hline
108 & TMO & Thermo Fisher Scientific & Health Care & HC \\ \hline
109 & UNH & United Health Group Inc. & Health Care & HC \\ \hline
110 & VAR & Varian Medical Systems & Health Care & HC \\ \hline
111 & AVY & Avery Dennison Corp & Industrials & ID \\ \hline
112 & BA & Boeing Company & Industrials & ID \\ \hline
113 & CAT & Caterpillar Inc. & Industrials & ID \\ \hline
114 & CMI & Cummins Inc. & Industrials & ID \\ \hline
115 & CSX & CSX Corp. & Industrials & ID \\ \hline
116 & CTAS & Cintas Corporation & Industrials & ID \\ \hline
117 & DE & Deere \& Co. & Industrials & ID \\ \hline
118 & DHR & Danaher Corp. & Industrials & ID \\ \hline
119 & DNB & The Dun \& Bradstreet Corporation & Industrials & ID \\ \hline
120 & DOV & Dover Corp. & Industrials & ID \\ \hline
121 & EMR & Emerson Electric Company & Industrials & ID \\ \hline
122 & ETN & Eaton Corporation & Industrials & ID \\ \hline
123 & EXPD & Expeditors International & Industrials & ID \\ \hline
124 & FDX & FedEx Corporation & Industrials & ID \\ \hline
125 & FLS & Flowserve Corporation & Industrials & ID \\ \hline
126 & GD & General Dynamics & Industrials & ID \\ \hline
127 & GE & General Electric & Industrials & ID \\ \hline
128 & GLW & Corning Inc. & Industrials & ID \\ \hline
129 & GWW & Grainger (W.W.) Inc. & Industrials & ID \\ \hline
130 & HON & Honeywell Int'l Inc. & Industrials & ID \\ \hline
131 & IR & Ingersoll-Rand PLC & Industrials & ID \\ \hline
132 & ITW & Illinois Tool Works & Industrials & ID \\ \hline
133 & JEC & Jacobs Engineering Group & Industrials & ID \\ \hline
134 & LMT & Lockheed Martin Corp. & Industrials & ID \\ \hline
135 & LUV & Southwest Airlines & Industrials & ID \\ \hline
136 & MAS & Masco Corp. & Industrials & ID \\ \hline
137 & MMM & 3M Company & Industrials & ID \\ \hline
138 & ROK & Rockwell Automation Inc. & Industrials & ID \\ \hline
139 & RTN & Raytheon Co. & Industrials & ID \\ \hline
140 & TXT & Textron Inc. & Industrials & ID \\ \hline
141 & UNP & Union Pacific & Industrials & ID \\ \hline
142 & UTX & United Technologies & Industrials & ID \\ \hline
143 & AAPL & Apple Inc. & Information Technology & IT \\ \hline
144 & ADI & Analog Devices, Inc. & Information Technology & IT \\ \hline
145 & ADP & Automatic Data Processing & Information Technology & IT \\ \hline
146 & AMAT & Applied Materials Inc & Information Technology & IT \\ \hline
147 & AMD & Advanced Micro Devices Inc & Information Technology & IT \\ \hline
148 & CA & CA, Inc. & Information Technology & IT \\ \hline
149 & HPQ & HP Inc. & Information Technology & IT \\ \hline
150 & HRS & Harris Corporation & Information Technology & IT \\ \hline

\end{tabular}
\end{table}
\begin{table}[]
\centering
\begin{tabular}{|l|l|l|l|l|}
\hline
151 & IBM & International Business Machines & Information Technology & IT \\ \hline
152 & INTC & Intel Corp. & Information Technology & IT \\ \hline
153 & KLAC & KLA-Tencor Corp. & Information Technology & IT \\ \hline
154 & LRCX & Lam Research & Information Technology & IT \\ \hline
155 & MSI & Motorola Solutions Inc. & Information Technology & IT \\ \hline
156 & MU & Micron Technology & Information Technology & IT \\ \hline
157 & TSS & Total System Services, Inc. & Information Technology & IT \\ \hline
158 & TXN & Texas Instruments & Information Technology & IT \\ \hline
159 & WDC & Western Digital & Information Technology & IT \\ \hline
160 & XRX & Xerox Corp. & Information Technology & IT \\ \hline
161 & AA & Alcoa Corporation & Materials & MT \\ \hline
162 & APD & Air Products \& Chemicals Inc & Materials & MT \\ \hline
163 & BLL & Ball Corp & Materials & MT \\ \hline
164 & BMS & Bemis Company, Inc. & Materials & MT \\ \hline
165 & CLF & Cleveland-Cliffs Inc. & Materials & MT \\ \hline
166 & DD & DuPont & Materials & MT \\ \hline
167 & ECL & Ecolab Inc. & Materials & MT \\ \hline
168 & FMC & FMC Corporation & Materials & MT \\ \hline
169 & IFF & Intl Flavors \& Fragrances & Materials & MT \\ \hline
170 & IP & International Paper & Materials & MT \\ \hline
171 & NEM & Newmont Mining Corporation & Materials & MT \\ \hline
172 & PPG & PPG Industries & Materials & MT \\ \hline
173 & VMC & Vulcan Materials & Materials & MT \\ \hline
174 & CTL & CenturyLink Inc & Telecommunication Services & TC \\ \hline
175 & FTR & Frontier Communications Corporation & Telecommunication Services & TC \\ \hline
176 & S & Sprint Nextel Corp. & Telecommunication Services & TC \\ \hline
177 & T & AT\&T Inc & Telecommunication Services & TC \\ \hline
178 & VZ & Verizon Communications & Telecommunication Services & TC \\ \hline
179 & AEP & American Electric Power & Utilities & UT \\ \hline
180 & CMS & CMS Energy & Utilities & UT \\ \hline
181 & CNP & CenterPoint Energy & Utilities & UT \\ \hline
182 & D & Dominion Energy & Utilities & UT \\ \hline
183 & DTE & DTE Energy Co. & Utilities & UT \\ \hline
184 & ED & Consolidated Edison & Utilities & UT \\ \hline
185 & EIX & Edison Int'l & Utilities & UT \\ \hline
186 & EQT & EQT Corporation & Utilities & UT \\ \hline
187 & ETR & Entergy Corp. & Utilities & UT \\ \hline
188 & EXC & Exelon Corp. & Utilities & UT \\ \hline
189 & NEE & NextEra Energy & Utilities & UT \\ \hline
190 & NI & NiSource Inc. & Utilities & UT \\ \hline
191 & PNW & Pinnacle West Capital & Utilities & UT \\ \hline
192 & SO & Southern Co. & Utilities & UT \\ \hline
193 & WEC & Wec Energy Group Inc & Utilities & UT \\ \hline
194 & XEL & Xcel Energy Inc & Utilities & UT \\ \hline

\end{tabular}
\end{table}

\begin{table}[]
\centering
\caption{List of all stocks of Japan market (Nikkei-225) considered for the analysis. The first column has the serial number,
the second column has the abbreviation, the third column has the full name of the stock, and the
fourth column specifies the sector as given in the Nikkei-225.}
\label{JPN_Table}
\begin{tabular}{|l|l|l|l|l|}
\hline
\textbf{S.No.} & \textbf{Code} & \textbf{Company Name} & \textbf{Sector} & \textbf{Abbrv} \\ \hline
1 & S-8801 & MITSUI FUDOSAN CO., LTD. & Capital Goods & CG \\ \hline
2 & S-8802 & MITSUBISHI ESTATE CO., LTD. & Capital Goods & CG \\ \hline
3 & S-8804 & TOKYO TATEMONO CO., LTD. & Capital Goods & CG \\ \hline
4 & S-8830 & SUMITOMO REALTY \& DEVELOPMENT CO., LTD. & Capital Goods & CG \\ \hline
5 & S-7003 & MITSUI ENG. \& SHIPBUILD. CO., LTD. & Capital Goods & CG \\ \hline
6 & S-7012 & KAWASAKI HEAVY IND., LTD. & Capital Goods & CG \\ \hline
7 & S-9202 & ANA HOLDINGS INC. & Capital Goods & CG \\ \hline
8 & S-1801 & TAISEI CORP. & Capital Goods & CG \\ \hline
9 & S-1802 & OBAYASHI CORP. & Capital Goods & CG \\ \hline
10 & S-1803 & SHIMIZU CORP. & Capital Goods & CG \\ \hline
11 & S-1808 & HASEKO CORP. & Capital Goods & CG \\ \hline
12 & S-1812 & KAJIMA CORP. & Capital Goods & CG \\ \hline
13 & S-1925 & DAIWA HOUSE IND. CO., LTD. & Capital Goods & CG \\ \hline
14 & S-1928 & SEKISUI HOUSE, LTD. & Capital Goods & CG \\ \hline
15 & S-1963 & JGC CORP. & Capital Goods & CG \\ \hline
16 & S-5631 & THE JAPAN STEEL WORKS, LTD. & Capital Goods & CG \\ \hline
17 & S-6103 & OKUMA CORP. & Capital Goods & CG \\ \hline
18 & S-6113 & AMADA HOLDINGS CO., LTD. & Capital Goods & CG \\ \hline
19 & S-6301 & KOMATSU LTD. & Capital Goods & CG \\ \hline
20 & S-6302 & SUMITOMO HEAVY IND., LTD. & Capital Goods & CG \\ \hline
21 & S-6305 & HITACHI CONST. MACH. CO., LTD. & Capital Goods & CG \\ \hline
22 & S-6326 & KUBOTA CORP. & Capital Goods & CG \\ \hline
23 & S-6361 & EBARA CORP. & Capital Goods & CG \\ \hline
24 & S-6366 & CHIYODA CORP. & Capital Goods & CG \\ \hline
25 & S-6367 & DAIKIN INDUSTRIES, LTD. & Capital Goods & CG \\ \hline
26 & S-6471 & NSK LTD. & Capital Goods & CG \\ \hline
27 & S-6472 & NTN CORP. & Capital Goods & CG \\ \hline
28 & S-6473 & JTEKT CORP. & Capital Goods & CG \\ \hline
29 & S-7004 & HITACHI ZOSEN CORP. & Capital Goods & CG \\ \hline
30 & S-7011 & MITSUBISHI HEAVY IND., LTD. & Capital Goods & CG \\ \hline
31 & S-7013 & IHI CORP. & Capital Goods & CG \\ \hline
32 & S-7911 & TOPPAN PRINTING CO., LTD. & Capital Goods & CG \\ \hline
33 & S-7912 & DAI NIPPON PRINTING CO., LTD. & Capital Goods & CG \\ \hline
34 & S-7951 & YAMAHA CORP. & Capital Goods & CG \\ \hline
35 & S-1332 & NIPPON SUISAN KAISHA, LTD. & Consumer Goods & CN \\ \hline
36 & S-2002 & NISSHIN SEIFUN GROUP INC. & Consumer Goods & CN \\ \hline
37 & S-2282 & NH FOODS LTD. & Consumer Goods & CN \\ \hline
38 & S-2501 & SAPPORO HOLDINGS LTD. & Consumer Goods & CN \\ \hline
39 & S-2502 & ASAHI GROUP HOLDINGS, LTD. & Consumer Goods & CN \\ \hline
40 & S-2503 & KIRIN HOLDINGS CO., LTD. & Consumer Goods & CN \\ \hline
41 & S-2531 & TAKARA HOLDINGS INC. & Consumer Goods & CN \\ \hline
42 & S-2801 & KIKKOMAN CORP. & Consumer Goods & CN \\ \hline
43 & S-2802 & AJINOMOTO CO., INC. & Consumer Goods & CN \\ \hline
44 & S-2871 & NICHIREI CORP. & Consumer Goods & CN \\ \hline
45 & S-8233 & TAKASHIMAYA CO., LTD. & Consumer Goods & CN \\ \hline
46 & S-8252 & MARUI GROUP CO., LTD. & Consumer Goods & CN \\ \hline
47 & S-8267 & AEON CO., LTD. & Consumer Goods & CN \\ \hline
48 & S-9602 & TOHO CO., LTD & Consumer Goods & CN \\ \hline
49 & S-9681 & TOKYO DOME CORP. & Consumer Goods & CN \\ \hline
50 & S-9735 & SECOM CO., LTD. & Consumer Goods & CN \\ \hline

\end{tabular}
\end{table}

\begin{table}[]
\centering
\begin{tabular}{|l|l|l|l|l|}
\hline
51 & S-8331 & THE CHIBA BANK, LTD. & Financials & FN \\ \hline
52 & S-8355 & THE SHIZUOKA BANK, LTD. & Financials & FN \\ \hline
53 & S-8253 & CREDIT SAISON CO., LTD. & Financials & FN \\ \hline
54 & S-8601 & DAIWA SECURITIES GROUP INC. & Financials & FN \\ \hline
55 & S-8604 & NOMURA HOLDINGS, INC. & Financials & FN \\ \hline
56 & S-3405 & KURARAY CO., LTD. & Materials & MT \\ \hline
57 & S-3407 & ASAHI KASEI CORP. & Materials & MT \\ \hline
58 & S-4004 & SHOWA DENKO K.K. & Materials & MT \\ \hline
59 & S-4005 & SUMITOMO CHEMICAL CO., LTD. & Materials & MT \\ \hline
60 & S-4021 & NISSAN CHEMICAL IND., LTD. & Materials & MT \\ \hline
61 & S-4042 & TOSOH CORP. & Materials & MT \\ \hline
62 & S-4043 & TOKUYAMA CORP. & Materials & MT \\ \hline
63 & S-4061 & DENKA CO., LTD. & Materials & MT \\ \hline
64 & S-4063 & SHIN-ETSU CHEMICAL CO., LTD. & Materials & MT \\ \hline
65 & S-4183 & MITSUI CHEMICALS, INC. & Materials & MT \\ \hline
66 & S-4208 & UBE INDUSTRIES, LTD. & Materials & MT \\ \hline
67 & S-4272 & NIPPON KAYAKU CO., LTD. & Materials & MT \\ \hline
68 & S-4452 & KAO CORP. & Materials & MT \\ \hline
69 & S-4901 & FUJIFILM HOLDINGS CORP. & Materials & MT \\ \hline
70 & S-4911 & SHISEIDO CO., LTD. & Materials & MT \\ \hline
71 & S-6988 & NITTO DENKO CORP. & Materials & MT \\ \hline
72 & S-5002 & SHOWA SHELL SEKIYU K.K. & Materials & MT \\ \hline
73 & S-5201 & ASAHI GLASS CO., LTD. & Materials & MT \\ \hline
74 & S-5202 & NIPPON SHEET GLASS CO., LTD. & Materials & MT \\ \hline
75 & S-5214 & NIPPON ELECTRIC GLASS CO., LTD. & Materials & MT \\ \hline
76 & S-5232 & SUMITOMO OSAKA CEMENT CO., LTD. & Materials & MT \\ \hline
77 & S-5233 & TAIHEIYO CEMENT CORP. & Materials & MT \\ \hline
78 & S-5301 & TOKAI CARBON CO., LTD. & Materials & MT \\ \hline
79 & S-5332 & TOTO LTD. & Materials & MT \\ \hline
80 & S-5333 & NGK INSULATORS, LTD. & Materials & MT \\ \hline
81 & S-5706 & MITSUI MINING \& SMELTING CO. & Materials & MT \\ \hline
82 & S-5707 & TOHO ZINC CO., LTD. & Materials & MT \\ \hline
83 & S-5711 & MITSUBISHI MATERIALS CORP. & Materials & MT \\ \hline
84 & S-5713 & SUMITOMO METAL MINING CO., LTD. & Materials & MT \\ \hline
85 & S-5714 & DOWA HOLDINGS CO., LTD. & Materials & MT \\ \hline
86 & S-5715 & FURUKAWA CO., LTD. & Materials & MT \\ \hline
87 & S-5801 & FURUKAWA ELECTRIC CO., LTD. & Materials & MT \\ \hline
88 & S-5802 & SUMITOMO ELECTRIC IND., LTD. & Materials & MT \\ \hline
89 & S-5803 & FUJIKURA LTD. & Materials & MT \\ \hline
90 & S-5901 & TOYO SEIKAN GROUP HOLDINGS, LTD. & Materials & MT \\ \hline
91 & S-3865 & HOKUETSU KISHU PAPER CO., LTD. & Materials & MT \\ \hline
92 & S-3861 & OJI HOLDINGS CORP. & Materials & MT \\ \hline
93 & S-5101 & THE YOKOHAMA RUBBER CO., LTD. & Materials & MT \\ \hline
94 & S-5108 & BRIDGESTONE CORP. & Materials & MT \\ \hline
95 & S-5401 & NIPPON STEEL \& SUMITOMO METAL CORP. & Materials & MT \\ \hline
96 & S-5406 & KOBE STEEL, LTD. & Materials & MT \\ \hline
97 & S-5541 & PACIFIC METALS CO., LTD. & Materials & MT \\ \hline
98 & S-3101 & TOYOBO CO., LTD. & Materials & MT \\ \hline
99 & S-3103 & UNITIKA, LTD. & Materials & MT \\ \hline
100 & S-3401 & TEIJIN LTD. & Materials & MT \\ \hline

\end{tabular}
\end{table}

\begin{table}[]
\centering
\begin{tabular}{|l|l|l|l|l|}
\hline
101 & S-3402 & TORAY INDUSTRIES, INC. & Materials & MT \\ \hline
102 & S-8001 & ITOCHU CORP. & Materials & MT \\ \hline
103 & S-8002 & MARUBENI CORP. & Materials & MT \\ \hline
104 & S-8015 & TOYOTA TSUSHO CORP. & Materials & MT \\ \hline
105 & S-8031 & MITSUI \& CO., LTD. & Materials & MT \\ \hline
106 & S-8053 & SUMITOMO CORP. & Materials & MT \\ \hline
107 & S-8058 & MITSUBISHI CORP. & Materials & MT \\ \hline
108 & S-4151 & KYOWA HAKKO KIRIN CO., LTD. & Pharmaceuticals & PH \\ \hline
109 & S-4503 & ASTELLAS PHARMA INC. & Pharmaceuticals & PH \\ \hline
110 & S-4506 & SUMITOMO DAINIPPON PHARMA CO., LTD. & Pharmaceuticals & PH \\ \hline
111 & S-4507 & SHIONOGI \& CO., LTD. & Pharmaceuticals & PH \\ \hline
112 & S-4519 & CHUGAI PHARMACEUTICAL CO., LTD. & Pharmaceuticals & PH \\ \hline
113 & S-4523 & EISAI CO., LTD. & Pharmaceuticals & PH \\ \hline
114 & S-7201 & NISSAN MOTOR CO., LTD. & Information Technology & IT \\ \hline
115 & S-7202 & ISUZU MOTORS LTD. & Information Technology & IT \\ \hline
116 & S-7205 & HINO MOTORS, LTD. & Information Technology & IT \\ \hline
117 & S-7261 & MAZDA MOTOR CORP. & Information Technology & IT \\ \hline
118 & S-7267 & HONDA MOTOR CO., LTD. & Information Technology & IT \\ \hline
119 & S-7270 & SUBARU CORP. & Information Technology & IT \\ \hline
120 & S-7272 & YAMAHA MOTOR CO., LTD. & Information Technology & IT \\ \hline
121 & S-3105 & NISSHINBO HOLDINGS INC. & Information Technology & IT \\ \hline
122 & S-6479 & MINEBEA MITSUMI INC. & Information Technology & IT \\ \hline
123 & S-6501 & HITACHI, LTD. & Information Technology & IT \\ \hline
124 & S-6502 & TOSHIBA CORP. & Information Technology & IT \\ \hline
125 & S-6503 & MITSUBISHI ELECTRIC CORP. & Information Technology & IT \\ \hline
126 & S-6504 & FUJI ELECTRIC CO., LTD. & Information Technology & IT \\ \hline
127 & S-6506 & YASKAWA ELECTRIC CORP. & Information Technology & IT \\ \hline
128 & S-6508 & MEIDENSHA CORP. & Information Technology & IT \\ \hline
129 & S-6701 & NEC CORP. & Information Technology & IT \\ \hline
130 & S-6702 & FUJITSU LTD. & Information Technology & IT \\ \hline
131 & S-6703 & OKI ELECTRIC IND. CO., LTD. & Information Technology & IT \\ \hline
132 & S-6752 & PANASONIC CORP. & Information Technology & IT \\ \hline
133 & S-6758 & SONY CORP. & Information Technology & IT \\ \hline
134 & S-6762 & TDK CORP. & Information Technology & IT \\ \hline
135 & S-6770 & ALPS ELECTRIC CO., LTD. & Information Technology & IT \\ \hline
136 & S-6773 & PIONEER CORP. & Information Technology & IT \\ \hline
137 & S-6841 & YOKOGAWA ELECTRIC CORP. & Information Technology & IT \\ \hline
138 & S-6902 & DENSO CORP. & Information Technology & IT \\ \hline
139 & S-6952 & CASIO COMPUTER CO., LTD. & Information Technology & IT \\ \hline
140 & S-6954 & FANUC CORP. & Information Technology & IT \\ \hline
141 & S-6971 & KYOCERA CORP. & Information Technology & IT \\ \hline
142 & S-6976 & TAIYO YUDEN CO., LTD. & Information Technology & IT \\ \hline
143 & S-7752 & RICOH CO., LTD. & Information Technology & IT \\ \hline
144 & S-8035 & TOKYO ELECTRON LTD. & Information Technology & IT \\ \hline
145 & S-4543 & TERUMO CORP. & Information Technology & IT \\ \hline
146 & S-4902 & KONICA MINOLTA, INC. & Information Technology & IT \\ \hline
147 & S-7731 & NIKON CORP. & Information Technology & IT \\ \hline
148 & S-7733 & OLYMPUS CORP. & Information Technology & IT \\ \hline
149 & S-7762 & CITIZEN WATCH CO., LTD. & Information Technology & IT \\ \hline
150 & S-9501 & TOKYO ELECTRIC POWER COMPANY HOLDINGS, I & Transportation \& Utilities & TU \\ \hline

\end{tabular}
\end{table}

\begin{table}[]
\centering
\begin{tabular}{|l|l|l|l|l|}
\hline
151 & S-9502 & CHUBU ELECTRIC POWER CO., INC. & Transportation \& Utilities & TU \\ \hline
152 & S-9503 & THE KANSAI ELECTRIC POWER CO., INC. & Transportation \& Utilities & TU \\ \hline
153 & S-9531 & TOKYO GAS CO., LTD. & Transportation \& Utilities & TU \\ \hline
154 & S-9532 & OSAKA GAS CO., LTD. & Transportation \& Utilities & TU \\ \hline
155 & S-9062 & NIPPON EXPRESS CO., LTD. & Transportation \& Utilities & TU \\ \hline
156 & S-9064 & YAMATO HOLDINGS CO., LTD. & Transportation \& Utilities & TU \\ \hline
157 & S-9101 & NIPPON YUSEN K.K. & Transportation \& Utilities & TU \\ \hline
158 & S-9104 & MITSUI O.S.K.LINES, LTD. & Transportation \& Utilities & TU \\ \hline
159 & S-9107 & KAWASAKI KISEN KAISHA, LTD. & Transportation \& Utilities & TU \\ \hline
160 & S-9001 & TOBU RAILWAY CO., LTD. & Transportation \& Utilities & TU \\ \hline
161 & S-9005 & TOKYU CORP. & Transportation \& Utilities & TU \\ \hline
162 & S-9007 & ODAKYU ELECTRIC RAILWAY CO., LTD. & Transportation \& Utilities & TU \\ \hline
163 & S-9008 & KEIO CORP. & Transportation \& Utilities & TU \\ \hline
164 & S-9009 & KEISEI ELECTRIC RAILWAY CO., LTD. & Transportation \& Utilities & TU \\ \hline
165 & S-9301 & MITSUBISHI LOGISTICS CORP. & Transportation \& Utilities & TU \\ \hline

\end{tabular}
\end{table}

\begin{table}[]
\caption{List of major crashes and bubbles for USA and JPN markets and their characterization~\cite{list,bullmarkets,ushousing,history_crashes,selloff}. 
}%
\begin{small}

\begin{tabularx}{\linewidth}{|s| b| m| m|}
		\hline
		\textbf{Sl. No} & \textbf{Major crashes and type-1}       & \textbf{Period Date} & \textbf{{Region Affected}} \\ \hline
		1              & Black Monday                             & 19-10-1987      & USA,JPN                 \\ \hline
		2              & Friday the 13th Mini Crash              & 13-10-1989      & USA                     \\ \hline
		3              & Early 90s Recession                      & 1990            & USA                     \\ \hline
		5              & Mini Crash Due To Asian Financial Crisis & 27-10-1997      & USA                     \\ \hline
		6              & Lost Decade                              & 2001-2010       & JPN                     \\ \hline
		7             & 9/11 Financial Crisis                    & 11-09-2001      & USA,JPN                 \\ \hline
		8             & Stock Market Downturn Of 2002            & 09-10-2002      & JPN,USA                 \\ \hline
		9             & US Housing Bubble                        & 2005-2007       & USA                     \\ \hline
		10             & Lehman Brothers Crash                    & 16-09-2008      & USA,JPN                 \\ \hline
		11             & DJ Flash Crash                           & 06-05-2010      & USA,JPN                 \\ \hline
		12             & Tsunami/Fukushima                        & 11-03-2011      & JPN                     \\ \hline
		13             & August 2011 Stock Markets Fall           & 08-08-2011      & USA,JPN                 \\ \hline
		14             & Chinese Black Monday and 2015-2016 Sell Off                        & 24-08-2015      & USA                     \\ \hline
	\end{tabularx}
	\end{small}
	\label{table:crashes}
\end{table}



\end{document}